\documentclass[11pt,a4paper]{article}
\pdfoutput=1
\usepackage{jheppub}

\usepackage{amsmath,epsfig}
\usepackage{amssymb,amsfonts}
\usepackage{latexsym}
\usepackage[latin1]{inputenc}
\usepackage{slashed}
\usepackage{empheq}
\numberwithin{equation}{section}
\usepackage{dsfont}
\usepackage{cancel}
\usepackage{overpic}
\usepackage{subcaption}
\usepackage{subeqnarray}
\usepackage{xcolor}

\usepackage{graphicx}
\usepackage{longtable}
\usepackage{color}

\newcommand{\exclude}[1]{}
\def\a#1{\alpha_{#1}}

\def\beq{\begin{equation}}
\def\eeq{\end{equation}}
\def\be{\begin{equation}}
\def\ee{\end{equation}}
\def\bea{\begin{eqnarray}}
\def\eea{\end{eqnarray}}
\def\bal{\begin{align}}
\def\eal{\end{align}}

\def\2b2[#1,#2][#3,#4]{\left( \begin{array}{cc} #1 & #2 \\ #3 & #4 \end{array}
\right)}
\def\3b3[#1,#2,#3][#4,#5,#6][#7,#8,#9]{\left( \begin{array}{ccc} #1 & #2 #3 \\
#4 & #5 & #6\\#7&#8&#9\end{array} \right)}

\newcommand\fverb{\setbox\pippobox=\hbox\bgroup\verb}
\newcommand\fverbdo{\egroup\medskip\noindent%
                        \fbox{\unhbox\pippobox}\ }
\newcommand\fverbit{\egroup\item[\fbox{\unhbox\pippobox}]}

\newcommand{\bear}{\begin{eqnarray}}

\newcommand{\eear}{\end{eqnarray}}
\newcommand{\de}{\partial}
\newcommand{\bsea}{\begin{subeqnarray}}
\newcommand{\esea}{\end{subeqnarray}}
\newbox\pippobox

\def\f{\varphi}

\def\g{\gamma}

\def\6{\partial}
\def\a{\alpha}

\def\half{\frac12}

\def\sq
\def\a{\alpha}

\title{De Sitter and Anti-de Sitter branes in self-tuning models}

\author[\natural]{J.~K.~Ghosh,}
\author[\flat, \natural]{E.~Kiritsis,}
\author[\natural]{F.~Nitti,}
\author[\natural]{L.~T.~Witkowski}

\affiliation[\natural]{\href{http://www.apc.univ-paris7.fr}{APC, AstroParticule et Cosmologie}, Universit\'e Paris Diderot, CNRS/IN2P3, CEA/IRFU, \\
Observatoire de Paris, Sorbonne Paris Cit\'e,\\
 10, rue Alice Domon et L\'eonie Duquet, 75205 Paris
Cedex 13, France}
\affiliation[\flat]{\href{http://hep.physics.uoc.gr}{Crete Center for Theoretical Physics},
Department of Physics,\\
University of Crete, 71003 Heraklion, Greece}

\preprint{CCTP-2018-9\\
\hphantom{AAAAAAAAAAAAAAAAAAAAAAAAAAAAAAAAAAAAAAAAllll} ITCP-IPP 2018/7}

\abstract{Maximally symmetric curved-brane solutions are studied in dilatonic
  braneworld models which realise the self-tuning of the effective
  four-dimensional cosmological constant. It is found that  no vacua
  in which the brane has de Sitter or anti-de Sitter geometry exist,
  unless one modifies the near-boundary asymptotics of the
  bulk fields. In the holographic dual picture, this
  corresponds to coupling the UV CFT to a curved metric (possibly with a defect). Alternatively, the same  may be  achieved in a flat-space QFT with
  suitable variable scalar sources. With these ingredients, it is found that maximally symmetric,
  positive and negative curvature solutions with a stabilised brane
  position generically exist. The space of such solutions is studied in two
  different types of realisations of the self-tuning framework.
  In some regimes we observe a large hierarchy between the curvature
  on the brane and the boundary UV CFT curvature. This  is a dynamical
  effect due to the self-stabilisation mechanism. This setup provides
  an alternative route to realising de Sitter space in string theory.}

\begin{document}
\maketitle
\section{Introduction}
Soon after its first  introduction in \cite{9711200,9802150,9802109},
it has been clear that the holographic gauge/gravity  duality is intimately linked to a new way of
thinking about modified gravity and beyond the standard model
phenomenology which was  being developed around the same time: the
idea of the braneworld \cite{AADD,Randall:1999ee,Randall:1999vf}. In these
models, several problems of the Standard Model or its high-energy
completion were addressed by postulating that the
observed particles and fields are confined to a four-dimensional
hypersurface ({\em brane}) embedded in a higher-dimensional space-time
({\em bulk}). The connection to holography, observed in
\cite{malda,Gubser,ArkaniHamed:2000ds}, stems from the fact that the
bulk was often taken to be (a portion of) Anti-de Sitter space, which
may be given  a dual interpretation in terms of a strongly coupled,
large-$N$, four-dimensional  field theory. Since then, holographic
duality and braneworld phenomenology have often been two complementary
sides of model building.

One of the earliest applications of braneworlds in this context was
aimed at addressing the various naturalness problems which afflict the
Standard Model and General Relativity, both from the particle physics
side (electroweak hierarchy problem) and from the cosmology side (cosmological
constant problem(s)).

On the one hand, cosmological applications  led  to
departures from solutions describing  a static flat brane in a static
bulk, and  prompted the study of braneworld cosmology
\cite{Binetruy:1999ut,CGS,Binetruy:1999hy}. The connection between braneworld
models and gravity modifications was proposed as a way to model the observed
current acceleration of the universe
\cite{Deffayet:2000uy,Deffayet:2001pu}.

On the other
hand, it was proposed to use braneworld models to resolve the clash between
the huge vacuum energy resulting from quantum effective  field
theory calculations and the smallness of the observed cosmological
constant of the current de Sitter-like epoch. These proposal aimed at
realising a {\em self-tuning}
mechanism, first proposed in \cite{RS1,RS2}, in which, contrary to purely four-dimensional models, the
vacuum energy from quantum loops has no effect on the curvature
of the brane, which is perceived as (almost) flat by four-dimensional observers \cite{ArkaniHamed:2000eg,Kachru:2000hf}.

Although  in principle appealing,
the models which were proposed at the time all had issues related to
the apparent inevitability of naked singularities in the  bulk and/or
an impossibility to have both a successful self-tuning mechanism and
the existence of an effective  four-dimensional gravity on the brane\cite{Csaki:2000wz}.

Recently, a novel framework  was developed, which revisits the self-tuning
braneworld approach \cite{1704.05075} and which uses  holography as a guiding
principle for model building.
It  consists  of a general
two-derivative Einstein-dilaton  bulk theory,  and a codimension-one
brane whose effective world-volume action contains all possible
two-derivative terms (namely a brane potential for the scalar, an
induced  kinetic term  for the scalar  and an Einstein-Hilbert term for the induced
metric) preserving four-dimensional diffeomorphism
invariance.
The bulk action is expected to be dual to a strongly coupled, large-N four-dimensional QFT, while the brane action is expected to contain the Standard Model fields as its localized fluctuations.

In the spirit of {\em semi-holography} (see e.g.~\cite{Faulkner:2010tq}),  asymptotically anti-de Sitter solutions of this theory
are interpreted as a purely
four-dimensional theory in which the bulk geometry is dual to a strongly interacting  UV
conformal field theory (CFT), deformed by a relevant operator (dual
to the
bulk scalar) and coupled  to a weakly interacting  Standard Model sector (the
brane), a setup whose dual version was advocated in \cite{smgrav}.

In the model described above, the brane separates the bulk geometry into
two regions:  one side  connects to an  asymptotically AdS conformal boundary  (UV
of the field theory dual). On the other side of the brane, the
geometry may flow to another, regular, asymptotically AdS region  (in which
case the field theory flows to an IR conformal fixed point).\footnote{It may also have a mild (resolvable) naked singularity, according to the Gubser criterion.} The two sides of the geometry must  obey the bulk
Einstein-dilaton equations and the connection across the brane must
satisfy Israel's junction conditions.
Induced  four-dimensional gravity on the brane is
recovered in a range of distance scales  via the DGP mechanism
\cite{Dvali:2000rv,Dvali:2000hr} thanks to the localized
Einstein-Hilbert term in the brane action.\footnote{This term is generated via quantum effects of the brane localized fields.}

The use of holography as a guideline for
model-building, in order  to organise the space of solutions, has allowed to solve
or alleviate some of the difficulties of the earlier models. In particular,
 holography can give  a consistent  meaning to certain kinds of
 bulk singularities \cite{0002160}. These  are indeed necessary to
 construct holographic duals of confining theories \cite{0707.1324,0707.1349,GN},
 and they may be consistently eliminated by uplifting to higher
 dimensions \cite{1107.2116,Gouteraux:2011qh}.

As it was shown in \cite{1704.05075}, for rather generic choices of
the bulk and brane potentials,  enforcing the holographic
interpretation of the model results in a self-tuning mechanism for the
four-dimensional  cosmological constant.  The model admits solutions
in which the geometry on the brane is flat, regardless of the vacuum energy
arising from quantum loops of the brane fields. The brane is
stabilized in the bulk at an equilibrium position, which  is
dynamically determined by the bulk geometry and brane potentials via Israel's
junction conditions. Under certain general conditions, all
fluctuations around the equilibrium position have positive energy.

The fact that the framework proposed in \cite{1704.05075} allows
self-tuning flat solutions opens new
questions, and at the same time offers new possibilities for model
building.  In \cite{1704.05075}, brane flatness and  four-dimensional
Poincar\'e invariance were imposed by design on the solution ansatz,
and the self-tuning mechanism corresponds to the existence of
stabilized solutions with this symmetry.
It is important however to explore, in the same context, other
solutions  in which the brane has non-zero curvature and/or has a
time-dependent (cosmological) induced metric. One reason is to
understand  how these solutions compete with the flat solution (which represents the
Poincar\'e-invariant vacuum). In addition, because we currently live
in an accelerating universe, obtaining a positively curved
(e.g.~de Sitter) metric  on a brane is phenomenologically important. Finally, it is important to clarify what is responsible, from the
dual field theory perspective, for obtaining a curved brane geometry.
Exploring these questions  is the purpose of the present work.

In this paper, we look for solutions of the self-tuning framework in
which the brane has a curved geometry.  Our first result can be
formulated as follows: no {\em vacuum} curved-brane solutions
generically exist\footnote{By ``vacuum'' solution here we mean one where the bulk
  has 4d Poincar\'e invariance,
    representing the ground state of the dual QFT.}. A non-trivial
  brane geometry can be obtained if  one modifies the UV boundary
  conditions on the bulk fields, such that they allow  domain wall solutions with
  constant curvature radial slices.   In the dual QFT language this amounts
  to changing the dual QFT. As we shall see, a de Sitter brane geometry
   will be possible when the domain wall solution is sliced by de
   Sitter slices. In this case, the  bulk geometry is dual to  a QFT defined on a constant positive
  curvature manifold. For an AdS brane geometry, the bulk geometry is
  a  domain wall with negative curvature  slices, and the UV field
  theory is a CFT with an additional defect
  \cite{skenderisham,Clark:2004sb}.

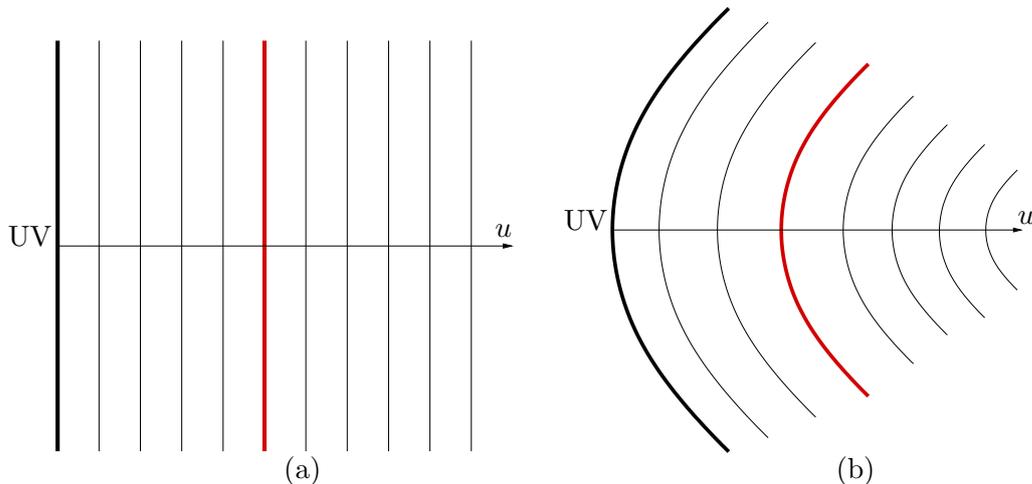
\begin{figure}[t]
\centering
\begin{overpic}
[width=0.4\textwidth]{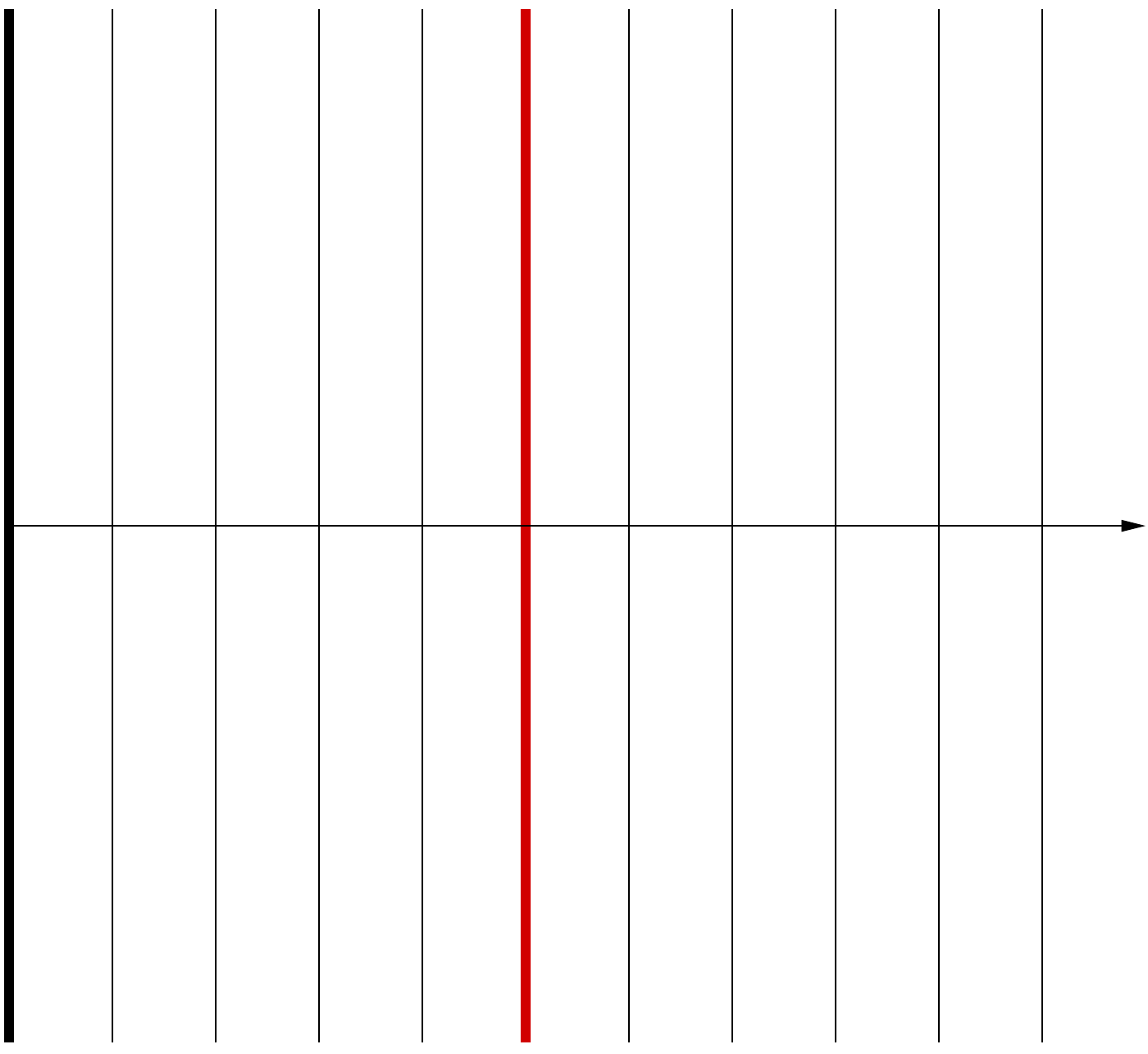}
\put(96,48){$u$}
\put(-10,46){UV}
\put(50,-5){(a)}
\end{overpic}
\hspace{1cm}
\begin{overpic}
[width=0.36\textwidth]{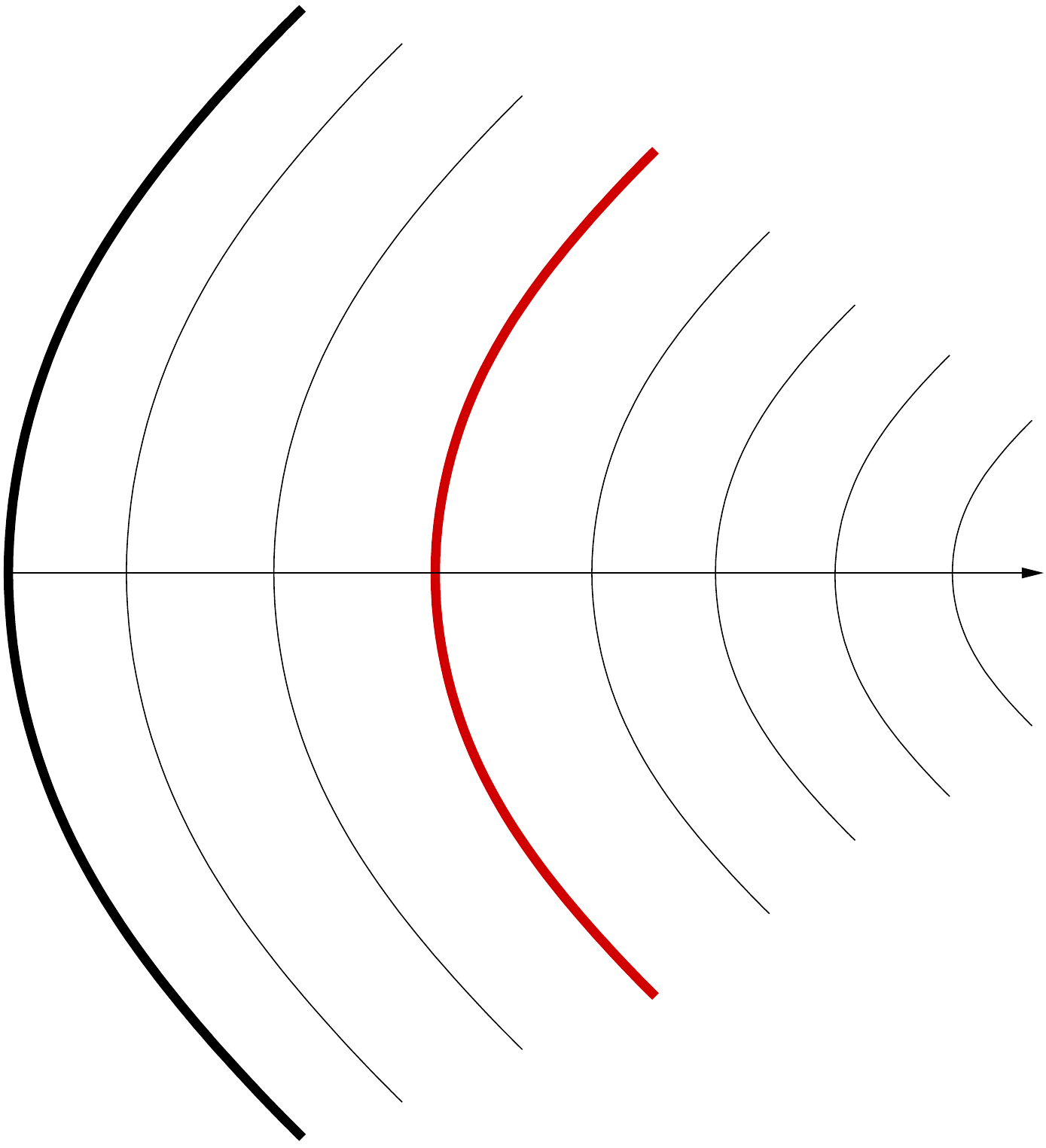}
\put(90,51.5){$u$}
\put(-10,50){UV}
\put(50,-5){(b)}
\end{overpic}
\hspace{0.5cm}
\caption{A sketch of the solutions allowing for a flat brane (a) and those
  allowing for a curved brane (b).  The thick black line represents the
  conformal boundary of AdS, whereas the brane is represented in
  red. The direction $u$ is the holographic direction, and the thin
  black lines are constant-$u$ hypersurfaces.
 The brane metric is inherited from the UV boundary metric (up to a
 rescaling).}
\label{fig:slicings}
\end{figure}

Prompted by the above result, we set forward to study  curved domain
wall geometries, and we ask the
question whether stabilized curved brane solutions do arise.
We  focus in particular on
maximally symmetric geometries,  in which both the bulk and the brane
preserve four-dimensional de Sitter (dS) or Anti-de Sitter (AdS)
invariance. The structure of these
solutions is sketched in figure \ref{fig:slicings}, in which we show a
comparison between the flat solutions studied in \cite{1704.05075}
and the curved embeddings we discuss in this paper.

As far as the  bulk is concerned, the corresponding
solutions describe holographic RG flows  on maximally symmetric
spaces, and they were studied extensively in \cite{Ghosh:2017big},
to which the reader is referred for details. The introduction of the
brane amounts to gluing together two bulk RG-flow solutions of the
type described in \cite{Ghosh:2017big}, to impose regularity on the
IR side of the solution and to solve at the same time for the brane position and the UV part of the geometry in such a way that Israel's
junction conditions are satisfied.

In these solutions, the
metric on the brane is the same (up to a scaling factor, which depends
on the position in  the bulk)  as the UV metric to which the dual CFT
is coupled. For example, a  de Sitter brane solution  can exist only
if the dual CFT is set on de Sitter space, and similarly for Anti-de
Sitter. In the latter case, the holographic interpretation is more
subtle because, in addition to the usual boundary conditions in the
radial direction, one has also to introduce boundary conditions at the
boundary of the slices. As discussed in \cite{Ghosh:2017big} (see also
\cite{skenderisham} for a previous similar discussion) this introduces a codimension-one defect  in the dual field theory.

Although the type of geometry of the brane is fixed by the boundary
conditions, the magnitude of the brane curvature and its position in
the bulk are determined dynamically by the field equations and the
junction conditions. We refer to this as {\em self-stabilisation},
which  is the curved counterpart of the self-tuning
mechanism found in  \cite{1704.05075}.\footnote{We reserve the phrase {\em self-tuning} for the
  flat solutions, in which the effective cosmological constant on the
  brane is dynamically set to zero.}

There is an alternative realisation of the same solutions,
which can be obtained via a bulk coordinate transformation, in
which the leading UV asymptotics correspond instead to a flat metric,
but the scalar sources are varying in space or time.
This leads to an inequivalent description in
terms of the dual QFT: instead of a boundary QFT living
on a curved space-time, we have a flat-space  QFT driven by a time-varying (in the dS case) or
space-dependent (in the AdS case) source.
The two descriptions however result in the same brane
geometry. Although in the bulk the two solutions are related by a
coordinate transformation, the latter acts non-trivially on the
boundary, and it leads therefore to  an inequivalent theory with
different boundary sources. In most of the paper we will work with  the
curved-CFT description with constant scalar sources.

The solutions we study are extensions to general dilatonic braneworld models,
with general induced gravity and kinetic terms, of the
curved braneworld solutions first introduced by Kaloper in
\cite{Kaloper:1999sm} and  by
Karch and Randall in
\cite{KarchRandall}, in which the embedding of a
curved brane whose tension was de-tuned from the RS \cite{Randall:1999vf} value was first studied. Further generalisation,
which considered asymmetric setups and/or included an induced
Einstein-Hilbert term, were explored e.g.~in
\cite{Padilla:2004tp,Padilla:2004mc,Gergely:2004ax,Koyama:2005br,Charmousis:2007ji}
and  recently revisited  in \cite{Banerjee:2018qey}.
The main difference between those models and the ones studied
here is the effect of self-stabilisation: both the bulk solution and the
brane geometry are determined dynamically,  by the interplay between
the running of the bulk scalar and the brane  potential and kinetic
terms. As a  consequence,  any hierarchy which is produced between the
CFT curvature (the UV  boundary conditions)  and the brane curvature
has a dynamical origin.
Also, in previous works,
the two sides of the  bulk geometry were often taken  to be pure AdS
space-times with different curvatures.  In our case, instead, both sides of the geometry are different solutions
of the same bulk theory. This  allows for  a consistent holographic
interpretation.
Finally, the presence of the scalar field and the self-stabilisation mechanism
are  important for perturbative stability, which was shown to be a
problem for the pure-gravity models  \cite{Charmousis:2007ji,Koyama:2009cm}.

After a general discussion, we focus on two specific models, which were
those presented in \cite{1704.05075} as concrete realisations of the
self-tuning mechanism. They share a similar UV structure but differ in
the IR, as the first one admits a  regular IR fixed point where the
dilaton reaches a finite value, while the
second one has a ``good'' IR singularity where the dilaton diverges.  Although the former model has the nice property of having a
regular interior, it is the latter which is more promising for
phenomenology, as it is more suitable to implement the self-tuning
mechanism without violating the conditions for stability of the flat equilibrium
solution. In the curved case we study here, the difference between
these IR geometries gives rise to different phenomenology of the
stabilisation mechanism, and the corresponding scaling relations
between brane, bulk, and boundary curvatures.

The phenomenology of self-stabilisation in the various examples may
differ from case to case, though a few features are worth
mentioning. Generically, we find that one or more solutions exist for
both positive and negative curvature, with one branch always
connected to the flat self-tuning solution. Along this branch, the brane
curvature can be continuously tuned to zero by decreasing the value of
the UV curvature.  Interestingly, we find situations where the brane
curvature is bounded by a maximum value,  no matter how large  one takes the
boundary curvature.

From the phenomenological perspective, our approach is relevant  to the
general discussion of how to find  de Sitter (or more generally,
accelerating) solutions in string theory,
which recently has received renewed attention. The reason is that, as argued first in \cite{HV} the braneworld setups described here can arise in warped compactifications of string theory.
  On the one hand,
traditional methods based on compactifications from ten dimensions,
with various combinations of brane and fluxes, have a hard time realizing dS space, \cite{Danielsson:2018ztv,Andriot}. On  the other
hand, it has been recently argued that de Sitter solutions are
forbidden in a quantum gravity theory by  generalisations of the
weak gravity conjecture \cite{Obied:2018sgi}.
The class of solutions we find here in the context of holography offer
an alternative way of producing de Sitter, on the brane,
compared to methods based on engineering  bulk solutions, as was
recently advocated in \cite{Danielsson:2018ztv}.  In  this sense our results agree with
the general arguments that de Sitter cannot arise as a vacuum solution:
rather, as it is found here, it must be driven by a non-trivial UV coupling of a CFT to
curved  metric  sources.

As a final remark, we note that more general curved-brane
  solutions in which, in the bulk, both the scale factor and the dilaton have a
  non-trivial time-dependence are expected to exist, even without
  departing from UV asymptotics  corresponding to a flat UV CFT
  metric. These solutions however are expected to
  describe a cosmological, non-vacuum state. Generically, these will
  not describe exact de
  Sitter space, although in some regions of parameter space they may have a quasi-de Sitter regime
  (similar to slow-roll inflation or quintessence).
 Their study is needed to answer the
  important question regarding time-evolution  (and approach to
  equilibrium) and cosmology of the self-tuning model. This  will be
  the subject of future work \cite{Branecosmo}.

This paper  is organized as follows. In  Section \ref{sec:general} we
introduce the setup, present the general results,  review the bulk
geometries corresponding to curved holographic RG flows, discuss
the junction conditions for curved branes, and the connection with
the flat CFT description. The rest of the paper is
devoted to exploring the space of constant positive or negative
curvature solutions in specific models, and the results are mostly obtained numerically. In Section  \ref{caseI} we look for curved
brane solutions
in a  model with a conformal IR fixed point. In  Section
\ref{sec:numexpV} we turn to models with an asymptotically exponential
dilaton potential, and study solutions in which the brane is
stabilized at a large value of the dilaton field. Several technical
details are left to the Appendices.

\section{A curved brane in a warped bulk} \label{sec:general}
We will consider Einstein-scalar theory in $d+1$ dimensions, coupled
to a $d$-dimensional dynamical hypersurface ({\em brane}).  The bulk
space-time is parametrized by coordinates $(u,x^\mu)$ and we consider
both the Euclidean and Lorentzian metric, for which we take the
signature $(-,+,+,\cdots,+)$ for the $x^\mu$ coordinates.  We will
work with the  most general two derivative action for this set-up:
\begin{equation}
S=S_{bulk}+S_{brane}\ ,
\end{equation}
where
\begin{equation}
S_{bulk}[g,\f]= M_P^{d-1} \int du \ d^dx \sqrt{-g}
\left(R^{(g)}-\frac{1}{2} \partial_A \f \partial^A \f -V(\f)\right)+
S_{GHY} + S_{ct}\, , \label{sbulk}
\end{equation}
\begin{equation}
S_{brane}=M_P^{d-1} \int d^dx \sqrt{-\gamma} \left(-W_B(\f)-\frac{1}{2} Z(\f) \gamma^{\mu\nu} \partial_\mu \f \partial_\nu \f+U(\f) R_{B} \right). \label{sbrane}
\end{equation}
Here,  $g_{AB}$ is the bulk metric, $R^{(g)}$  the bulk Ricci
scalar, $\gamma_{\mu\nu}$  the induced metric on  the brane, $R_{B}$
the corresponding Ricci scalar,  $V(\f)$ is the bulk potential, $S_{GHY}$ is the Gibbons-Hawking-York
term, and $S_{ct}$ a boundary counterterm action needed for
holographic renormalisation, and whose details here are
unimportant.
The  bulk solution and the  brane embedding in the bulk are determined
by the bulk Einstein's equations and by imposing
Israel's junction conditions across the co-dimension-one brane.

The framework above was considered in \cite{1704.05075} as a way to describe, in
a gravity dual language, the interaction between  weakly coupled
physics (e.g. the Standard Model) localized on the brane, and a
strongly coupled, large $N$ CFT, described by the bulk geometry. In this context, the
functions $W_B(\f), Z(\f)$ and $U(\f)$ may be thought as
 generated by integrating out  the brane-localized fields.
In particular the function $W_B(\f)$ contains contributions from the brane vacuum energy.

It was shown in \cite{1704.05075} that,
rather generically, this kind of models  allow self-tuning solutions,
in which the brane geometry is flat, regardless of the value of the brane vacuum
energy. In that work, the bulk geometry enjoyed
four-dimensional Poincar\'e invariance of constant-$u$ hypersurfaces,
which was inherited by the brane.

In this work, we will  move beyond flat brane solutions, and ask the
question, what kind of non-trivial brane geometries one can obtain
within the same framework. After some general considerations, we will
then restrict our attention to {\em constant curvature} brane
geometries, i.e.~either de Sitter or Anti-de Sitter. We will not study
the most general solution of the model specified by the action \eqref{sbulk}--\eqref{sbrane}, but we will
restrict to situations in which the bulk is static, leaving more
general time-dependent geometries for future work.

The simplest possibility to move in this direction is to look for a
solution in which the curvature is due solely to the  embedding of the
brane, and the  bulk geometry retains its four-dimensional Poincar\'e
invariance. In this ansatz, the boundary conditions (which define the
dual CFT data) are the same\footnote{In
the dual CFT language such a solution  would  constitute an
alternative state of the same theory which gave the self-tuning
vacuum. } as those studied in \cite{1704.05075}.
As we will show in section \ref{no-flat} however, for generic
bulk and brane potentials,   no solution of this kind exists. This
leads us to generalize the bulk ansatz, in a way described  in section
\ref{RG}.

In order to find a curved brane
embedding in a static bulk we will need to modify the metric
asymptotics. In the dual CFT language this means that  we have to
couple the boundary field theory to a non-trivial metric
$\zeta_{\mu\nu}^{UV}$. In the gravity dual,  the simplest ansatz
describing this situation while keeping the solution
static\footnote{In a sense which we will specify more precisely below.} takes the form
\begin{equation} \label{metric}
\f=\f(u), \quad \quad ds^2= du^2+ e^{2A(u)}\zeta_{\mu\nu}dx^\mu dx^\nu\ .
\end{equation}
where $A(u)$ is the warp factor and $\zeta_{\mu\nu}$ a fiducial
$d$-dimensional metric.   The ansatz (\ref{metric})  is the
simplest static solution such that a brane embedded at a fixed  $u$ has a
non-trivial world-volume curvature. Even though the metric
$\zeta_{\mu\nu}$ may depend explicitly on $t$ (e.g.~it may be a
$d$-dimensional FRW metric), we still call this metric static because
the functions to be solved for (namely $A(u)$ and $\f(u)$) depend only
on the holographic coordinates and not on time.
More complicated bulk solutions are possible, and we will comment briefly on them.

We will assume that the bulk  has an asymptotic near-boundary region
(which we can choose to be reached as $u\to-\infty$) where the
solution takes the form of a Fefferman-Graham expansion, in which the
leading term defines the boundary QFT metric $\zeta_{\mu\nu}^{UV}$,
\be \label{asympt}
ds^2 \simeq du^2 + e^{-{2u\over \ell} }\left[\zeta_{\mu\nu}^{UV} +
  \ldots\right] dx^\mu dx^\nu, \qquad u\to -\infty
\ee
If $\zeta_{\mu\nu}^{UV}=\eta_{\mu\nu}$, the equation above is the
leading near-boundary behavior of the metric of the Poincar\'e patch of
$(d+1)$-dimensional  Anti-de Sitter space. For (\ref{asympt})   to hold, it is enough that the bulk potential has a local maximum (say
at $\f=0$) where it takes on a negative  value, $V(0) = -
d(d-1)/\ell^2$. We will give a more detailed description of both the
near-boundary  (UV) and the interior (IR) regions in section \ref{RG}.

 By adjusting a free additive
 constant in $A(u)$  we can always
 identify the fiducial slice metric $\zeta_{\mu\nu}$ in (\ref{metric})  with the
 metric $\zeta_{\mu\nu}^{UV}$ in (\ref{asympt}). Therefore from now on
 we will assume
\be \label{fid}
 \zeta_{\mu\nu} = \zeta_{\mu\nu}^{UV}.
\ee

In general, a co-dimension-one brane
configuration  preserving space-rotations is  described by an
embedding of the form\footnote{For definiteness, here we focus on a
  ``cosmological'' brane, whose induced metric depends non-trivially
  on the time-coordinate $\tau$. Similar considerations apply  a static curved
brane, after trading $\tau$ for one of the space coordinates.} $F(u,\tau) =0$ for some function $F$, or more
explicitly by giving  a trajectory in the holographic direction,
\be
u = u_{\star}(\tau).
\ee
The induced metric on the brane takes the form
\be\label{ind1}
ds^2 = \left[\left({d u_{\star} \over d\tau}\right)^2 + e^{2A(u_{\star}(\tau))}\zeta_{\tau\tau} \right]  d\tau^2 +
e^{2A(u_{\star}(\tau))} \zeta_{ij} dx^idx^j.
\ee

The situation considered in \cite{1704.05075} was the  case of  a
static, flat   brane located   at  $u=u_{\star}$ and separating  two different geometries
(one for $u< u_{\star}$, one for $u>u_{\star}$) of the form (\ref{metric})
with flat slices, $\zeta_{\mu\nu} = \eta_{\mu\nu}$, and different
scale factors. Here, we want to look at more general solutions which
allow for a  {\em curved} brane.

\subsection{(No) Curved brane in a flat-sliced bulk} \label{no-flat}

First, we address the question  whether a constant curvature brane can be embedded
in a flat-sliced bulk, i.e. we take the bulk geometry to be the same
as in  \cite{1704.05075},
\be \label{flatm}
ds^2 =  du^2 + e^{2A(u)}\eta_{\mu\nu}dx^\mu dx^\nu, \quad \f= \f(u)
\ee
 but look for a more general  brane embedding, specified by a
 non-trivial function $u_\star(\tau)$. Such a solution would result in a curved
 ``cosmological'' brane. The bulk scale factor and scalar field
 on each side of the brane are a priori different solutions of the  the bulk
 Einstein's equation,
\be\label{flat1}
(A, \f) = \left\{\begin{array}{ll} \left(A_-(u), \f_-(u)\right) &
    \quad u <
      u_{\star}(\tau) \\  & \\ \left( A_+(u), \f_+(u)\right) & \quad u > u_{\star}(\tau) \end{array}\right.
\ee
Israel's junction conditions then  dictate how the left and right solutions
must be glued  across the brane.  The question  is whether, for a
given bulk theory and a given choice of brane potentials,  it is
possible to  find a non-trivial embedding function such that Israel's
junction conditions are satisfied.

 In general, the answer  is negative: as we show in detail in Appendix \ref{Israel2}, for generic choices of the
 brane potentials, no solutions  to the junction conditions may be
 found:  for a non-trivial embedding function $u_\star(\tau)$ the junction
 conditions  require that  all world-volume terms in
 the brane action (\ref{sbrane}) must vanish\footnote{This may still
   lead to interesting physics if we can treat the brane as a
   probe, as is the case in the so called {\em mirage cosmology}
   \cite{mirage-1,mirage-2}.  In this case the trajectory $u_{\star}(t)$ is determined by
   extremizing the world-volume action in a fixed background, ignoring
   the backreaction on the bulk. However this is not our goal here, as
 we want to keep the backreaction intact. It will be studied in another publication}.
As we
   discuss in Appendix \ref{Israel2}, for {\em special}
 choices of the brane potentials however, non-trivial solutions may be
 found. These have the curious property that, although these solutions
 are exact, there is no  backreaction
 of the brane onto the bulk ({\em evanescent branes}).  For example,
 if $W_B$, $U$  and $Z$ are positive constants, then a de Sitter embedding
 of the brane exists if the  bulk is  Poincar\'e-AdS
 space-time with constant scalar field. More generally, it may be possible to tune the brane potentials so
 that a certain FWR embedding is possible in a given bulk
 solution. These special cases however require the brane potentials to
 be tuned to specific functions so that the corresponding brane
 embedding is compatible with the bulk solution. Although this is
 be interesting,  this goes against our general
 philosophy, which consists in taking the bulk and brane data as
 unrelated and as generic as possible.

Given the result above,  we have two possibilities
for obtaining a curved brane embedding:
\begin{enumerate}

\item We can keep a flat UV metric $\zeta^{UV}_{\mu\nu} = \eta_{\mu\nu}$
  but   generalise the bulk ansatz (\ref{metric}),  embedding  a
  non-trivial brane trajectory $u_\star(\tau)$ in a time-dependent  bulk of
  the general form
\be \label{t-dep}
ds^2 = n^2(u,\tau) du^2  + \beta^2 (u,\tau) d\tau^2 + \gamma^2 (u,\tau) \delta_{ij}dx^i x^j
\ee
We can then impose boundary conditions such
  that the metric reduces asymptotically to  the form (\ref{asympt})
  with $\zeta^{UV}_{\mu\nu}=\eta_{\mu\nu}$, with time-dependent corrections
  entering only at  subleading
  orders.

\item Alternatively, we can study  solutions in which the brane is static, and at a fixed
 position $u=u_{\star}$,  in a static  bulk solution  like
 (\ref{metric}) where each slice is curved. The induced metric on the
 brane will inherit the curvature of the corresponding constant-$u$
 slice.   This choice necessarily leads to a
 near-boundary expansion of the metric like in  (\ref{asympt}), with
 a curved UV metric, which amounts to couple the  dual UV CFT to a
 curved background\footnote{The same considerations applies to a
    more general class of bulk metrics than (\ref{metric}), of the form
\be
ds^2 = du^2 - \beta^2(u) d\tau^2 + a^2(\tau) \gamma_{ij}(u) dx^i dx^j.
\ee
These metrics have the property that any surface at $u=u_{\star}$ has  a fixed FRW
geometry, which can be brought in standard form after a world-volume
coordinate transformation acting on $(\tau,x_i)$. However different
constant-$u$ slices differ by more than just an overall rescaling. One can easily show that, also in this case, the near-boundary
expansion leads to a non-trivial time-dependent UV metric.}.
Note that this involves different AdS boundary conditions and is
therefore not in the same class of solutions  as asymptotically flat
conditions. However, as we will see in subsection \ref{ssec:flat}, it
is possible to rewrite these solutions in terms of a flat space CFT,
which however is coupled to time- or space-dependent external
sources.
\end{enumerate}

The first choice,
  appropriate for the important question of the the cosmological
  evolution of the self-tuning model, is not the road we will pursue
  here,  and we  it will be analysed in a separate paper, \cite{Branecosmo}.

In the rest of this  paper we will explore  the second option, and
embed the brane as a static hypersurface in a bulk metric of the form
(\ref{metric}), where $\zeta_{\mu\nu}$ is identified with the metric
of the dual UV CFT. However these solutions are not unrelated to the
first option described above: as we will discuss in more detail  in
section \ref{ssec:flat}, a coordinate transformation can bring a
solution of the form (\ref{metric}) to one of the form (\ref{t-dep}) with
flat asymptotic conditions, at the cost of introducing a
time- or space-dependence in the scalar field at leading order in the
near-boundary expansion. In the holographic dual language, this
situation describes a CFT living on flat space, but driven by a  varying
scalar source.

\subsection{Review of Holographic RG flows for curved QFTs } \label{RG}

From now on we focus on a setup in which the brane is  the interface between two different geometries
 of the type (\ref{metric}), each one characterized by a warp factor
 $A(u)$ and a scalar
 field profile $\f(u)$. One of them connects with the boundary of AdS (UV), the
 other should have a regular interior (IR).

The induced metric  on the brane is, up  to  a constant scaling, the
same as the slice metric (and the UV metric),
\be\label{RG1}
\gamma_{\mu\nu} = e^{2A(u_{\star})} \zeta_{\mu\nu},
\ee
and the  induced curvature scalar is
\be \label{brane_c}
R_B  = e^{-2A(u_{\star})} R^{(\zeta)}.
\ee
Because we are working with $\zeta_{\mu\nu} = \zeta_{\mu\nu}^{UV}$,
the brane geometry  is the same  (up to an overall scaling)
 as the  UV metric of the space where the CFT
lives.

For simplicity, from now on we will restrict to bulk geometries (\ref{metric}) whose constant-$u$ slices
are {\em maximally symmetric} $d$-dimensional hypersurfaces, whose
metric $\zeta_{\mu\nu}$ has the property
\begin{equation} \label{RG3}
 R_{\mu\nu}^{(\zeta)}=\kappa \zeta_{\mu\nu}, \quad R^{(\zeta)}=d \kappa .
 \end{equation}
 The constant $\kappa$ is given by
 \begin{equation}
  \kappa = \left\{
  \begin{array}{c l}
   \hphantom{-} \frac{(d-1)}{\alpha^2} & \quad \textrm{dS}_d\\
   0 & \quad \mathcal{M}^d\\
- \frac{(d-1)}{\alpha^2} & \quad \textrm{AdS}_d\\
  \end{array} \right. \, \ ,
 \end{equation}
 where $\alpha$ is the radius of curvature.\footnote{Here we have
   included the case $\kappa =0$ for completeness, but as we have
   explained in the previous subsection this can only lead to flat
   brane embeddings.}
Whether $\kappa$ is
 positive, negative or zero is fixed by the UV boundary
 conditions, i.e.~by the metric to which the CFT is coupled.

One of the reasons we restrict to maximally symmetric slices  is practical: in
this case the bulk geometry was studied in detail in
\cite{Ghosh:2017big} and we can use the results of that
work. However, the case of constant positive curvature  is already
interesting for phenomenology, as it can relate both to primordial
inflation and to late time cosmology, both described by quasi-de
Sitter space-time.  Extensions to a more
general FRW-like brane can be obtained using similar  techniques as
those presented here, but will be left for future work.

Before studying solutions  including a curved brane,
we briefly review the features of bulk solutions with curved slicing,
which were studied in depth in \cite{Ghosh:2017big}.

Throughout this paper, a dot will denote a derivative with respect to
the $u$ coordinate and a prime will denote a derivative with respect
to the $\f$ coordinate, e.g.
 \begin{equation} \label{RG2}
 \dot{f}(u) \equiv \frac{df(u)}{du}, \quad F'(\f)\equiv \frac{d F(\f)}{d\f}.
 \end{equation}

With the restriction (\ref{RG3}), the bulk Einstein's  equations are:
 \begin{align}
\label{eq:EOM1} 2(d-1) \ddot{A} + \dot{\f}^2 + \frac{2}{d} e^{-2A} R^{(\zeta)} &=0 \, , \\
\label{eq:EOM2} d(d-1) \dot{A}^2 - \frac{1}{2} \dot{\f}^2 + V - e^{-2A} R^{(\zeta)} &=0 \, , \\
\label{eq:EOM3} \ddot{\f} +d \dot{A} \dot{\f} - V' &= 0 \, .
\end{align}

In \cite{Ghosh:2017big}, a formalism  was developed by introducing
three {\em scalar functions}  of the bulk scalar field, $W(\f)$, $S(\f)$ and $T(\f)$,
which results in  flow-like equations for the warp
factor and scalar field,
\begin{align}
&\dot{A}(u) = -{1\over 2(d-1)}  W(\f(u)), \label{defW} \\
\nonumber \\
&\dot{\f}(u) = S(\f(u)), \label{defS}\\
\nonumber \\
& R^{(\zeta)}e^{-2 A(u)} = T(\f(u)).  \label{defT}
\end{align}
The scalar functions  $W(\f),S(\f),T(\f)$ may be  defined  piecewise
in any region in which the scalar field $\f(u)$ is monotonic, and can
eventually be glued together (imposing regularity) at extrema of the function
$\f(u)$ (see \cite{exotic,Ghosh:2017big} for details).

We note that, for $R\neq 0$,  our rewriting of the bulk Einstein's
equation in terms of  scalar functions differs from the first-order
formalism obtained using Hamilton-Jacobi theory applied to holography \cite{0404176}.  In the flat case, the function $W(\f)$ is in one-to-one correspondence
with a solution of the radial Hamilton-Jacobi equation for Hamilton's
principal function ${\cal S}(A,\f)$. For $R=0$ Hamilton-Jacobi's  equation is separable and the solution can be written in
the form $S = e^{dA} W(\f)$, where the superpotential $W$ satisfies
equation (\ref{Wflat}).  For $R\neq 0$ however, this is not the case: the
Hamilton-Jacobi equation is non-separable, and a true first-order
formalism can  only be obtained  starting from a non-trivial function
${\cal  S}(A,\f)$. The reader is referred to
\cite{Papadimitriou:2007sj} for a discussion of the
first order  Hamilton-Jacobi formalism for curved domain-walls in
holography. Similar considerations apply to black-hole geometries
\cite{Lindgren:2015lia}.  A connection between our scalar functions and
Hamilton-Jacobi formalism can be obtained proceeding along the  lines of
appendix B in \cite{Caldarelli:2016nni}.

Using the scalar functions $W,S,T$, we can write
eqs.~\eqref{eq:EOM1}--\eqref{eq:EOM3} as a system of first order differential
equations in which $\f$ is the independent variable:
\begin{align}
\label{eq:EOM4} S^2 - SW' + \frac{2}{d} T &=0 \, , \\
\label{eq:EOM5} \frac{d}{2(d-1)} W^2 -S^2 -2 T +2V &=0 \, , \\
\label{eq:EOM6} SS' - \frac{d}{2(d-1)} SW - V' &= 0 \, .
\end{align}

Flat space-time holographic RG flows are recovered by  setting
$T=0$ and $S = W'$, resulting  in the usual superpotential
equation for $W(\f)$,
\begin{equation}
 \kappa =0: \qquad \frac{d}{4(d-1)} W^2 -{1\over 2} W'^2  = -V \ , \label{Wflat}
  \end{equation}

Equation \eqref{eq:EOM5} is algebraic and it can be used to eliminate the function $T(\f)$ and we are left with the following two independent equations
  \begin{align}
\label{eq:EOM7} \frac{d}{2(d-1)} W^2 + (d-1) S^2 -d S W' + 2V &=0 \, , \\
\label{eq:EOM8} SS' - \frac{d}{2(d-1)} SW - V' &= 0 \, .
\end{align}

We will consider solutions in which we can identify a UV and and IR
region, whose features we summarize below.  Roughly, they can be
identified with the regions where  the scale factor $e^{A(u)}$ becomes large (UV) or  small (IR).

\subsubsection*{UV Region}
This is the region where the scale factor becomes large and the
geometry asymptotes to the conformal boundary of Anti-de Sitter
space. Generically, this is realised when $\f(u)$ approaches a maximum of the scalar potential (which we
set at $\f=0$ for simplicity). Around the maximum, the potential has an expansion of the form
\begin{equation}
 V(\f)= -\frac{d(d-1)}{\ell^2}+ \frac{m^2}{2}\f^2+\mathcal{O}(\f^3)
 \end{equation}
 where $m^2<0$. The boundary is approached as $u\to -\infty$,  and the
 solution takes the asymptotic form
\be\label{UV0}
 A(u) =  -\frac{u}{\ell} + \ldots , \quad  \f(u) = \f_- \ell^{\Delta_-}e^{\Delta_-u / \ell} +
\ldots \qquad  u \to -\infty,
\ee
where $\Delta_- = d/2-\sqrt{d^2/4 + m^2\ell^2}$ and    $\f_-$ is an
integration constant.  In the AdS/CFT dictionary, the constant   $\f_-$ is interpreted, in the dual field theory,
as the  value of the  source for a gauge-invariant  operator $\mathcal{O}$ whose
dimension is $\Delta_+ \equiv d-\Delta_-$. This
geometry is dual to an RG flow away from a UV conformal fixed point,
driven by a deformation by the relevant  operator $\mathcal{O}$. As $u$ increases,  $\f$ grows, backreaction starts
becoming important, and the geometry deviates more and more from
$AdS$.

A detailed description of the solution in the UV, both in terms of the
functions $W(\f)$, $S(\f)$ and $T(\f)$ and of the scale factor $A(u)$ and scalar field profile $\f(u)$, can
be found in Appendix \ref{app:UV}. Here we point out a few important
features:
\begin{itemize}
\item To leading order as $\f\to 0$, the behaviour of
$W$ and $S$  is universal and it is the same as in the zero-curvature
case,\footnote{Here we only give the behavior of the ``$W_-$'' type
function, which in AdS/CFT language (and with the standard quantisation) corresponds to a deformation of
the CFT by a source. For completeness, the reader will also find in
Appendix \ref{app:UV} the ``$W_+$'' type function, which
corresponds to a deformation by a vev with no source term.}
\be \label{Wlead}
W = {2(d-1) \over \ell} + {\Delta_- \over 2 \ell}\f^2 +\ldots ,
\quad S \simeq W' = \Delta_- \f + \ldots , \qquad \f \to 0.
\ee
The precise  behavior including subleading terms can be found in
appendix \ref{app:UV}.
\item
The two  first order differential equations
(\ref{eq:EOM7}-\ref{eq:EOM8})   contain two independent
integration constants $C$ and ${\mathcal R}$. They appear as
coefficients of  subleading (with respect  to those in (\ref{Wlead}))
terms  $W_{sub}(\f)$, $S_{sub}(\f)$ in the  expansion of $W$ and $S$
around the UV $\f=0$:
\be\label{Wsub}
W_{sub} =  {{\mathcal R} \over d\ell} |\f|^{2\over \Delta_-} + \ldots
+
{C\over \ell}|\f|^{d\over \Delta_-} + \ldots, \quad \f\to 0.
\ee
and similarly for $S_{sub}$.

Together with $\f_-$, the parameters ${\mathcal R}$ and $C$
exhaust the three integration constants of the third order system of
Einstein equations (\ref{eq:EOM1}-\ref{eq:EOM2}). The constants $C$
and ${\mathcal R}$  fix
the ``dimensionless'' (i.e. in units of $\f_-$)  boundary curvature and
vacuum expectation value of the dual operator,
\be \label{curlyR}
\langle \mathcal{O} \rangle_- = \frac{Cd}{\Delta_-} \,
|\f_-|^{\Delta_+ / \Delta_-} \, , \quad
\mathcal{R}=R^{(\zeta)} |\f_{-}|^{-2/\Delta_{-}},
\ee
\end{itemize}
\vspace{0.5cm}

As we have discussed, the properties of the UV part of the solution are
universal, and depend on the curvature only at subleading
orders. In contrast, properties of the solution in the region, where the scale factor
becomes small (corresponding  to the IR of the dual QFT) are very
different depending on the sign of the curvature. Below we  review the
structure of the regular IR region for zero, positive and negative
curvature.  More details can be found in \cite{exotic} and
\cite{Ghosh:2017big}.

\subsubsection*{IR Region, $\kappa = 0$}

\begin{enumerate}
\item {\bf Regular AdS interior}\\
We start by recalling the situation for zero curvature. In the flat
case, a regular interior can arise only as a new  asymptotically AdS
region, where $e^{A(u)} \sim e^{-2u/\ell_{IR}}$ vanishes as $u\to
+\infty$. This can occur as $\f$ approaches a minimum $\f_{IR}$
 of the bulk potential, around which
\be
V(\f) \simeq - {d(d-1) \over \ell^2_{IR}} + {m^2_{IR} \over 2}
(\f-\f_{IR})^2 + \ldots
\ee
with  $m^2_{IR}>0$.
 Unlike the case for a UV maximum, where all
 $W$ solutions have the same behaviour (\ref{Wlead}), here only a single
 solution can reach a minimum. The
 regular solution reaching the minimum behaves close to $\f_{IR}$ as
\be
W_{IR}(\f) \simeq {2(d-1) \over \ell_{IR}} + {\Delta^{-}_{IR} \over
2\ell_{IR}} (\f-\f_{IR})^2 + \ldots
\ee
where $\Delta^{-}_{IR} =  d/2-\sqrt{d^2/4 + m^2\ell^2} < 0$. We can see
that when $V$ has a minimum, $W$ has a maximum. The corresponding
geometry approaches the Poincar\'e horizon of $AdS$ as $u\to +\infty$,
\be
A(u) = -{u \over \ell_{IR}} + \ldots, \qquad \f = \f_{IR} - \bar{\f}
\exp\left[{{\Delta^{-}_{IR} u \over \ell_{IR}}}\right],  \qquad u \to +\infty.
\ee
where $\bar{\phi}$ is a constant, interpreted as the (irrelevant) coupling
 in the IR CFT.

\item {\bf $\f\to \infty$: Good singularities}\\
If $\f$ does not reach a fixed point, it will flow all the way to
$\f \to \pm \infty$. In this case, the solution has a naked
singularity in the IR. Nevertheless, this can be an acceptable IR region  if it
satisfies certain requirements, which can be translated into the
asymptotic behaviour of the function $W$ as $\f\to \infty$. Below
we summarize these requirements for  a potential whose asymptotic
behaviour for large $\f$  is parametrized by a positive exponential
(we discuss the case  $\f \to +\infty$ for concreteness),
\be \label{exp1}
V\simeq -V_{\infty}e^{b \f} \qquad \f \to +\infty.
\ee
For such a potential, the asymptotic solutions  $W(\f)$ as $\f\to
\infty$ fall in two classes:
\begin{itemize}
\item {\bf Generic}\\
\be\label{exp2}
W(\f) \simeq W_{0} e^{Q\f} , \qquad Q\equiv  \sqrt{d\over 2(d-1)},
\ee
where $W_0$ is an arbitrary constant. Notice that the exponential
behavior does not depend on the parameter $b$ in the potential
in equation (\ref{exp1}).
Indeed the potential $V$ affects this class of solutions at subleading orders.
\item {\bf Special}
\be\label{exp3}
W(\f) \simeq \sqrt{8V_{\infty} \over 4Q^2 - b^2} e^{{b\over 2} \f}.
\ee
This solution exists only for $b< 2Q$, is isolated, as it has no
tunable integration constants, and has a softer exponential growth
than all of the solutions of type (\ref{exp2}).
\end{itemize}
One can argue that  {\em only the isolated special solution} has a meaningful
holographic interpretation. For example, one can impose  Gubser's
criterion \cite{0002160},
which requires that one can cloak the singularity by an arbitrarily
small horizon. This is true for the solution (\ref{exp3}), but not for
the solutions of type (\ref{exp2}). Notice that the existence of the
special solution puts an upper bound  $b < 2Q$ on the asymptotic
growth rate of the
bulk potential.
\end{enumerate}

\subsubsection*{IR Region, $\kappa > 0$}

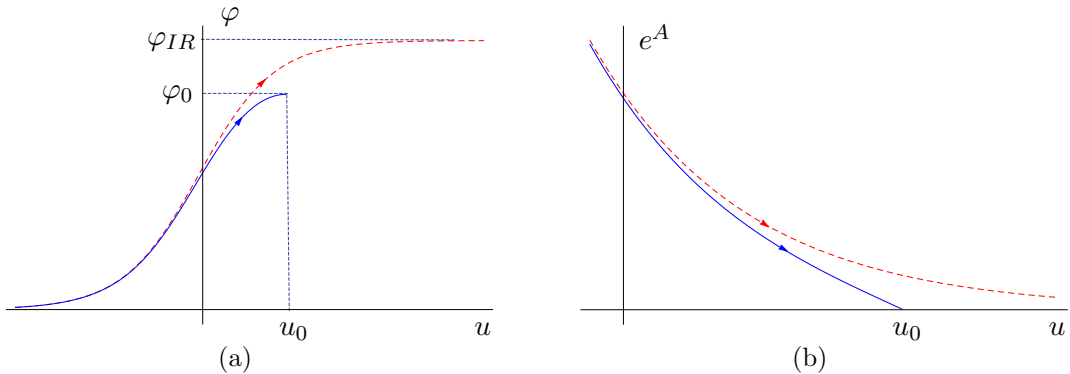
\begin{figure}[t]
\centering
\begin{subfigure}{.5\textwidth}
 \centering
   \begin{overpic}[width=0.85\textwidth,tics=10]{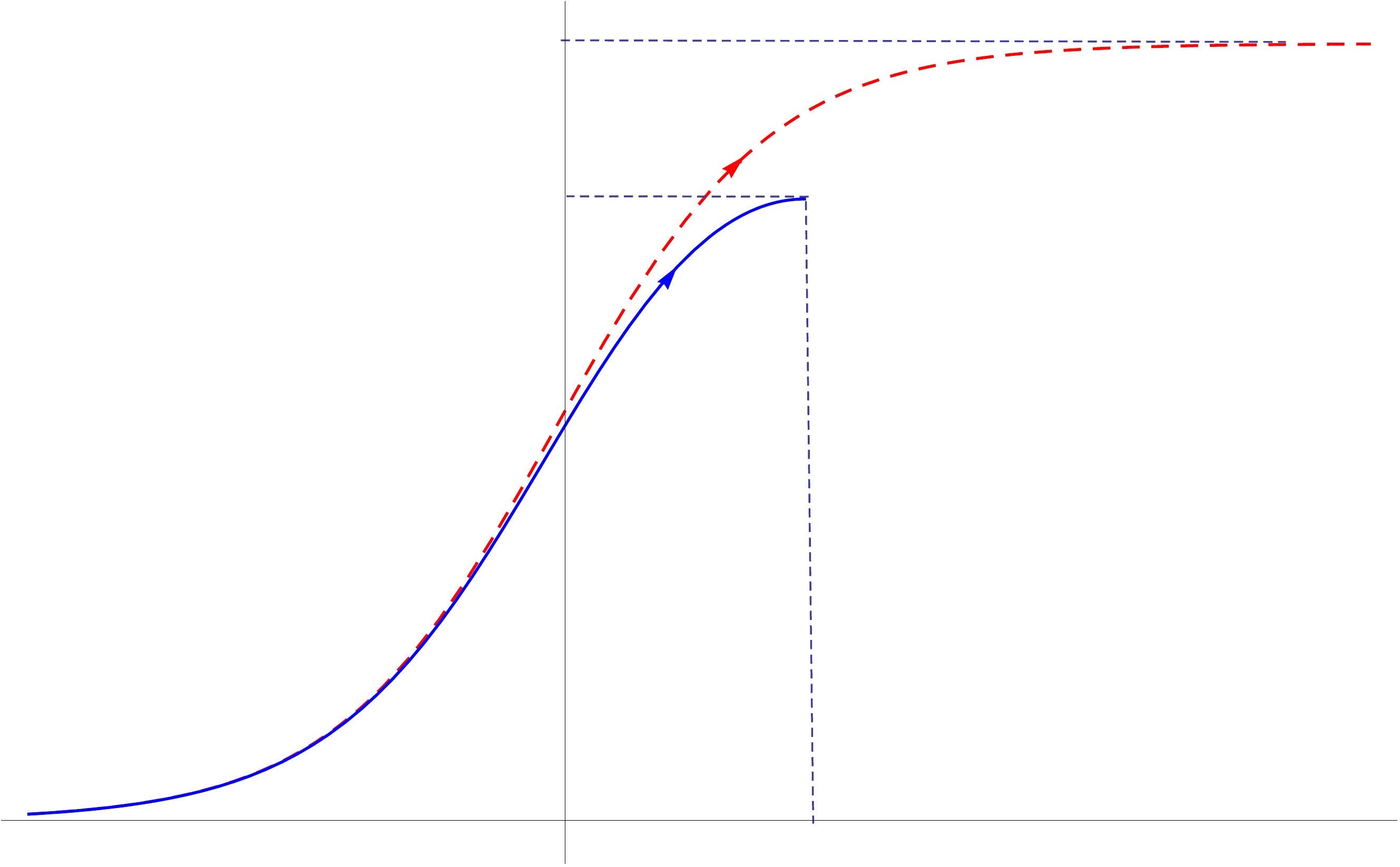}
\put (44,63) {$\f$} \put (29,58) {$\f_{IR}$} \put (32,47) {$\f_{0}$} \put (96,-2) {$u$} \put (56,-2) {$u_0$}
\end{overpic}
 \caption{\hphantom{A}}
  \label{positiveRflowphi}
\end{subfigure}%
\begin{subfigure}{.5\textwidth}
  \centering
 \begin{overpic}[width=0.85\textwidth,tics=10]{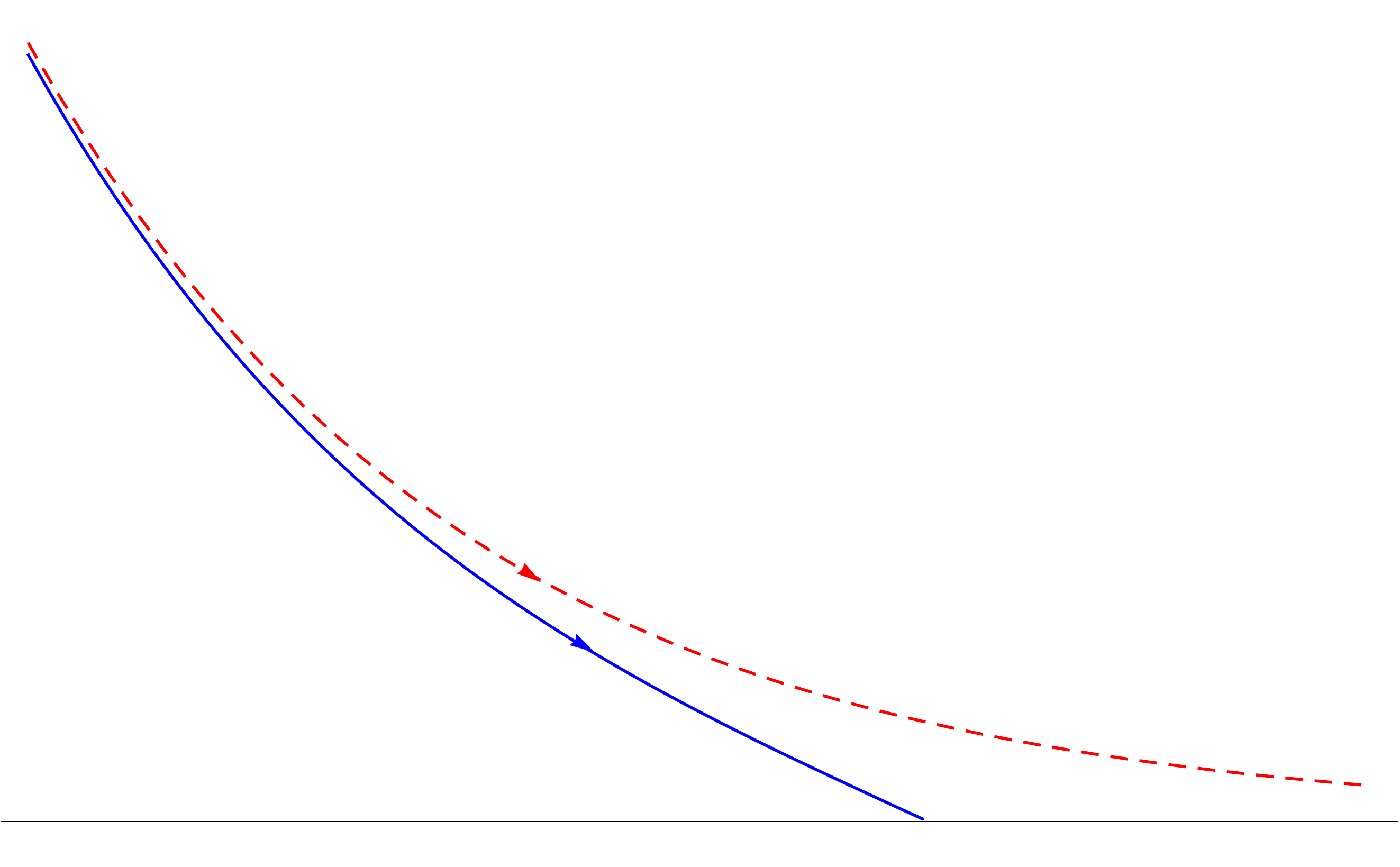}
\put (12,57) {$e^A$} \put (96,-2) {$u$} \put (64,-2) {$u_0$}
\end{overpic}
\caption{\hphantom{A}}
  \label{positiveRflowA}
\end{subfigure}
\caption{The scalar field (a) and scale factor (b) profiles of a
   positive curvature RG flow geometry (solid lines). Both the scale
   factor and the scalar field have an IR end point at $u=u_0$, $\f =
  \f_0$. The corresponding zero-curvature solutions extending to the conformal
   fixed point $\f=\f_{IR}$ (dashed lines) are shown for comparison.}
\label{positiveRflow}
\end{figure}

 For $R^{(\zeta)} > 0$, it was shown in \cite{Ghosh:2017big} that regular
  geometries reach a point  $u=u_0$,
  at which $\dot\f=0$ and $e^{2A}$ vanishes as $(u_0 - u)^2$.  This is
  a regular endpoint in the Euclidean case, and a horizon in the
  Lorentzian signature. This point corresponds to the extreme
  infrared, since the whole space-time shrinks to zero size. Figure
  \ref{positiveRflow} shows a
    comparison between a positive curvature flow and the corresponding
    regular zero-curvature
    solution, in a case where the latter arrives at an  IR fixed point
    at finite $\f$.

A few properties of these solutions are listed below:
\begin{itemize}
  \item The dimensionless parameters of these solutions are  fixed by specifying the
  value of the scalar field $\f_0$ at the IR endpoint, and imposing
  regularity. This determines the dimensionless UV parameters
  ${\mathcal R}$ and $C$.
\item Fixing $\f_0$ leaves a one-parameter family of solutions
  parametrized by $\f_-$. The latter  can be taken as
  setting the scale for the  all dimensionful features of the solution
  (curvature, operator vev). Therefore, fixing $\f_-$ and  varying the endpoint in field
  space $\f_0$, one can scan over the range of positive curvature
  solutions.
\item For strictly positive $R^{(\zeta)} > 0$, the endpoint $\f_0$
  cannot coincide with an extremum of the potential: this would lead
  to either infinite curvature (maximum) or zero curvature (minimum).
\item At the IR endpoint,  $W(\f)$ diverges as $|\f_0 -
  \f|^{-1/2}$ and $S(\f)$ vanishes as $|\f_0 -
  \f|^{1/2}$.
\end{itemize}

\subsubsection*{IR Region, $\kappa < 0$}
In this case, the scale factor never shrinks to zero. Instead, the
geometry reaches a throat of finite size where both $\dot\f=0$ and $\dot{A}=0$, and the scale factor
  takes a minimum value $A(u_0) \neq 0$. The two sides of the throat
  connect to different regions of the asymptotic AdS boundary. As
  explained in \cite{skenderisham} in the case of constant potential,
  and noted also in \cite{Ghosh:2017big} in the general case, this corresponds to
  the fact that the UV theory has a codimension-one defect. 
 These features of negatively curved flows can be seen in Figure
  \ref{negativeRflow}: 

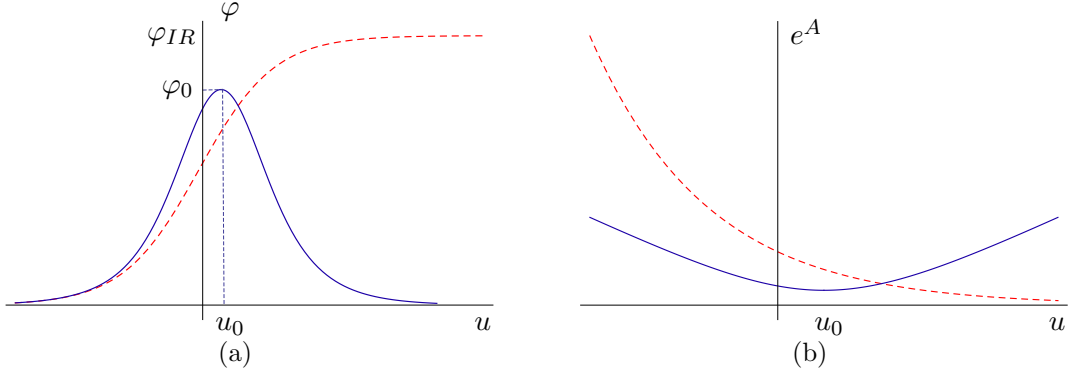
\begin{figure}[t]
\centering
\begin{subfigure}{.5\textwidth}
 \centering
   \begin{overpic}[width=0.85\textwidth,tics=10]{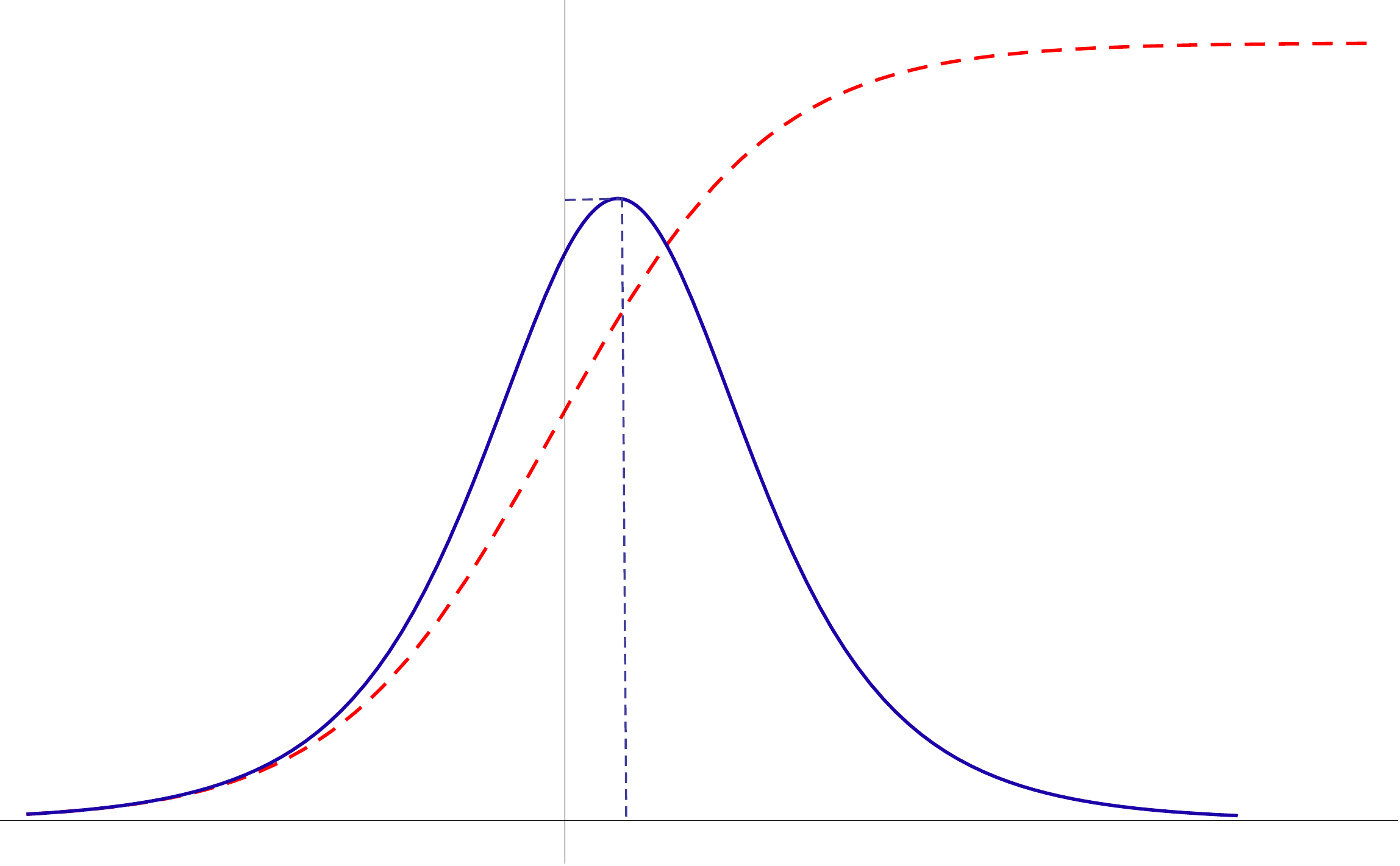}
\put (44,63) {$\f$} \put (29,58) {$\f_{IR}$} \put (32,47) {$\f_{0}$} \put (96,-2) {$u$} \put (43,-2) {$u_0$}
\end{overpic}
 \caption{\hphantom{A}}
  \label{negativeRflowphi}
\end{subfigure}%
\begin{subfigure}{.5\textwidth}
  \centering
 \begin{overpic}[width=0.85\textwidth,tics=10]{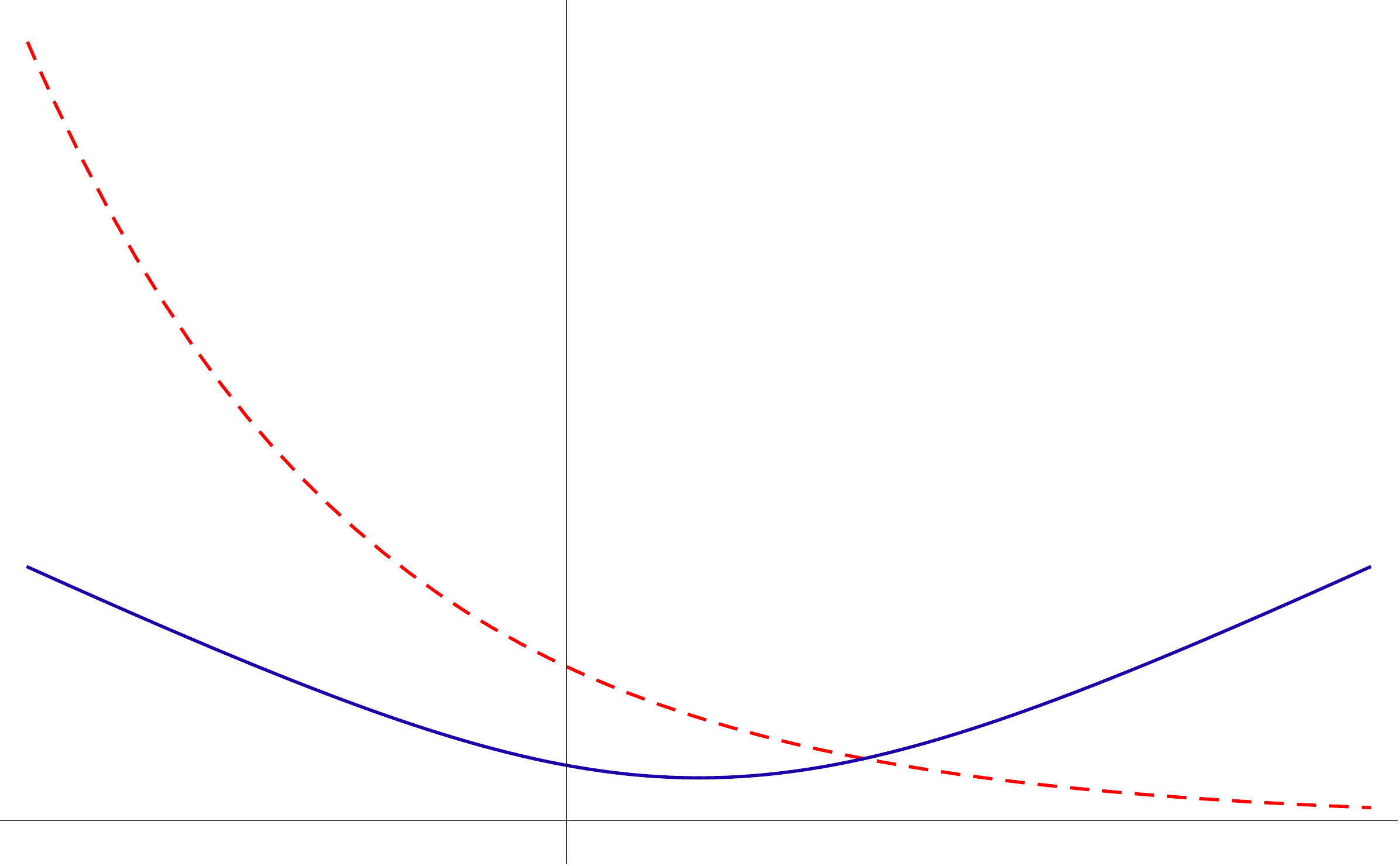}
\put (43,57) {$e^A$} \put (96,-2) {$u$} \put (48,-2) {$u_0$}
\end{overpic}
\caption{\hphantom{A}}
  \label{negativeRflowA}
\end{subfigure}
\caption{The scalar field (a) and scale factor (b) profiles of a
   negative curvature RG flow geometry (solid lines). Both the scale
   factor and the scalar field have a turning point at the bottom of
   the throat, $u=u_0$. The
   corresponding zero-curvature solutions extending to the conformal
   fixed point $\f=\f_{IR}$ (dashed lines) are shown for comparison.
 The solution is symmetric around $u_0$. }
\label{negativeRflow}
\end{figure}

\begin{itemize}
  \item As in the $\kappa>0$ case, the value of the scalar field
    $\f_0$ at the bottom of the throat fixes all  integration constants
    in the solution (except $\f_-$ which can always be chosen independently).
\item Also in this case, the  turning point  $\f_0$
  cannot coincide with an extremum of the potential.
\item At the IR endpoint, both   $W(\f)$ and $S(\f)$ vanish as $|\f_0 -
  \f|^{1/2}$.
\end{itemize}
\vspace{0.5cm}

\begin{figure}[h]
\centering
\begin{overpic}
[width=0.75\textwidth]{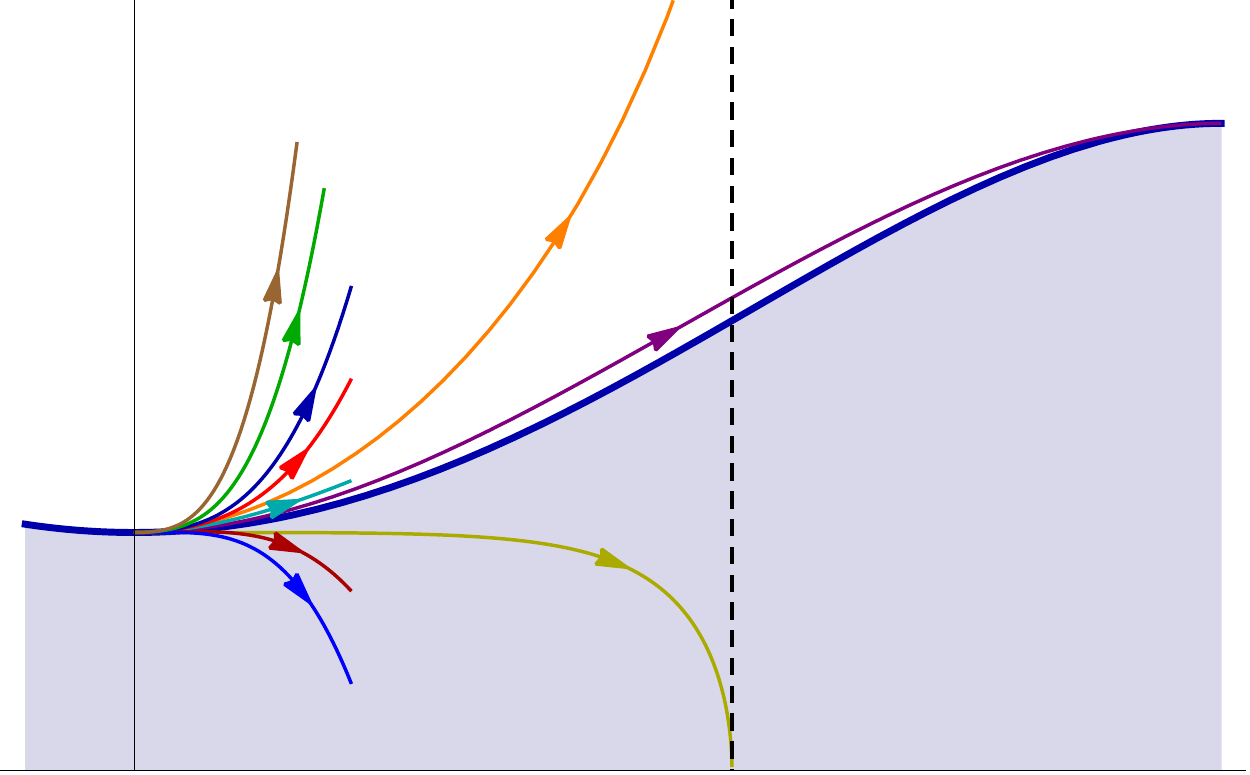}
\put(4,7){UV}
\put(60,3){$\f_0$}
\put(60,58){{\footnotesize IR endpoint (curved)}}
\put(85,52.7){{\footnotesize IR fixed point $\f_{IR}=1$ (flat)}}
\put(99,3){$\f$}
\put(-2,58){$W(\f)$}
\put(42,33.8){{\footnotesize $R=0$}}
\put(42,55){{\footnotesize $R>0$}}
\put(42,14){{\footnotesize $R<0$}}
\put(24,49){{\footnotesize $W_{C, \mathcal{R}}(\f)$}}
\put(71.5,40){$\sqrt{-\tfrac{4(d-1) V(\f)}{d}}$}
\end{overpic}
\caption{Sketch of different solutions for $W(\f)$, for various
  curvatures, in a theory with a UV and an IR fixed point at $\f=0$
  and $\f=1$, respectively. Of all the
  solution starting from the UV fixed point, three are followed to the
  IR endpoint, which can be at the conformal fixed point ($R=0$) or
  an intermediate point $0< \f_0< 1$. For $R^{(\zeta)}\geq 0$ the shaded region below the blue
  curve cannot be accessed.}
\label{fig:flowsintro}
\end{figure}
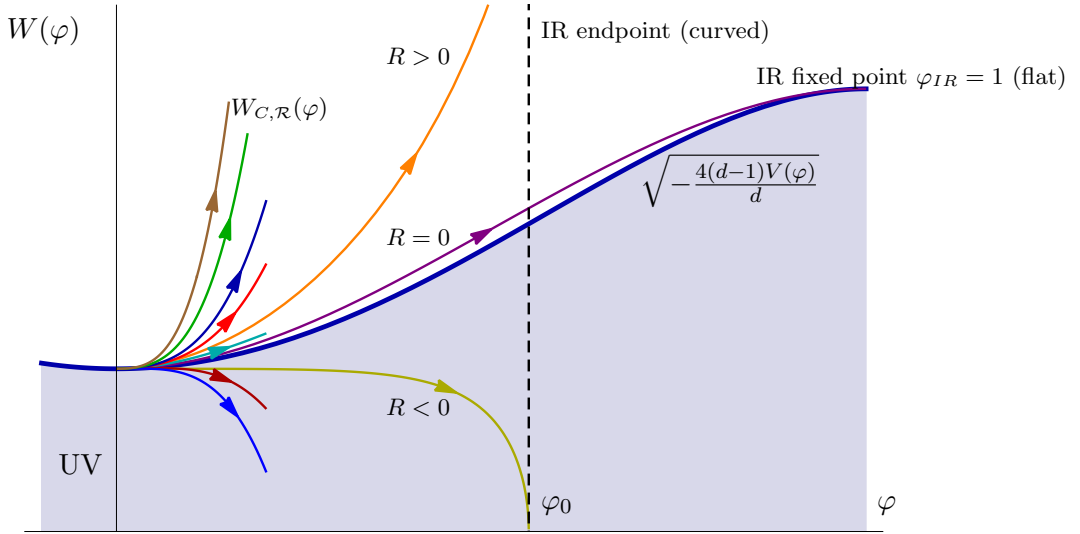

The properties of the solutions we have described in this subsection
are summarized in figure \ref{fig:flowsintro}, where we sketch the
behavior of different solutions for the  scalar function $W(\f)$ in a model in which the
potential $V(\f)$ has a maximum at $\f=0$ and a minimum at $\f=1$. As
one can see, all solutions  connect  to the UV fixed point at $\f=0$.  The flat
solution reaches the IR conformal fixed at $\f_{IR}=1$; The regular curved RG
flow solutions reach an endpoint or a throat at $0< \f_0 < 1$. An
important role is played by the curve $B(\f) \equiv \sqrt{-4(d-1)
  V(\f)/d}$, displayed in dark blue in the figure: for zero and
positive curvature, this curve mark the boundary of a forbidden region
(shaded area in the figure) which $W(\f)$ cannot reach.  On the other
hand, for $R<0$ the solution can reach into the forbidden region. \\

We end this review section with the remark that solutions may exhibit one or several
{\em bounces}, i.e. regular points where $\dot{\f}=0$ but $\dot{A}\neq
0$ \cite{exotic,1711.10969}. These can occur in principle at any point where $V'\neq 0$, and
they where found to be a generic feature of curved RG-flow
solutions\cite{Ghosh:2017big}. The possibility of bounces  has to be
taken into account when we introduce the brane.

 The details of the
behavior of the functions $W$, $S$ and $T$ near an IR endpoint
or a bounce are given in Appendix \ref{app:IR}.

\subsection{The junction conditions} \label{junc}
We are now in the position  to introduce the brane as an interface
between two geometries of the form
(\ref{metric}).
The dynamics of the brane is encoded in the junction conditions, as we explain below.

We consider solutions where the geometry to one side of the brane
connects to an UV-type region, and the other to an
IR-type region. To distinguish between the two sides, we label the metric and scalar field on the two sides of the brane by $g^{UV}_{ab}, g^{IR}_{ab}$ and $\f^{UV}, \f^{IR}$. When a quantity $X$ exhibits a jump across the position of the brane, this will be written as $\big[ X\big]^{UV}_{IR}$. The Israel matching conditions then result in the following two requirements:
\begin{enumerate}
\item The metric and scalar field are continuous across the brane:
\be\label{FE3}
\Big[g_{ab}\Big]^{UV}_{IR} = 0 \, ,   \qquad \Big[\f\Big]^{IR}_{UV} =0 \, .
\ee
\item The extrinsic curvature as well as the normal derivative of $\f$ are discontinuous:
\be\label{FE4}
\Big[K_{\mu\nu} - \gamma_{\mu\nu} K \Big]^{IR}_{UV} =   {1\over \sqrt{-\gamma}}{\delta S_{brane} \over \delta \gamma^{\mu\nu}}  ,  \qquad \Big[n^a\de_a \f\Big]^{IR}_{UV} =- {1\over \sqrt{-\gamma}}{\delta S_{brane} \over \delta \f} .
\ee
Here $\gamma_{\mu\nu}=e^{2 A(u)}\zeta_{\mu\nu}$ is the induced metric, $K_{\mu\nu}$ is the extrinsic curvature of the brane with $K = \gamma^{\mu\nu}K_{\mu\nu}$ the trace, and $n^a$ is a unit vector normal to the brane with orientation towards the IR.
\end{enumerate}
For our setup given in \eqref{sbrane} the equations \eqref{FE4} become
\bea
&&
\Big[K_{\mu\nu} - \gamma_{\mu\nu} K \Big]^{IR}_{UV} = \left[\half
  W_B(\f) \gamma_{\mu\nu} + U(\f) G^B_{\mu\nu} - \half Z(\f)
  \left(\de_\mu \f \de_\nu \f - \half \g_{\mu\nu} (\de \f)^2
  \right)\right. \nonumber \\
&&  \qquad \qquad \qquad \qquad \quad \ \ + \left(\gamma_{\mu\nu}\gamma^{\rho\sigma}\nabla_\rho
    \nabla_\sigma - \nabla_\mu\nabla_\nu\right)U(\f)
\bigg]_{\f_{\star}\ \ \ },  \label{FE5}\\
&& \nonumber \\
&& \Big[n^a\de_a \f\Big]^{IR}_{UV}  = \left[{d W_B \over d \f} - {d U\over d \f} R_B + \half {d Z \over d \f }(\de \f)^2 - {1\over \sqrt{-\g}}\de_\mu \left( Z(\f) \sqrt{-\g} \g^{\mu\nu}\de_\nu \f \right) \right]_{\f_{\star}},  \label{FE6}
\eea
where $\nabla_\mu$, $G^B_{\mu\nu}$ and $R_B$ are  the covariant
derivative, Einstein tensor and Ricci scalar computed from the induced metric and  $\f_{\star}(x^{\mu}) \equiv \f(u_{\star}, x^{\mu})$ is the scalar field at the position of the brane. Furthermore, for our setting we find:
\be
K_{\mu\nu} = \dot{A} \gamma_{\mu\nu}, \qquad K_{\mu\nu} - \gamma_{\mu\nu} K = -(d-1) \dot{A} \gamma_{\mu\nu} = \half W \gamma_{\mu\nu}, \qquad  n^a\de_a \f = \dot{\f} =S\, .
\ee
Further, recall that $R^B_{\mu \nu} = R^{(\zeta)}_{\mu \nu}$ and $R_B = e^{-2A} R^{(\zeta)}$, hence
\be
G_{\mu \nu}^B = G_{\mu \nu}^{\zeta} =  \frac{1}{2} (2-d) \kappa \, e^{-2A} \gamma_{\mu \nu} = \frac{2-d}{2d} \, T \gamma_{\mu \nu} \, ,
\ee
where we have used the definitions \eqref{defW}--\eqref{defT}. We introduce $W_{UV}, S_{UV}$ and $W_{IR}, S_{IR}$ as the functions $W$ and $S$ for the UV and the IR regions, respectively. Using these quantities we can then write the junction conditions \eqref{FE5}--\eqref{FE6} as:
\begin{align}
{\left. W_{IR} - W_{UV} \right|}_{\f_*} &= {\left. W_B + \frac{(2-d)}{d}  \, U \, T \right|}_{\f_\star}  \, , \label{JC1}\\
{\left. S_{IR} - S_{UV} \right|}_{\f_*} &= {\left. {W_B}' - U' \, T \right|}_{\f_\star} \, .\label{JC2}
\end{align}
From the continuity of the metric \eqref{FE3}, we can infer that
the scale factor is continuous across the brane and the same is true
for the  function $T(\f)$,
\begin{equation}
T_{UV}(\f_\star)=T_{IR}(\f_\star) \ .
\end{equation}

Using the continuity of $T$ and $\f$ across the brane,  it
follows from \eqref{eq:EOM5} that
\be
\frac{d}{2(d-1)} W_{UV}^2 - S_{UV}^2\Big |_{\f_{\star}}=\frac{d}{2(d-1)} W_{IR}^2 - S_{IR}^2\Big |_{\f_{\star}}
\ee
We can write the conditions \eqref{JC1}--\eqref{JC2} as
\begin{align}
{\left. W_{UV} \right|}_{\f_*}&={\left. W_{IR}-W_{B}-\frac{2-d}{d} UT_{IR}\right|}_{\f_\star} \label{WUV}\\
{\left. S_{UV}\right|}_{\f_*} &={\left. S_{IR}-W_B'+U' T_{IR} \right|}_{\f_\star}\label{SUV}\ .
\end{align}
From the equation of motion we can write
\begin{align}
Q^2 W_{UV}^2-S_{UV}^2-2 T_{UV}+2 V=0
\end{align}
where $Q^2=\frac{d}{2(d-1)}$. Using eqs.~\eqref{WUV}--\eqref{SUV} and using the fact that $T$ and $\f$ are continuous, we can express everything in terms of IR quantities. Using also $Q^2 W_{IR}^2-S_{IR}^2-2 T_{IR}+2 V=0  $ and after a bit of algebra we obtain the condition:
\begin{align}
\Bigg[ -2Q^2 W_{IR} \left(W_B+\frac{2-d}{d}U T_{IR} \right) &+Q^2 \left( W_B+\frac{2-d}{d}U T_{IR} \right)^2   \nonumber \\
& +2 S_{IR}\left(W_B' -U' T_{IR} \right) - \left(W_B' -U' T_{IR} \right)^2 \Bigg]_{\f_{\star}}=0 \ . \label{JC}
\end{align}

Notice that all functions of $\f$ involved in this equation are
in principle known, in terms of a few input quantities:  $V$, $W_B$
and $U$ are fixed by the choice of the action; $W_{IR}$, $S_{IR}$ and
$T_{IR}$ are determined by regularity, plus the choice of the endpoint
$\f_0$ of the IR solution. Therefore, once the underlying model and
$\f_0$ are chosen, (\ref{JC}) provides a transcendental equation for the brane position  $\f_\star$, which generically has a
finite number of solutions (including the possibility of no
solution).

Once $\f_\star$ is determined, we can use equations
(\ref{WUV})--(\ref{SUV}) as initial conditions for $W_{UV}$ and $S_{UV}$,
to be used in the system of differential
equations (\ref{eq:EOM7})--(\ref{eq:EOM8}) which determines the solution for
$W$ and $S$ in the UV and the
corresponding values of $\mathcal{R}$ and $C$ (the dimensionless
curvature and vev parameters).

To summarize, one can use the following algorithmic procedure  to
solve the system from the IR, across the brane, to the UV:
\be
\text{choice of $\f_0$} \; \rightarrow\;  W_{IR}, S_{IR}
\; \rightarrow \; \f_\star \; \rightarrow \;  W_{UV}, S_{UV} \; \rightarrow
\; {\mathcal R}, C
\ee
The only control parameter here is $\f_0$, which determines everything
else. In particular, the choice of $\f_0$ at an IR extremum of the
potential would result in the flat-sliced solution with $\mathcal{R} =
0$. For the case when the flat solution IR is reached as $\f\to
\infty$, things are more subtle, as we will see in section \ref{sec:numexpV}.

\subsection{Junction rules}
Here we discuss what are the geometric rules
to  patch together two geometries across the brane, and  which types
of junctions give rise to a sensible holographic interpretation.

In the positive or zero curvature case, the flow of $A(u)$ is
monotonically decreasing from the UV to the IR. Since $\dot{A} \propto
-W$,  the scalar function $W(\f)$ cannot change sign.\footnote{Since the overall sign of $W$ can be changed by sending $u\to -u$, we
will always choose $W>0$ for definiteness.} At the junction,  we
must  require that one side of the brane actually connects to a UV
region, and the other to an IR region.  This implies that the flow of $A(u)$ must not change direction,
i.e.~$\dot{A}$ should not change sign, across the brane. Since at the
brane position $\dot{A}_{UV} \propto - W_{UV}(\f_{\star})$ and
$\dot{A}_{IR} \propto - W_{IR}(\f_{\star})$, we must discard solutions in
which $W_{UV}(\f_{\star})$ and $W_{IR}(\f_{\star})$ have opposite signs. If that
were the case, we would be joining two UV or two IR regions.

The above constraint does not apply to solutions with
negatively curved slices: in this case $\dot{A}$ (and $W$) can change
sign in the bulk, and there is no reason why it should not change sign
across the brane. In fact, in this case, both sides of the brane
eventually reach a UV region.

\begin{figure}[t]
\centering
\includegraphics[width=0.45\textwidth]{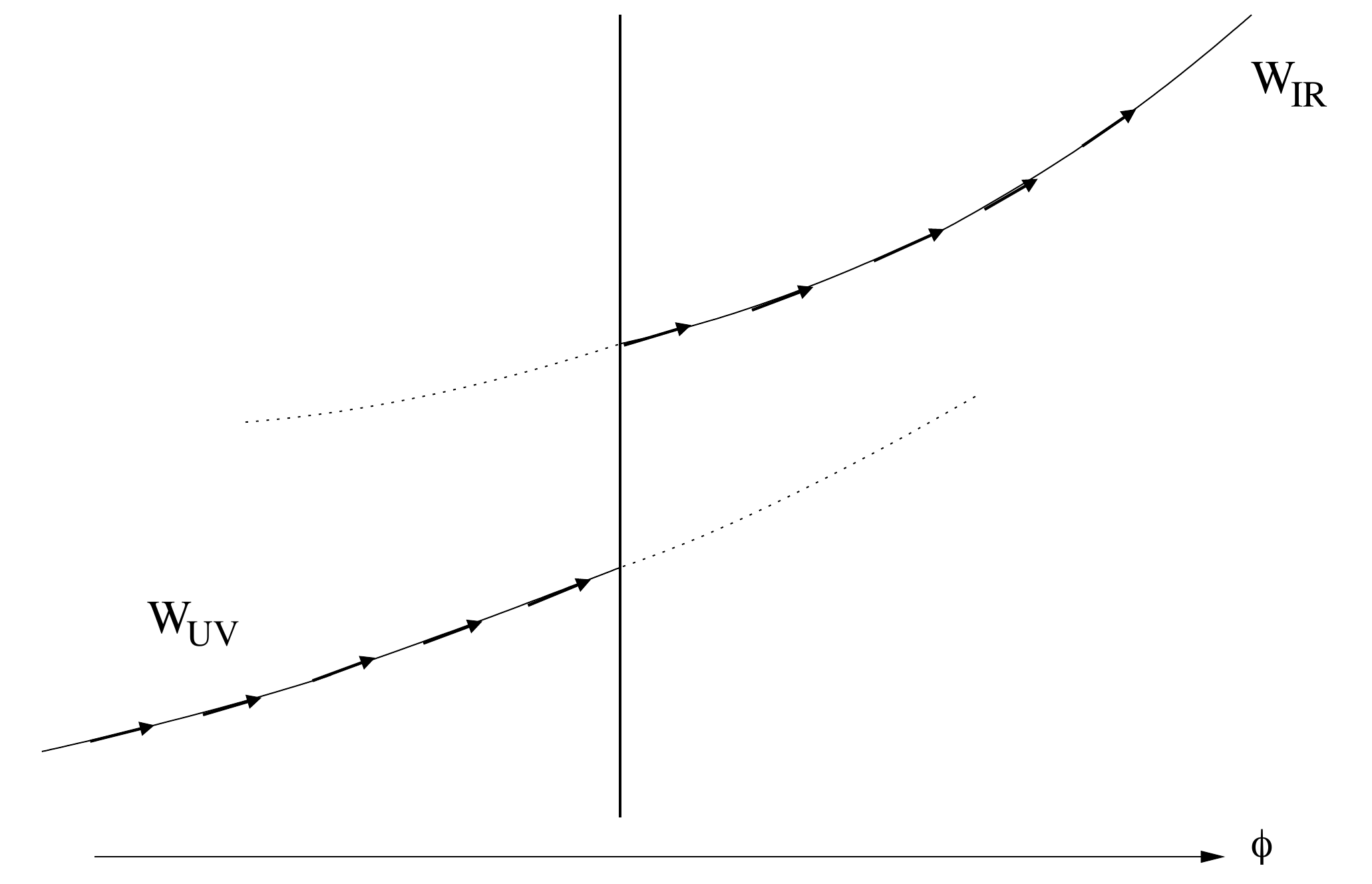}
\includegraphics[width=0.45\textwidth]{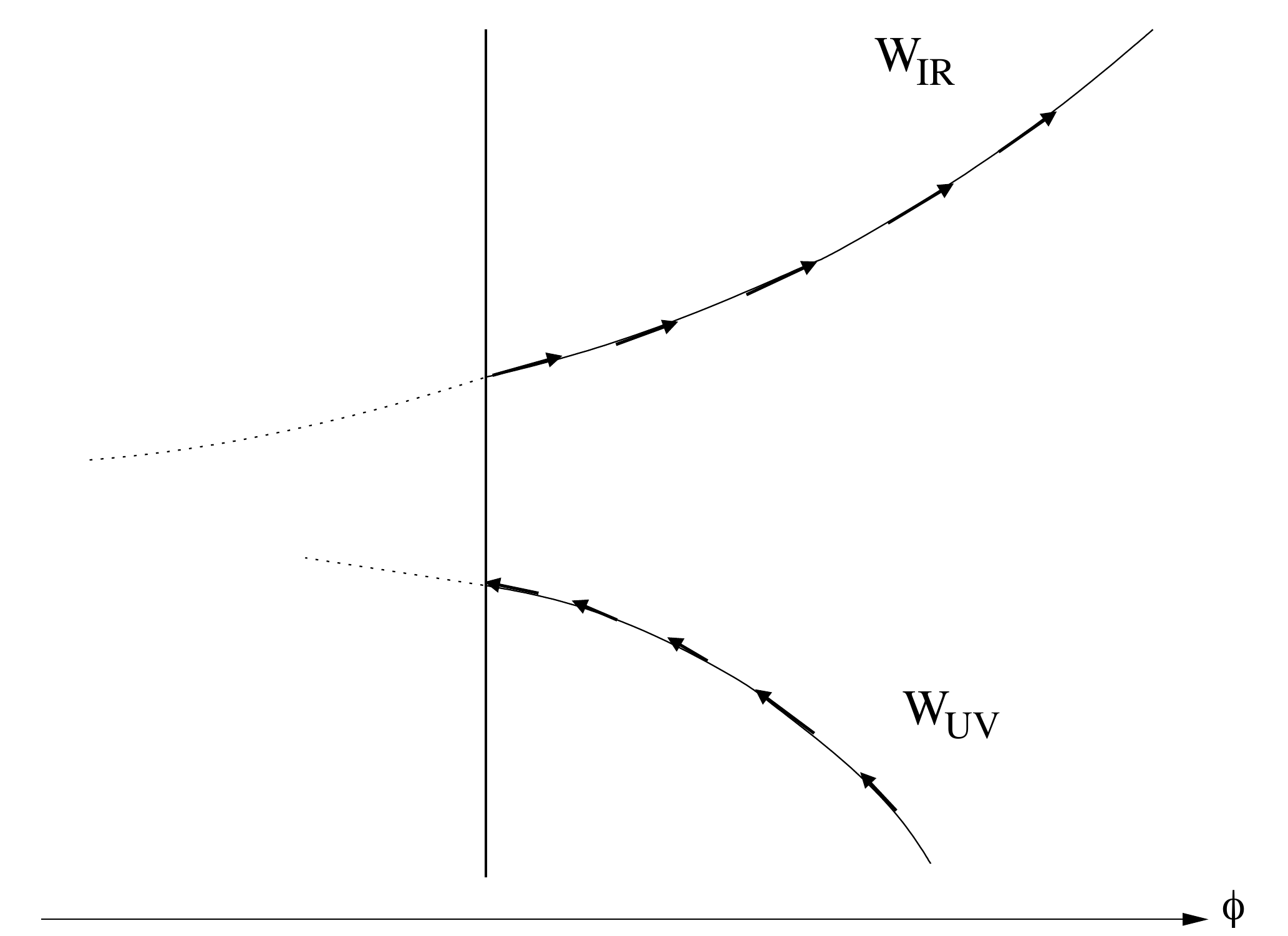}\\
$S_{UV}(\f_\star)>0$ \hspace{3cm} $S_{UV}(\f_\star) <0$
\caption{Junction rules in field space. The vertical solid line indicates the locus of the brane.}
\label{fig:junctionrules}
\end{figure}

Next, since  we will solve the matching conditions in field space, rather than
in coordinate space, we need to understand towards which side (i.e.~direction of increasing
or decreasing $\f$ away from $\f_{\star}$) one should follow the solution
$W_{UV}(\f)$ on the UV side of the brane. As we
discussed at the end of the previous subsection, if we start from the
IR side of the solution, the junction conditions determine the pair of
initial conditions  $(S_{UV}(\f_\star),W_{UV}(\f_\star))$ for the system
\eqref{WUV}--\eqref{SUV}, and we need to know if we should keep  the
solution for $\f >\f_{\star}$ or
$\f< \f_{\star}$. To understand what  the correct choice is, recall that in
our conventions the coordinate $u$ runs in  the same direction on both
sides of the brane, and we  take it to be  increasing from the UV to
the IR. therefore if the brane is at $u_\star$, the IR side is $u>u_\star$,
and the UV side is $u< u_\star$. Then, to be consistent with this choice,
it is the sign of $\dot{\f}(u_\star)\equiv S_{UV}(\f_\star)$ which decides
which one is the
right direction to follow on the UV side:
\begin{itemize}
\item If  $S_{UV}(\f_\star)>0$, then
$\dot{\f}(u_\star) >0$ at the brane, and we should take the UV solution
such that $\f$ increases towards the brane, i.e.~the solution
$W_{UV},S_{UV}$ for  $\f_{UV} <
\f_\star $.
\item  Conversely, if  $S_{UV}(\f_\star)<0$, we should take the other part
of the solution, the one with $\f_{UV} >
\f_\star $.
\end{itemize}
This junction rule is summarized graphically in figure \ref{fig:junctionrules}.

Finally, due to stability requirements of the solution, some care is
needed when choosing the combination of bulk and brane potentials
appearing in the action.  Because we will not pursue phenomenological
applications here, and in general we will not worry about whether the
functions chosen can lead to physics compatible with observation (e.g.~the presence of four-dimensional gravity in the brane), we will try to
require that at least the flat solutions  be free of ghosts and
tachyonic instabilities. Although this does not straightforwardly
guarantee stability of the curved solutions, there are strong
indications that this is the case, at least for the positive curvature
solutions, as discussed in section 4.5 of \cite{Ghosh:2017big}.

That there are no instabilities in the bulk is automatically
guaranteed if the scalar field kinetic term has the correct sign, and
there are no violations of the BF bound in the UV or in the
IR. However, some unstable modes can still arise due to the brane
fluctuations.  For the flat case, the analysis of \cite{1704.05075}
 showed that there are very
simple {\em sufficient} conditions which guarantee the absence of
ghosts and tachyonic instabilities for the self-tuning flat brane
solutions. These are:
\be\label{inst1}
U(\f_{\star}) > 0,  \qquad Z(\f_{\star}) >0
\ee

\be\label{inst2}
{W_B(\f_{\star}) \over W_{IR}(\f_{\star}) W_{UV}(\f_{\star})}> \frac{U(\f_{\star})}{3} ,
\ee

\be \label{inst3}
Z(\f_{\star})\left({W_B(\f_{\star}) \over W_{IR}(\f_{\star}) W_{UV}(\f_{\star})}- \frac{U(\f_{\star})}{3}\right)  >  \left({dU \over d\f}\right)_{\f_{\star}}^2,
\ee

\be\label{inst4}
W_{IR}''(\f_{\star}) - W_{UV}''(\f_{\star}) < W_{B}''(\f_{\star}).
\ee
Conditions (\ref{inst1}) are also \emph{necessary}: violating either of them would
guarantee the presence of either  a spin-2 or a spin-0 ghost. On the
other hand, the other conditions are only sufficient, but if they are
violated the stability analysis becomes much more involved, and
requires a detailed perturbation analysis of the full bulk solution.

Since here we will not aim to build explicit, realistic phenomenological
models, we will  not always strictly enforce  the conditions
\eqref{inst2}--\eqref{inst4}, not to limit too much the scope of the
examples we  study.

\subsection{Curved CFT boundary metrics vs.~variable scalar sources} \label{ssec:flat}

In this section we briefly comment on a possible alternative
realisation of the solutions described so far, in terms of a dual CFT
living in {\em Minkowski} space, but coupled to a non-trivial
time-dependent (for dS branes) or space-dependent (for AdS) external
source.

In the UV region, the bulk metric (\ref{metric}) takes the asymptotic
form (\ref{asympt}), where $\zeta^{UV}_{\mu\nu}$ is the metric to which the
dual CFT is coupled, and in the examples described above is
(A)dS$_d$. It is well known (see e.g.~\cite{KarchRandall}) that one can foliate AdS$_{d+1}$
by either Minkowski, dS$_d$ or AdS$_d$. From the bulk point of view,  these choices only differ by a
coordinate transformation.  From the dual field theory point of
view, however, different coordinate choices lead to different physical theories,
as the appropriate coordinate transformation acts non-trivially on the
conformal boundary and it changes both the  metric and the scalar
 sources.\footnote{A well-known example of this phenomenon is the
 difference between global and Poincar\'e AdS coordinates, which give
 a different structure of the conformal boundary metric and
 describe a dual field theory on $R\times S^{d-1}$, and  $R^{1,d-1}$,
 respectively.} Therefore, we can use these coordinate
transformations to find new  solutions with a different holographic
interpretation. In the rest of this section we consider the case of a
dS brane for definiteness.  The case of  AdS-slicing is more subtle
because the slices are non-compact and one can reach the boundary
along each radial slice. This means that we need boundary conditions
also at the boundary of the slices. This is related to the fact that,
as we explained in Section \ref{RG}, when
discussing the case $\kappa<0$,  there are two (disconnected)
boundary regions, which correspond to the field theory dual  theory having a
defect. These issues are discussed in \cite{Ghosh:2017big} and are
similar to those arising in the case of the Janus solution, \cite{skenderisham}.

Let us therefore consider dS-sliced domain walls. Going from a curved to a flat foliation by a coordinate transformation
is only possible if the bulk is {\em exactly} AdS$_{d+1}$, not only
asymptotically. However, we can still perform a coordinate
transformation  to new radial and time coordinates $(\rho,t)$ of the form
\be\label{flat1}
u = f(\rho,t), \qquad \tau = g(\rho,t),
\ee
such that close to the UV boundary it changes a dS slicing into a flat
one.  If we take  $\tau$ to be the conformal time coordinate in the de Sitter metric
ansatz, i.e. $\zeta_{\mu\nu}^{UV} = {H^2 \over \tau^2} \eta_{\mu\nu}$,
$f$ and $g$ can be  two arbitrary smooth functions
constrained only by  demanding that their asymptotic form leads to a
flat UV metric in the new coordinates ($\rho,t$),
$\zeta^{UV}_{\mu\nu} = \eta_{\mu\nu}$, in the limit $u\to -\infty$.
This imposes the following constraints,
\be\label{flat2}
f(\rho,t) \to \rho - \ell \ln\left(-{t\over\ell}\right), \qquad
g(\rho,t) \to t ,
  \qquad \rho \to -\infty ,
\ee
where $-\infty < t< 0$ (see e.g. section 4.4 of \cite{Ghosh:2017big} for
the full coordinate transformation, from which the limits (\ref{flat2}) can
be easily obtained).

In the original coordinates $(u,\tau)$ the brane was located at the equilibrium
position $u=u_\star$. In the new coordinates this will result in a
non-trivial trajectory,
\be\label{flat3}
f(\rho, t) = u_\star \quad  \Rightarrow  \quad \rho = \rho_\star(t).
\ee
The important point is that, by construction,
\begin{enumerate}
\item The junction conditions are still satisfied, since they have
  tensorial nature;
\item The induced metric on the brane is diffeomorphic to the original
  one before the change of coordinates.
\end{enumerate}
This implies that we have an alternative embedding of the same dS
brane, which is now moving
in an asymptotically AdS space-time whose asymptotic boundary has a
flat metric source. The dual field theory lives therefore in flat space.

This is not the end of the story however: recall that the bulk has
also a non-trivial  scalar field profile. In the old coordinates $(u,\tau)$ this is
described by a function $\f(u)$, which becomes a
time-dependent function $\f(\rho,t)$ in the new coordinates. This has
an important implication: writing  the near-boundary scalar field
asymptotics (\ref{UV0}) in the new coordinates using (\ref{flat2}),
we find
\be\label{flat4}
\f(\rho,t)\simeq 
\f_-
 \left(\ell \over | t|\right)^{\Delta_-} \, \ell^{\Delta_-} e^{\Delta_-
  \rho/\ell } + \ldots \qquad \rho \to -\infty.
\ee
In the holographic dictionary, this implies that the CFT  is coupled
to a
{\em time-dependent} external source
\be
j =  \left({\ell \over  |t|}\right)^{\Delta_-} \f_-
\ee
The source is switched on from $j=0$ at early times and increases  in
time as a power-law. Thus, in this language, cosmological de Sitter
expansion of the brane is driven by a time-dependent source in a
flat-space CFT. This gives an alternative (and, from the CFT
standpoint, inequivalent)  description of the
solutions we are discussing.

Notice that the only case in which one can embed a dS brane in a flat
CFT with no sources is the Karch-Randall-like setup, where the bulk is AdS
(in any coordinates), with no scalar
field. In this case a non-zero brane tension and/or induced Einstein
term generically require patching
together two AdS spaces with different curvatures, as was the case in
\cite{Padilla:2004tp,Padilla:2004mc,Charmousis:2007ji}.


\section{Solutions with IR fixed points} \label{caseI}
We now present implementations of the self-stabilisation mechanism in several example models. Thus, here and in what follows we set $d=4$. A particular model will be characterised by a choice of bulk potential $V(\f)$ and the brane quantities $W_B(\f)$ and $U(\f)$. While the functions $V$, $W_B$ and $U$ should be determined from a microscopic model, this goes beyond the scope of this paper. Instead, the functions will be chosen by hand and the consequences for self-stabilisation studied. Also, we will not be interested in constructing phenomenologically viable models, as this also goes beyond the scope of this investigation. The main goal of this section is to study the viability and efficacy of self-stabilisation in this holographic setting with non-zero UV curvature. In particular, we wish to answer the following questions:
\begin{enumerate}
\item How do self-stabilising solutions with non-zero UV curvature differ from the self-tuning solutions with vanishing UV curvature studied in \cite{1704.05075}?
\item How does the brane curvature $R_B$ depend on the UV curvature $R^{(\zeta)}$? E.g.~can $R_B$ be small while $R^{(\zeta)}$ is large (in suitable units) and vice versa.
\item Can the brane curvature $R_B$ be small in units of the 4d Planck mass $M_4$ on the brane without the need of tuning of model parameters?
\end{enumerate}
Finally, since analytical solutions are out of the question for generic setups, the results presented in this section come from numerical studies.

In the first part of the numerical investigation, we choose a generic bulk potential and focus on a region containing a maximum and bounded by its two neighbouring minima. To be specific, we choose
\begin{equation}
\label{eq:bulk1} V(\f)=\frac{1}{\ell^2} \left(-12+\frac{\Delta_-(4-\Delta_-)}{2}\f^2+\frac{\Delta_-(4-\Delta_-)}{4}\f^4 \right) \, ,
\end{equation}
where $\ell$ is the UV AdS length. In the following, we will set $\ell=1$ to remove clutter. The potential exhibits a maximum at $\f_{\textrm{max}}=0$ and two minima at $\f_{\textrm{min}}= \pm 1$. In the context of holography, the maximum is associated with a UV CFT perturbed by a scalar operator of dimension $\Delta_+ = 4 - \Delta_-$. The minima are associated with IR fixed points for flows with vanishing UV curvature, while flows for finite UV curvature end at generic points $\f_0$ with $- |\f_{\textrm{min}}| < \f_0 < |\f_{\textrm{min}}|$.

\subsection{IR-AdS with constant $U(\f)$}
To complete the model we also need to specify the brane potential $W_B$ and the function $U$. For the first numerical study we choose
\begin{equation}
\label{eq:brane1} W_B(\f)= \omega  \exp (\gamma  \f ) \, , \qquad U(\f)=1 \, ,
\end{equation}
with $\omega$ and $\gamma$ numerical parameters. This choice for $W_B$ ensures that it exhibits significant variation when the brane is displaced in $\f$. Furthermore, scalar fields with exponential potentials also arise frequently in effective (supergravity) theories obtained from string compactifications. In contrast, in this first example we choose $U(\f)$ to be unaffected with the position of the brane and hence take $U(\f)$ to be constant. This will be relaxed in the next numerical study in sec.~\ref{sec:expU}. To be specific, the numerical parameters used in this section are given by
\begin{align}
\label{eq:para1} \Delta_- = 1.2 \, , \qquad \omega=-0.015 \, , \qquad \gamma=5 \, .
\end{align}
Note that the phenomena we will describe do not hinge on this precise choice of parameters, which are in no way special.

Before we describe our findings, we describe scope and method of our numerical analysis. Here, all solutions will originate from the UV fixed point at the maximum $\f_{\textrm{max}}=0$. The UV theory at the fixed point is characterised by the values of the UV curvature $R^{(\zeta)}$ and the the coupling $\f_-$ of the dual theory, and we are free to adjust these parameters. The idea is to scan over these parameters, or equivalently over $\mathcal{R} = R^{(\zeta)} |\f_-|^{-2 / \Delta_-}$ and the sign of $\f_-$. For every choice of UV sources we then check whether the flow admits a solution of the junction conditions at some value $\f_{\star}$ sec.~\ref{junc} where the brane can be located. In addition, we need to make sure that on the IR side of the brane the solution is completed to a non-singular flow ending at some $\f_0$.

In practice, it is more convenient to work backwards, i.e.~to begin at an IR end/turning point $\f_0$ and follow the flow backwards. The reason is that solutions that are regular in the IR only constitute a small subset compared to solutions which are singular in the IR. Hence, when starting from the UV it is numerically hard to pick out IR regular solutions. In contrast, no such problem arises when starting with a regular solution in the IR. This can typically be followed back to the UV without any problems as the UV fixed point is an attractor. Thus, our strategy is as follows:
\begin{enumerate}
\item We begin at some IR end/turning point $\f_0$ and solve for $W_{IR}$, $S_{IR}$ and $T_{IR}$ with end or turning point at $\f_0$ with regular boundary conditions as listed in appendix \ref{app:IR}.
\item Given this solution, we then turn to the junction conditions \eqref{JC} to search for possible brane loci $\f_{\star}$. Note that the junction conditions may have several solutions.
\item For every tentative brane locus $\f_{\star}$ we then check whether the solution can be continued on the UV side. To be acceptable, a solution on the UV side has to connect to the UV fixed point at $\f_{\textrm{max}}=0$.
\item Once we have an acceptable solution we can then read off the UV data $\mathcal{R}$ and the sign of $\f_-$ as well as the corresponding brane curvature $R_B = T(\f_{\star})$.
\item To ensure that we capture all solutions associated with $\f_{\textrm{max}}=0$ we employ the following strategy. By performing this analysis for any value of the end/turning point $\f_0$ in the interval $- |\f_{\textrm{min}}| < \f_0 < |\f_{\textrm{min}}|$ we can make sure that no solutions are missed. This is equivalent to scanning over all possible values of $\mathcal{R}$ and $\textrm{sign}(\f_-)$ in the UV.
\item Last, note that both on the UV or IR side of the brane the RG flow solutions can in principle exhibit one or multiple bounces, i.e.~reversals of the flow in $\f$ as described at the end of sec.~\ref{RG}.
\end{enumerate}

\begin{figure}[t]
\centering
\begin{overpic}
[width=0.75\textwidth]{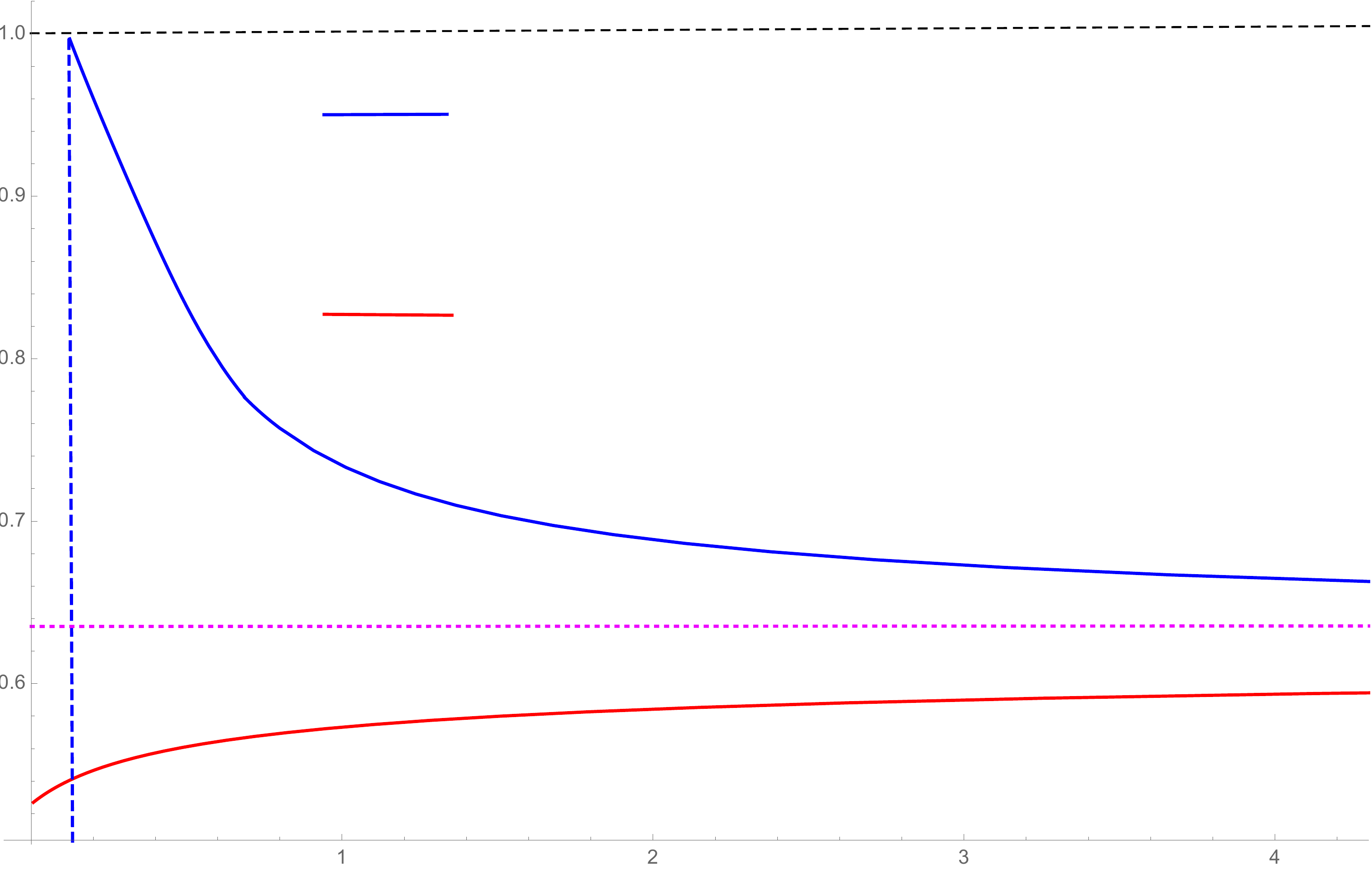}
\put(100,2){$\mathcal{R}$}
\put(1,64){$\f_\star$}
\put(34,54){$\f_-<0$}
\put(34,40){$\f_-> 0$}
\put(4,-1){$\mathcal{R}_c$}
\end{overpic}
\caption{Equilibrium brane position $\f_\star$ vs.~$\mathcal{R}$ for bulk potential \protect\eqref{eq:bulk1}, brane quantities \protect\eqref{eq:brane1} and parameter values \protect\eqref{eq:para1}. The blue line corresponds to solutions with $\f_- < 0$ while the red line denotes solutions with $\f_->0$. The brane position cannot exceed the position of the minimum of the potential at $\f_{\textrm{min}}=1$, which is indicated by the black dashed line. For $\mathcal{R} \rightarrow \infty$ both the red and blue branch asymptote to the value for $\f_{\star}$ indicated by the dotted magenta line.}
\label{phistarvsRdS}
\end{figure}
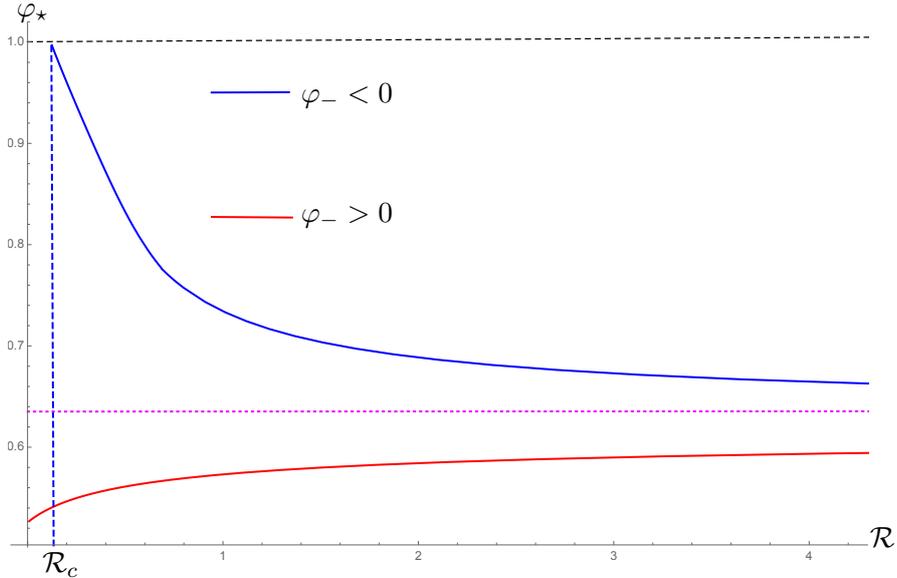

\subsubsection*{Solutions with $\mathcal{R} > 0$}
As we will see, the space of solutions is very rich. One way of organising our results is then to distinguish between solutions with positive brane (and UV) curvature and negative brane (and UV) curvature. Here we begin with the former case.

To capture all solutions we solve numerically, scanning over all possible end points $\f_0$ with $- |\f_{\textrm{min}}|< \f_0 < |\f_{\textrm{min}}|$. For every possible end point we check for solutions and, if they exist, we record the equilibrium brane position $\f_{\star}$, the brane curvature $R_B$ and UV data (dimensionless curvature $\mathcal{R}$ and sign of the source $\f_-$).

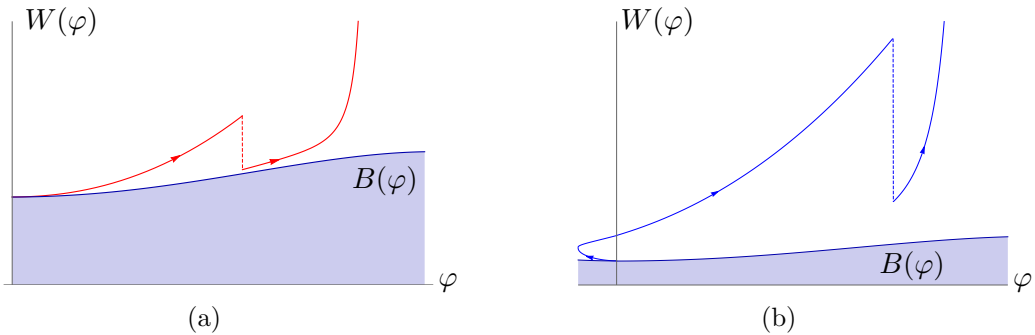
\begin{figure}[t]
\centering
\begin{subfigure}{.5\textwidth}
 \centering
  \begin{overpic}
[width=0.75\textwidth]{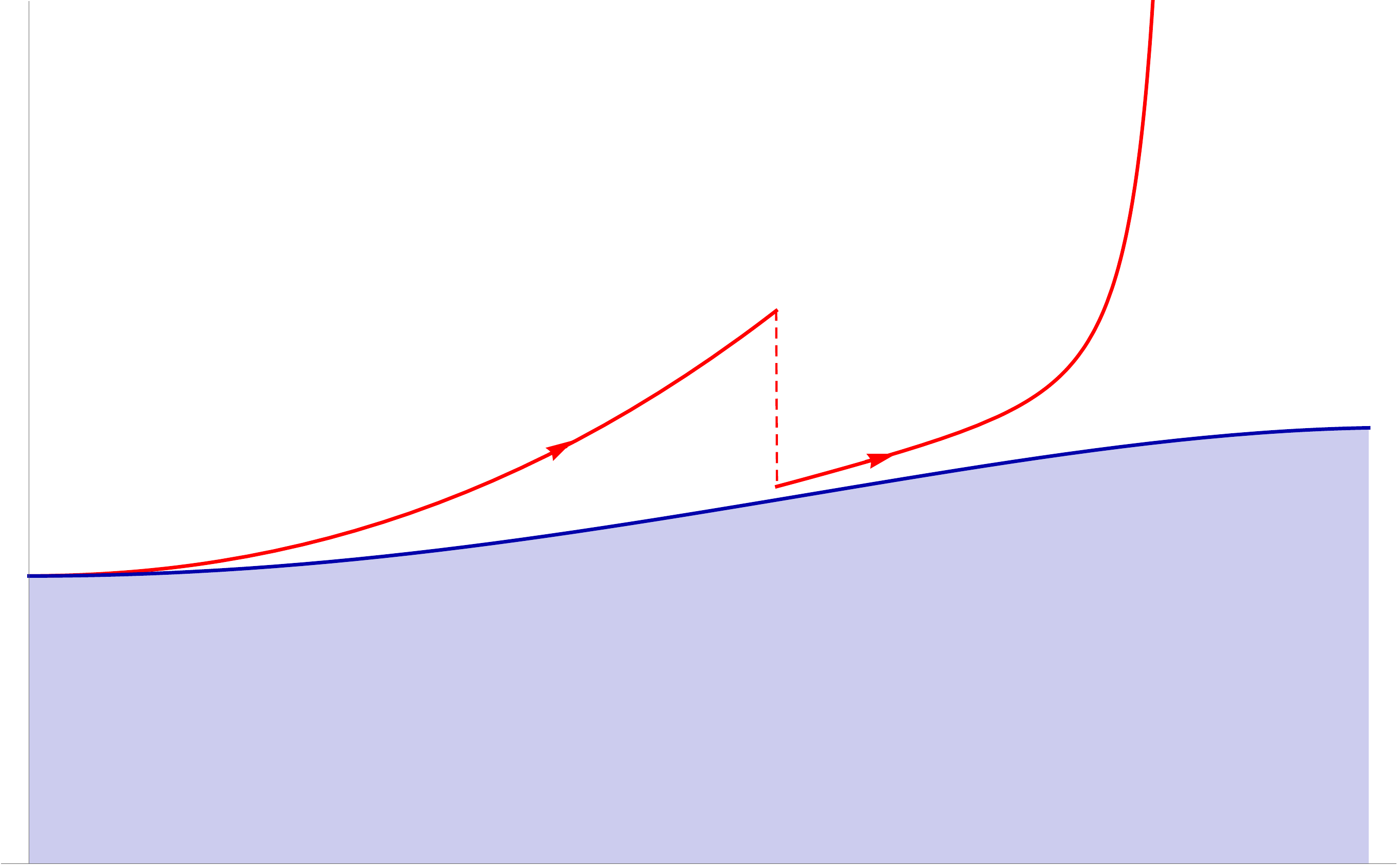}
\put(101,0){$\f$}
\put(5,60){$W(\f)$}
\put(81,23){$B(\f)$}
\end{overpic}
 \caption{\hphantom{A}}
  \label{fig:VpolUconstWnb}
\end{subfigure}%
\begin{subfigure}{.5\textwidth}
  \centering
\begin{overpic}
[width=0.75\textwidth]{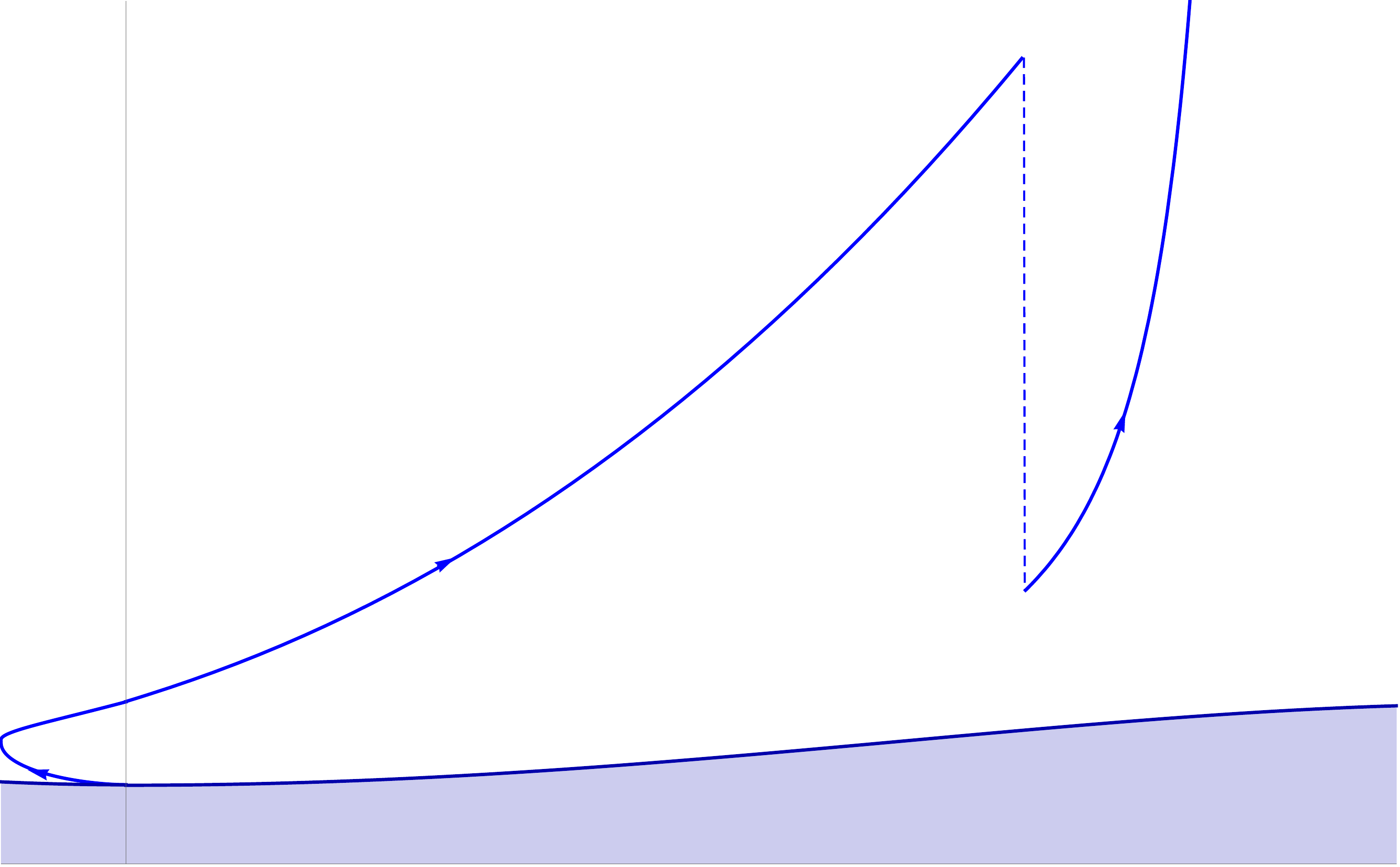}
\put(101,0){$\f$}
\put(10,60){$W(\f)$}
\put(70,3){$B(\f)$}
\end{overpic}
\caption{\hphantom{A}}
\label{fig:VpolUconstWb}
\end{subfigure}
\caption{Two solutions for $W(\f)$ with the same value of $\mathcal{R}$ but different sign of $\f_-$. The results were obtained for bulk potential \protect\eqref{eq:bulk1}, brane quantities \protect\eqref{eq:brane1} and parameter values \protect\eqref{eq:para1}. The jump in $W$ is the discontinuity across the brane. In fig.~(a) we plot a solution with $\f_- >0$, while in (b) a solution with $\f_- < 0$ is shown. Note that the solution with $\f_- <0$ exhibits a bounce, i.e.~a reversal of direction in $\f$.}
\label{fig:VpolUconstW}
\end{figure}

One convenient way of presenting the space of all solutions, is to plot brane quantities ($\f_{\star}, R_B$) vs.~UV data. For example, in fig.~\ref{phistarvsRdS} we plot the brane equilibrium position $\f_{\star}$ vs.~the dimensionless curvature $\mathcal{R}$. We can then make the following observations.
\begin{itemize}
\item We find that for a given value of $\mathcal{R}$ there can be up to two solutions, one with $\f_->0$ (red) and one with $\f_- <0$ (blue). Recall that $\f_-$ is the UV value of the coupling of the theory dual to this geometry and thus a parameter that is fixed once the dual theory is specified..
\item Solutions on the red branch in fig.~\ref{phistarvsRdS} exist for all values of $\mathcal{R} >0$, i.e.~there is no gap in $\mathcal{R}$. One can further check that for $\mathcal{R} \rightarrow 0$ this branch of solutions continuously connects to a solution with $\mathcal{R}=0$ and a flat brane. In contrast, solutions on the blue branch only exist for $\mathcal{R}$ larger than some particular value $\mathcal{R}_c$, with $\mathcal{R}_c \simeq 0.16$.
\item We examine the solutions on the two branches in some more detail. In figure \ref{fig:VpolUconstW} we plot $W(\f)$ for two representative solutions with $\f_->0$ and $\f_-<0$ corresponding to the same value of $\mathcal{R}$. In both cases (fig.~\ref{fig:VpolUconstWnb} and fig.~\ref{fig:VpolUconstWb}) the brane equilibrium position is at some positive value $\f_{\star}$ and $W$ is discontinuous there. The main difference is that for $\f_-<0$ shown in fig.~\ref{fig:VpolUconstWb} the `flow' leaves the UV fixed point to the left. The solution then reverses direction in $\f$ (i.e.~it `bounces') and continues towards positive values of $\f$. In the the case $\f_->0$ shown in fig.~\ref{fig:VpolUconstWnb} the flow leaves the UV to the right and no subsequent reversal of direction occurs.
\item From fig.~\ref{phistarvsRdS} we can understand why solutions with $\f_-<0$ (the blue branch) stop existing for $\mathcal{R} < \mathcal{R}_c$. When approaching $\mathcal{R} \rightarrow \mathcal{R}_c$ the brane equilibrium position on the blue branch is pushed onto the minimum of the potential $\f_{\textrm{min}}=1$. As the brane cannot move beyond the minimum this branch of solutions ends.
\item We now return to the discussion of the full space of solutions in fig.~\ref{phistarvsRdS}. Another observation is that the equilibrium position for the brane over the whole solution space is limited to a particular band in $\f$. Here we find that the brane can only be located in the range $\f_{\star,0} \leq \f_{\star} < |\f_{\textrm{min}}|$, with $\f_{\star,0} \simeq 0.525$. For example, this implies that for this model the brane can never sit at values $\f_{\star} <0$, even though we did not exclude this in the analysis.
\item Finally, note that the two branches of solutions ($\f_->0$  and
  $\f_-<0$) are connected. However, to go from $\f_->0$ to $\f_-<0$
  the source has to cross $\f_-=0$ and the dimensionless curvature
  $\mathcal{R}= R^{(\zeta)} |\f_-|^{-2/ \Delta_-}$ diverges. Thus,
  the point where the two branches of solutions meet is displaced to
  $\mathcal{R} \rightarrow \infty$ in fig.~\ref{phistarvsRdS}. The
  corresponding value of $\f_{\star}$ is indicated by the dotted
  magenta line. The solution with $\f_-=0$ corresponds to a vev flow
  in the gravity dual language, as the RG flow is induced not by the
  source $\f_-$ of an operator $\mathcal{O}$, but by its vev $\langle
  \mathcal{O} \rangle$. This solution has with a fixed value of $\chi
  \equiv R^{(\zeta)} |\langle \mathcal{O} \rangle|^{-2 /
    \Delta_+}$. Thus, it is actually a one-parameter space of solutions spanned by all
  values of $R^{(\zeta)}$ and $\langle \mathcal{O} \rangle$ such
  that $\chi$ remains constant. For more details we refer readers to
  \cite{Ghosh:2017big}. A similar situation has been extensively discussed in
  \cite{Gursoy:2018umf} for  thermal states in holographic theories,
  where the role of curvature is taken by temperature.
\end{itemize}

\begin{figure}[t]
\centering
\begin{subfigure}{.55\textwidth}
 \centering
\begin{overpic}
[width=0.75\textwidth]{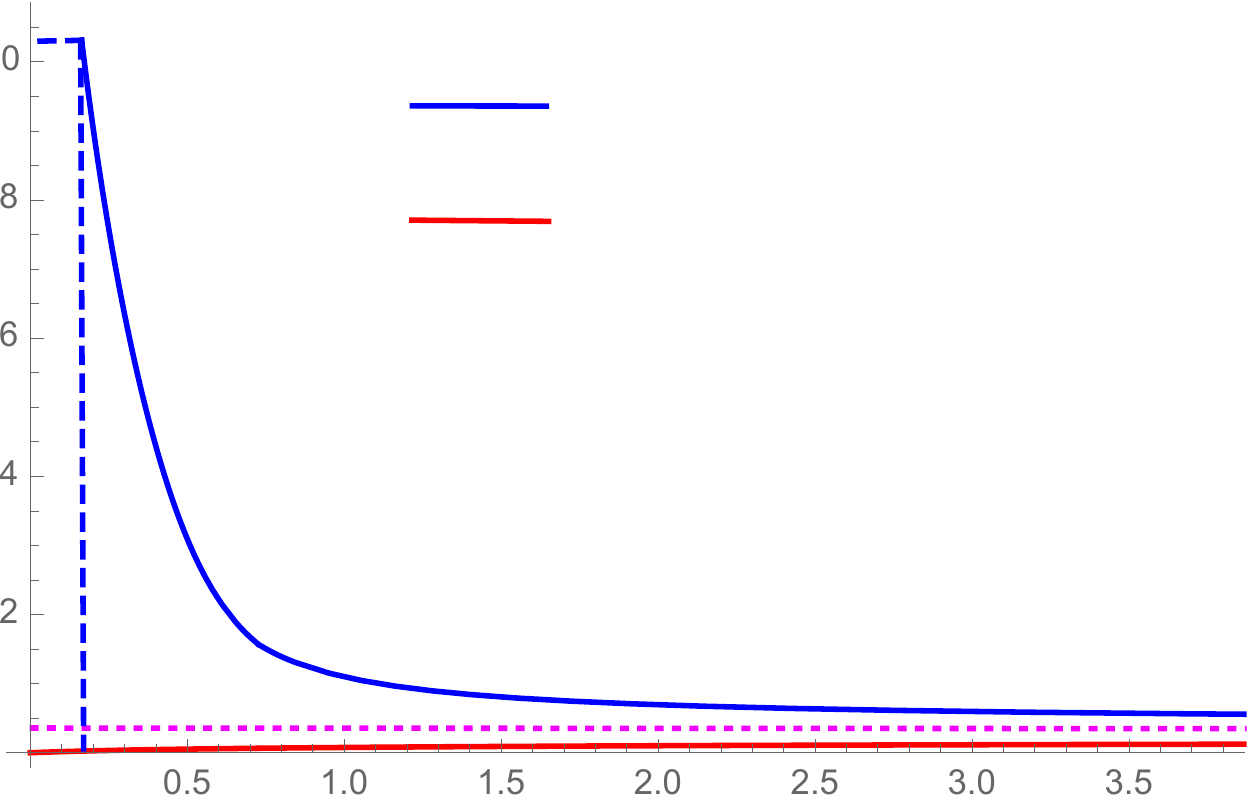}
\put(100,2){$\mathcal{R}$}
\put(-1,67){$R_B M_4^{-2}$}
\put(3,-2){$.16$}
\put(-5,4){$.2$}
\put(45,53){$\f_-<0$ }
\put(45,43){$\f_->0$ }
\end{overpic}
 \caption{\hphantom{A}}
  \label{RbvsRdSnozoom}
\end{subfigure}%
\begin{subfigure}{.55\textwidth}
  \centering
\begin{overpic}
[width=0.75\textwidth]{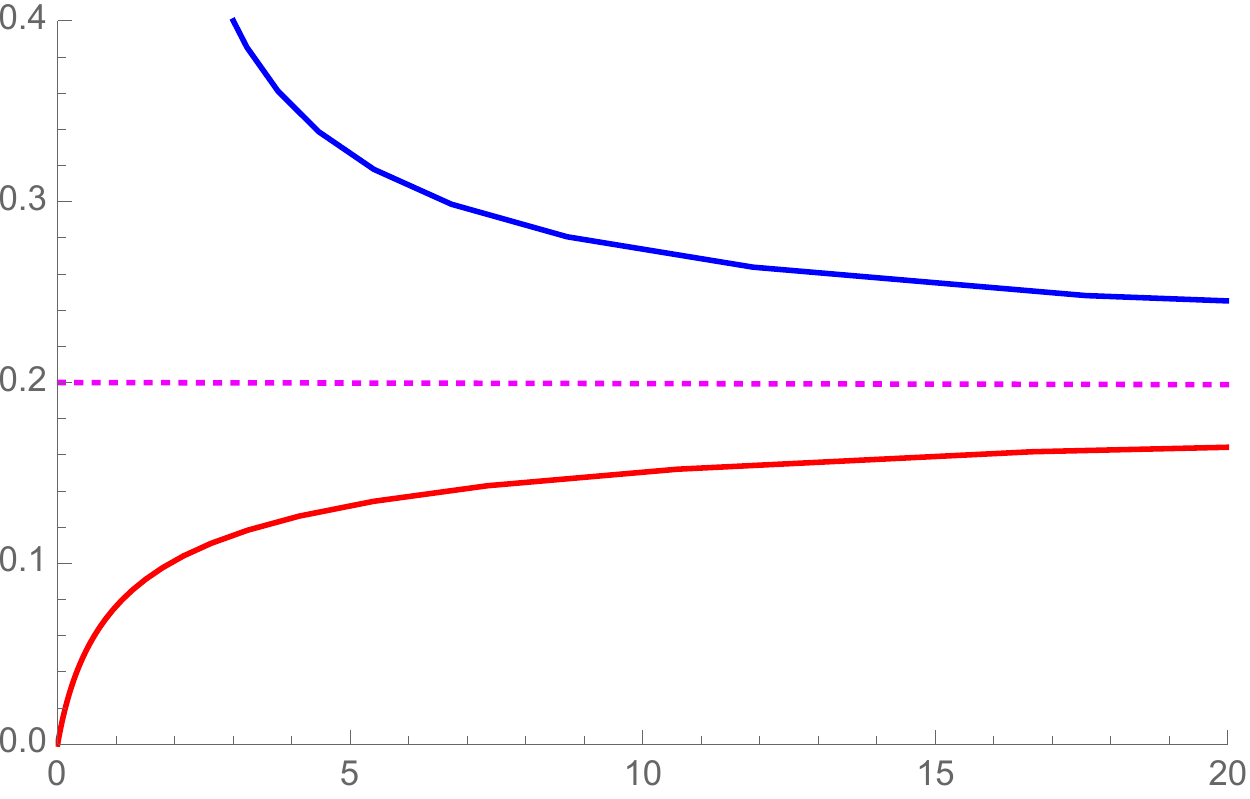}
\put(100,2){$\mathcal{R}$}
\put(-3,66){$R_B M_4^{-2}$}
\end{overpic}
\caption{\hphantom{A}}
\label{RbvsRdSzoom}
\end{subfigure}
\caption{Brane curvature in units of the 4d Planck mass $R_B \, M_4^{-2}$ vs.~$\mathcal{R}$ for bulk potential \protect\eqref{eq:bulk1}, brane quantities \protect\eqref{eq:brane1} and parameter values \protect\eqref{eq:para1}. In fig.~(a) both solutions with $\f_-<0$ and $\f_->0$ are shown. In fig.~(b) we adjusted the plot range to improve visibility of the result with $\f_- >0$. For $\mathcal{R} \rightarrow \infty$ both the red and blue branch asymptote to the value for $R_B \, M_4^{-2}$ indicated by the dotted magenta line.}
\label{RbvsRdS}
\end{figure}

\begin{figure}[t]
\centering
\begin{overpic}
[width=0.75\textwidth]{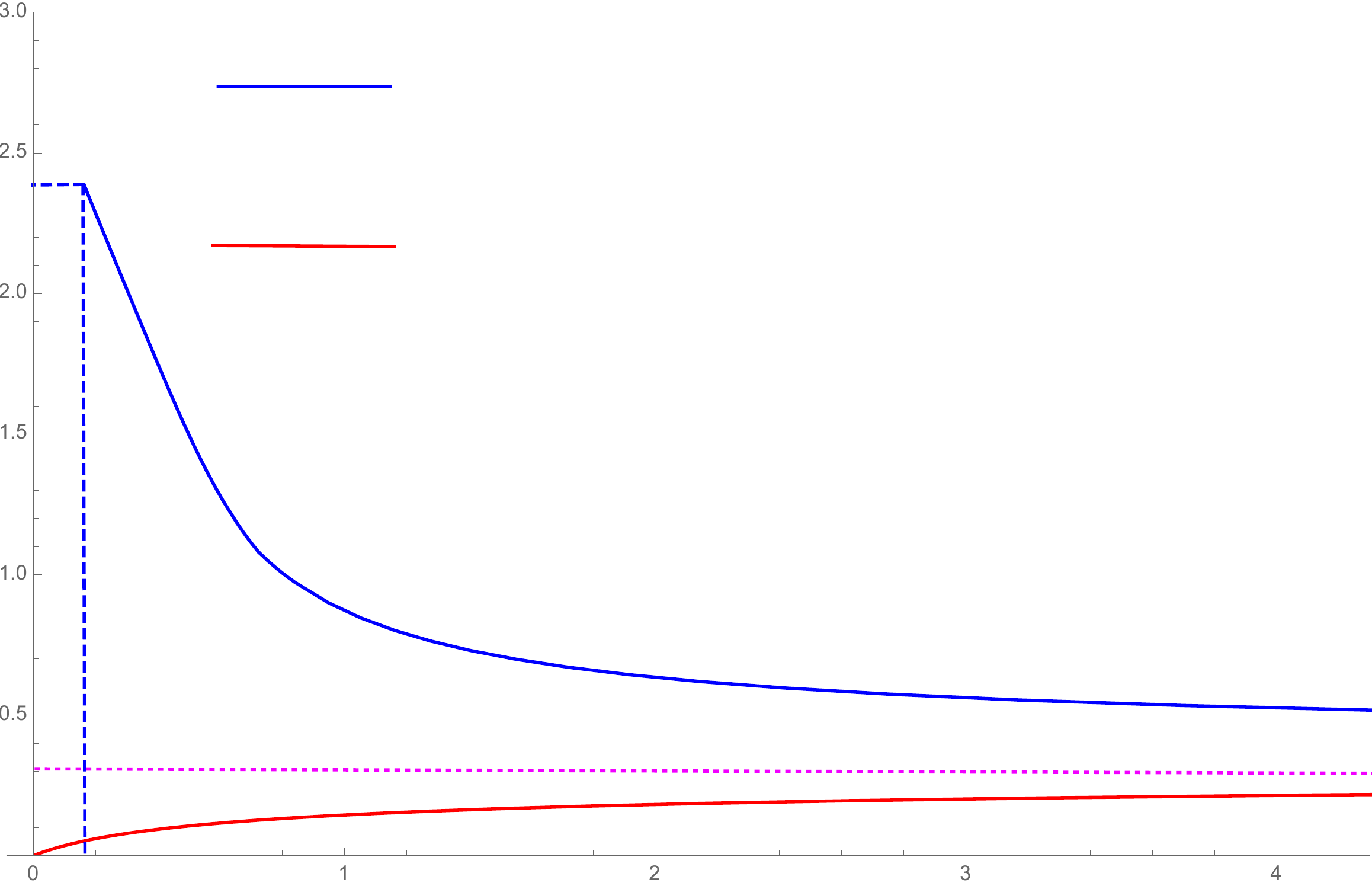}
\put(101,0){$\mathcal{R}$}
\put(1,65){$R_B/|R_\star|$}
\put(30,58){$\f_-<0$}
\put(30,46){$\f_-> 0$}
\put(4,-1){$.16$}
\end{overpic}
\caption{$R_B/|R_\star|$ as a function of $\mathcal{R}$ for bulk potential \protect\eqref{eq:bulk1}, brane quantities \protect\eqref{eq:brane1} and parameter values \protect\eqref{eq:para1}. $|R_{\star}|$ is defined as the `expected' brane curvature related to the brane cosmological constant $W_B(\f_{\star})$ in the absence of a 5d bulk. For $\mathcal{R} \rightarrow \infty$ both the red and blue branch asymptote to the value for $R_B / |R_{\star}|$ indicated by the dotted magenta line.}
\label{RboverRstarU1dS}
\end{figure}
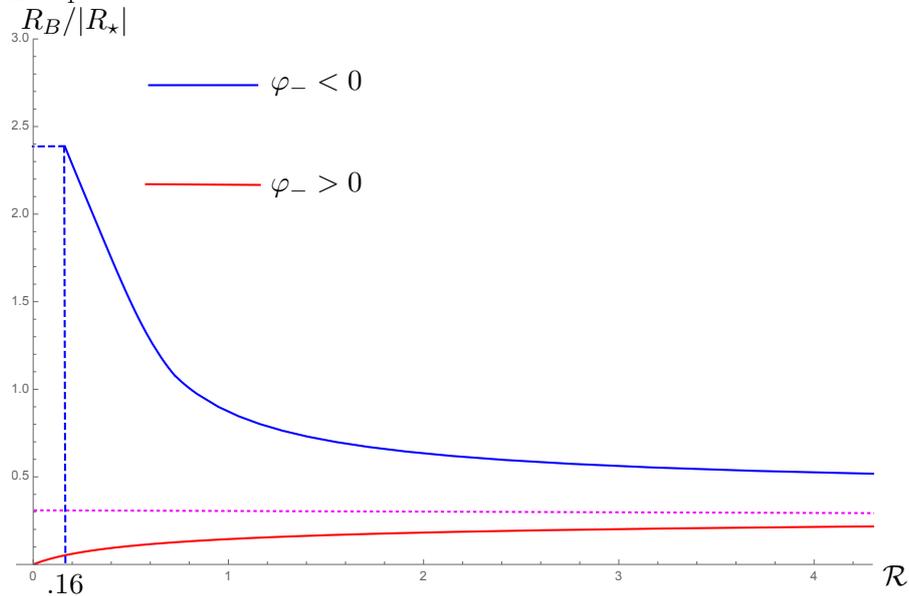

Having discussed the space of solutions, we will now describe how the brane curvature $R_B$ varies across the solution space. In particular, for 4-dimensional observers the main quantity of interest is brane curvature $R_B$ in units of the 4-dimensional Planck scale $M_4$. This is defined as
\begin{align}
\label{eq:M4def} M_4^2 \equiv M_P^2 \, U_B(\f_{\star}) \, ,
\end{align}
where $M_P$ is the 5d Planck scale as introduced in \eqref{sbulk} and \eqref{sbrane}.
Thus, in fig.~\ref{RbvsRdS} where we display $R_B \, M_4^{-2}$ vs.~$\mathcal{R}$. We again use the colours blue and red to distinguish between the two branches of solutions with $\f_- <0$ and $\f_- > 0$ respectively. We make the following observations.

\begin{figure}[t]
\centering
\begin{overpic}
[width=0.75\textwidth]{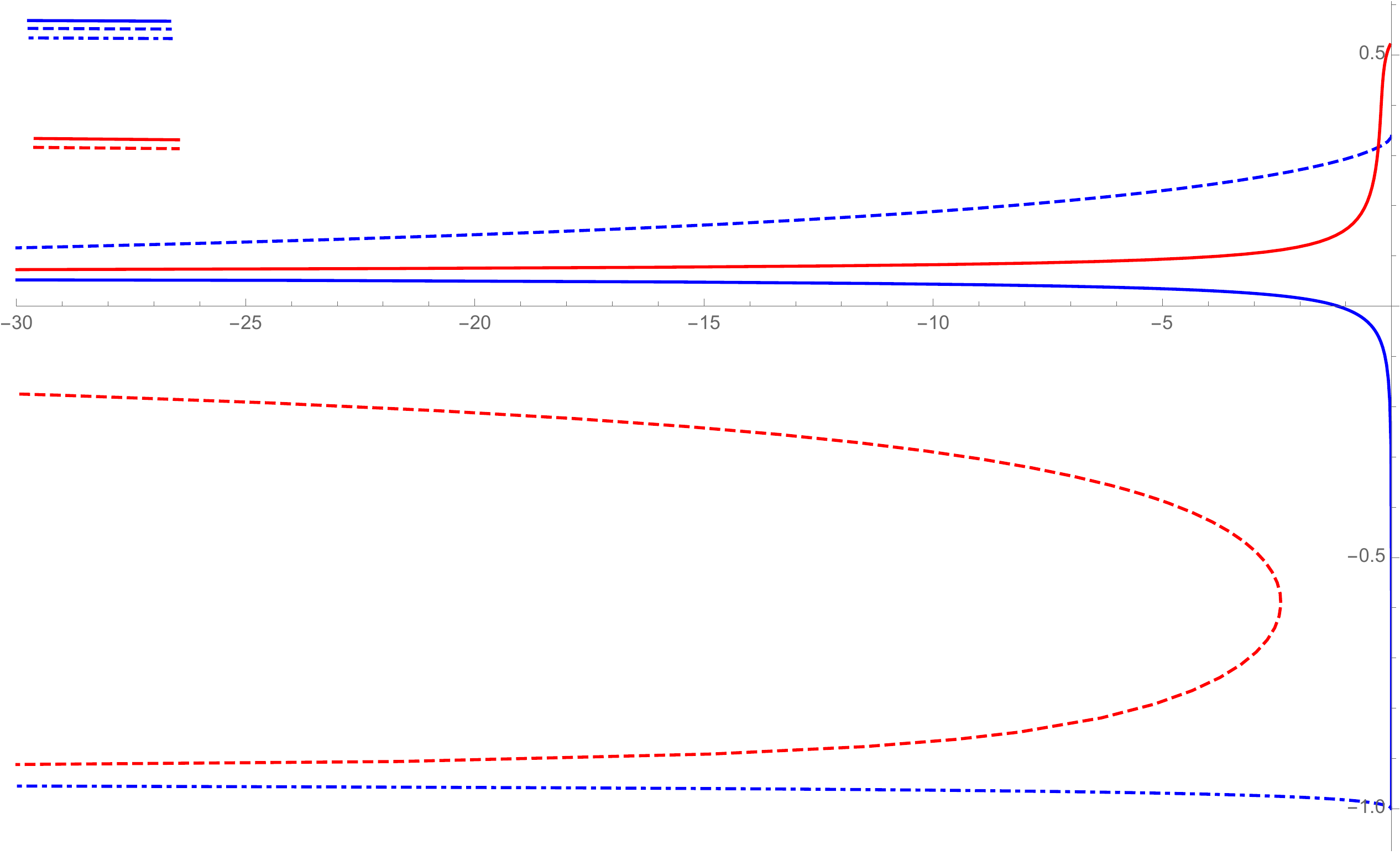}
\put(-3,37){$\mathcal{R}$}
\put(98,64){$\f_\star$}
\put(13,58){$\f_-<0$}
\put(13.2,50){$\f_->0$}
\end{overpic}
\caption{Equilibrium brane position $\f_\star$ vs.~$\mathcal{R}$ for $\mathcal{R} < 0$. Results are obtained for bulk potential \protect\eqref{eq:bulk1}, brane quantities \protect\eqref{eq:brane1} and parameter values \protect\eqref{eq:para1}. Blue lines corresponds to solutions with $\f_- < 0$ while red line denotes solutions with $\f_->0$.}
\label{phistarvsRAdS}
\end{figure}
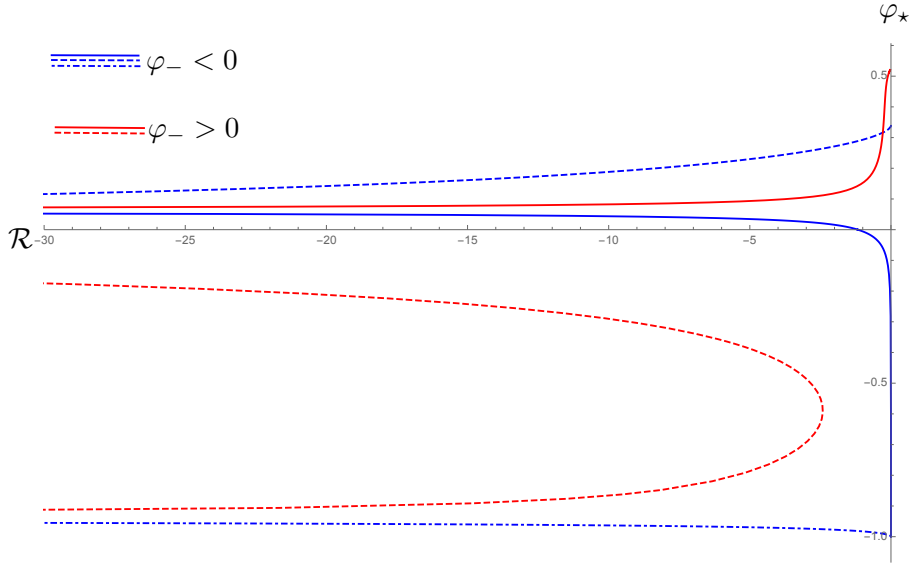

\begin{itemize}
\item We begin by describing the branch of solutions with $\f_- > 0$, which is seen most clearly in fig.~\ref{RbvsRdSzoom}. This branch of solutions exists for all values of $\mathcal{R}$. Note that for $\mathcal{R} \rightarrow 0$ we find $R_B \rightarrow 0$. Thus, in this limit we recover a solution with a flat brane in a setup with no UV curvature, a configuration studied before in \cite{1704.05075}. The brane curvature $R_B$ then grows monotonically with $\mathcal{R}$. For $\mathcal{R} \rightarrow \infty$ the brane curvature then asymptotes to the value $R_{B,0} M_4^{-2} \simeq 0.2$ from below.
\item The branch of solutions with $\f_- < 0$ is depicted most clearly in fig.~\ref{RbvsRdSnozoom}. Recall that solutions of this type only exist for $\mathcal{R} \geq \mathcal{R}_c \simeq 0.16$. At this point the brane curvature takes the maximum value $R_{B,\textrm{max}} M_4^{-2} \simeq 10.6$. It then falls monotonically with increasing $\mathcal{R}$, asymptoting to $R_{B,0} M_4^{-2} \simeq 0.2$ from above for $\mathcal{R} \rightarrow \infty$.
\end{itemize}

In the following, we will also quantify to what extent the embedding of the brane in the bulk modifies the value for $R_B$ compared to the value expected from the 4d cosmological constant alone. In particular, a brane at position $\f_\star$ exhibits a cosmological constant $W_B(\f_\star)$. If 4d gravity was the only physics governing the geometry of the brane, this would result in a brane curvature
\begin{align}
R_\star = 2 \, \frac{W_B(\f_\star)}{U(\f_\star)} \, .
\end{align}
Hence, in fig.~\ref{RboverRstarU1dS} we plot $R_B/ |R_\star|$ vs.~$\mathcal{R}$. We put an absolute value sign as $R_{\star} <0$ in this example. We make the following observations. Generically one finds $R_B/ |R_\star| \sim \mathcal{O}(1)$. That is, while the presence of the bulk modifies the value of $R_B$ somewhat, it is is no way `tuned' small generically. The only regime where $R_B/ |R_\star| \ll 1$ is on the red branch when $\mathcal{R} \rightarrow 0$. That is, we find that $R_B/ |R_\star|$ is small exactly when $R_B M_4^{-2}$ is also small, as can be seen by comparing with fig.~\ref{RbvsRdS}. We will make the same qualitative observations for $R_B/ |R_\star|$ vs.~$\mathcal{R}$ for all following models in this work. Hence we refrain from plotting $R_B/ |R_\star|$ vs.~$\mathcal{R}$ in the remainder of this work.

Before moving on to solutions with $\mathcal{R}<0$, we summarise the main findings. There are four points to be taken away from studying this example:
\begin{enumerate}
\item The range of possible values that $R_B$ can take is bounded in the model studied here. In particular, we find that $R_B$ is constrained to lie in the interval $0 \leq R_B < R_{B,\textrm{max}}$.
\item The solution with $R_B =0$ also exhibits $\mathcal{R}=0$. Thus, in this model a flat brane is only obtained if the UV curvature also vanishes and vice versa.
\item We observe that a hierarchically low brane curvature $R_B M_4^{-2} \ll 1$ only occurs in the vicinity of the flat solution. More precisely, we only find solutions with $R_B M_4^{-2} \ll 1$ if we also choose $\mathcal{R} \ll 1$. For generic values of $\mathcal{R}$ we find that $R_B M_4^{-2} \sim \mathcal{O}(1)$.
\item Most interestingly, we find that even when the UV curvature diverges, $\mathcal{R} \rightarrow \infty$, the brane curvature stays finite.
\end{enumerate}

\begin{figure}[t]
\centering
\begin{subfigure}{.5\textwidth}
 \centering
  \begin{overpic}
[width=0.75\textwidth]{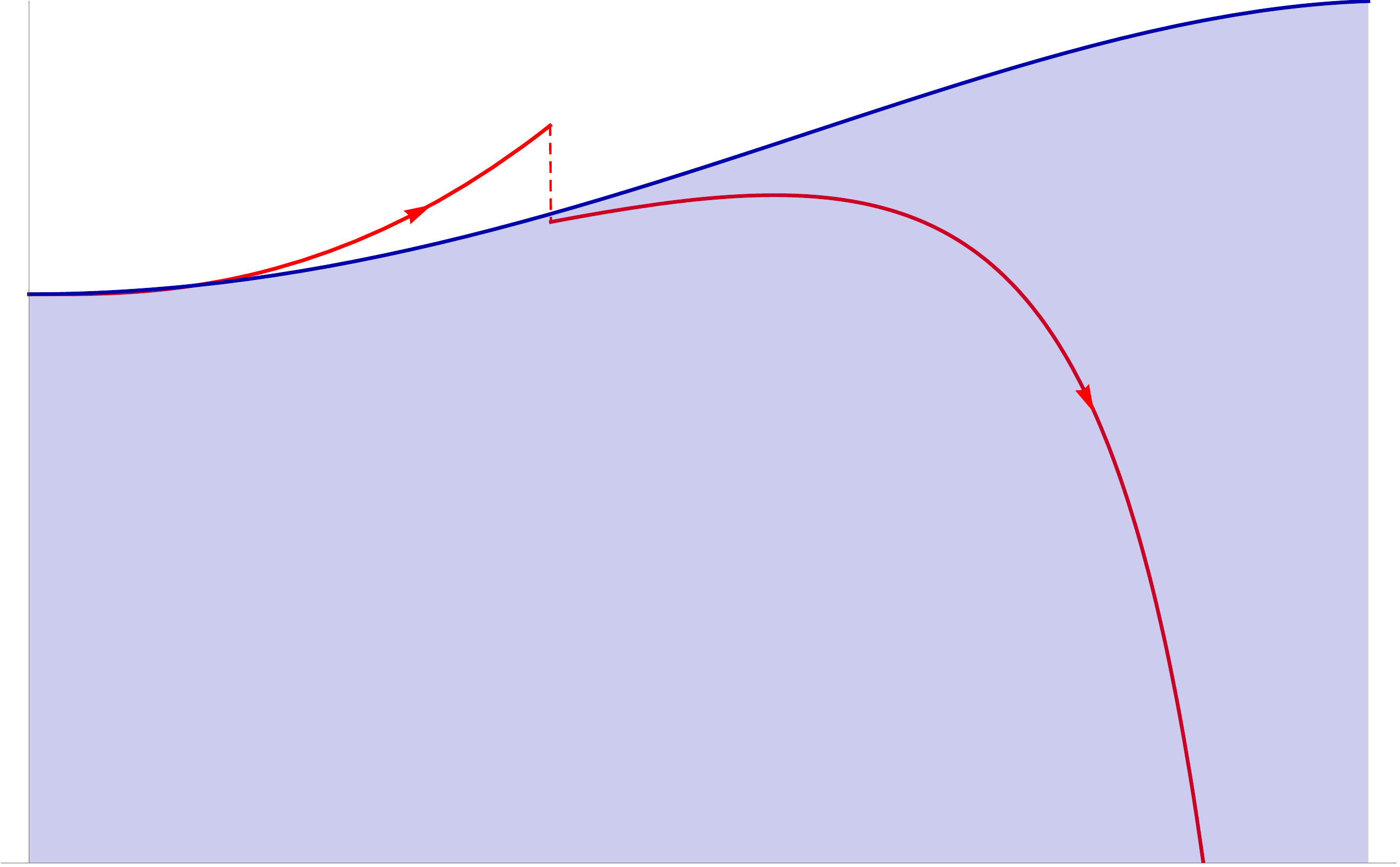}
\put(101,2){$\f$}
\put(8,60){$W(\f)$}
\put(79,50){$B(\f)$}
\end{overpic}
 \caption{\hphantom{A}}
  \label{fig:VpolUconstWAdSnb}
\end{subfigure}%
\begin{subfigure}{.5\textwidth}
  \centering
\begin{overpic}
[width=0.75\textwidth]{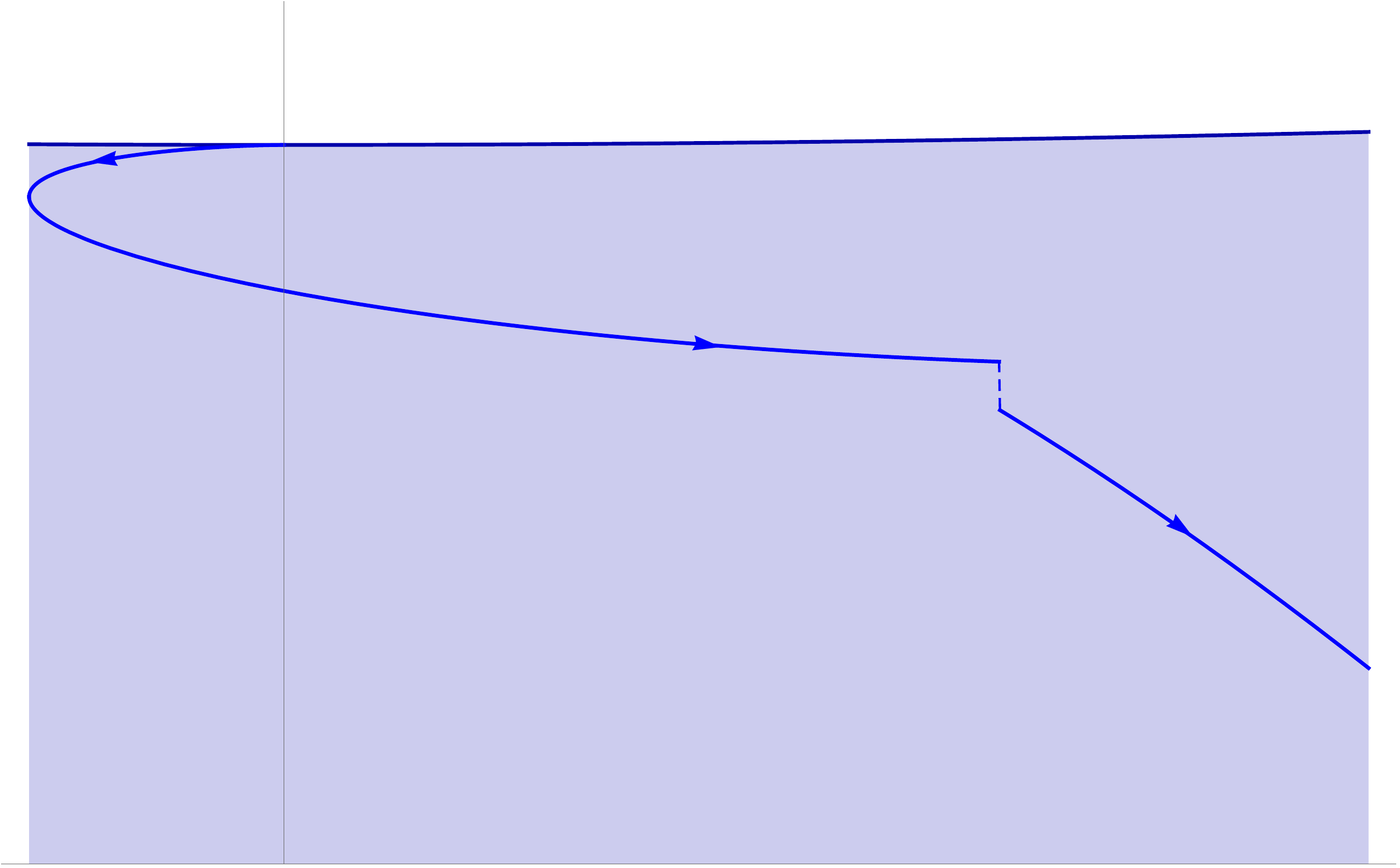}
\put(101,2){$\f$}
\put(25,60){$W(\f)$}
\put(78,42){$B(\f)$}
\end{overpic}
\caption{\hphantom{A}}
\label{fig:VpolUconstWAdSb}
\end{subfigure}
\caption{Two solutions for $W(\f)$ with the same (negative) value of $\mathcal{R}$ but different sign of $\f_-$. The results were obtained for bulk potential \protect\eqref{eq:bulk1}, brane quantities \protect\eqref{eq:brane1} and parameter values \protect\eqref{eq:para1}. The jump in $W$ is the discontinuity across the brane. In fig.~(a) we plot a solution with $\f_- >0$, while in (b) a solution with $\f_- < 0$ is shown. Note that the solution with $\f_- <0$ exhibits a bounce, i.e.~a reversal of direction in $\f$.}
\label{fig:VpolUconstWAdS}
\end{figure}

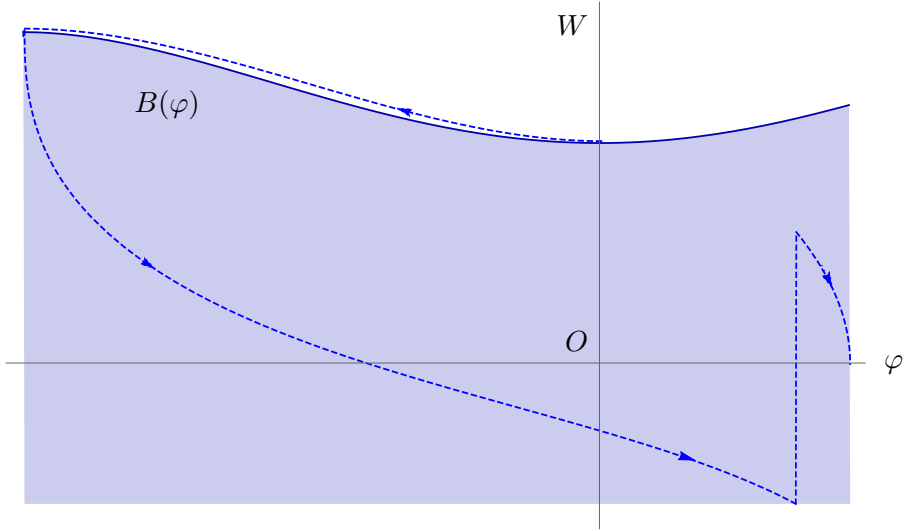
\begin{figure}[t]
\centering
\begin{overpic}
[width=0.75\textwidth]{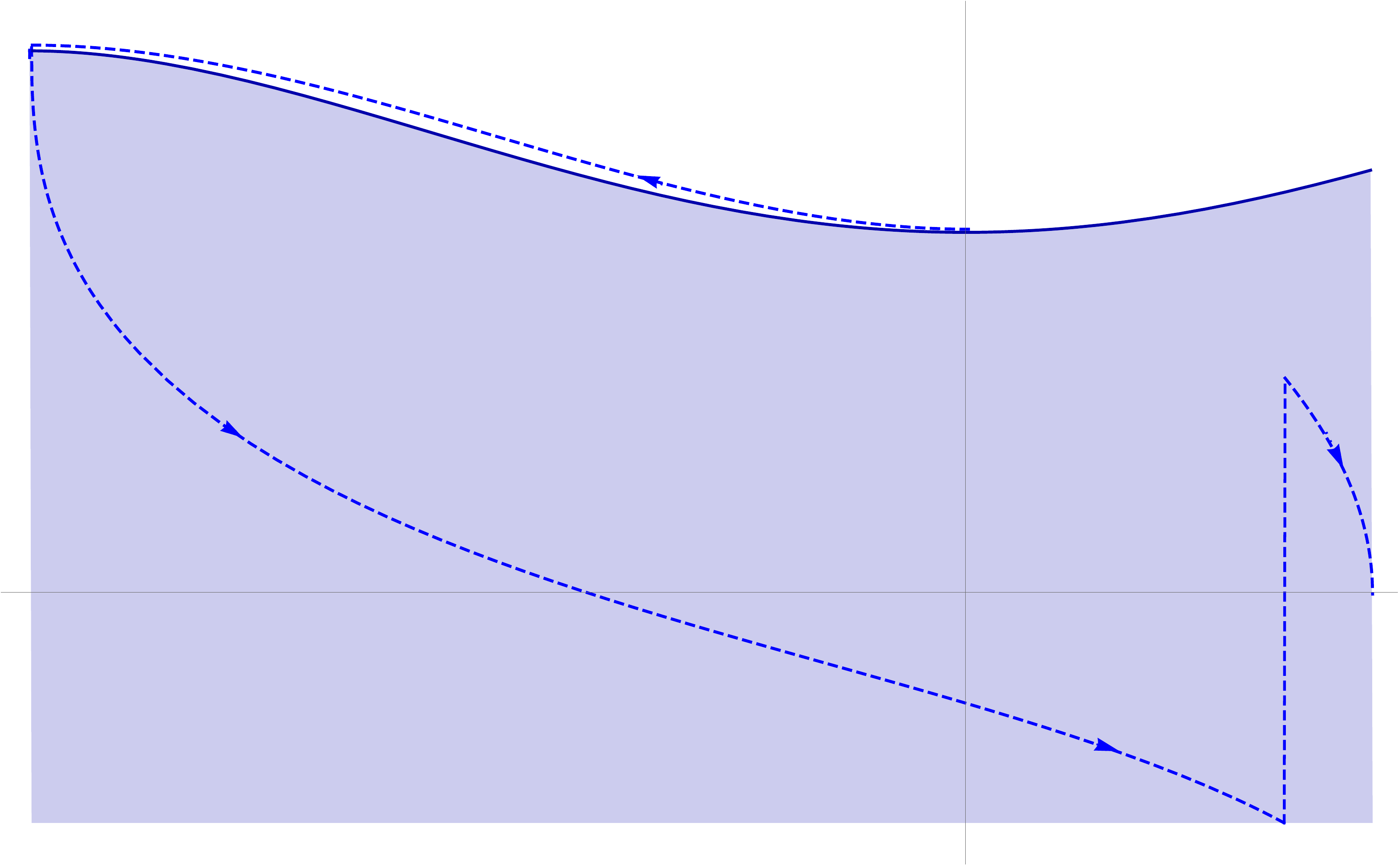}
\put(102,19){$\f$}
\put(64,58){$W$}
\put(15,49){$B(\f)$}
\put(65,21){$O$}
\end{overpic}
\caption{$W(\f)$ for $\mathcal{R} <0$. The results were obtained for bulk potential \protect\eqref{eq:bulk1}, brane quantities \protect\eqref{eq:brane1} and parameter values \protect\eqref{eq:para1}. This is a solution of the branch of solutions denoted by the dashed blue line in \protect\ref{phistarvsRAdS} with $\mathcal{R} \rightarrow 0$. The solution exhibits a reversal of direction, and for $\mathcal{R} \rightarrow 0$ the bounce locus is pushed towards a minimum of the potential. Note that this solution also exhibits a region with $W <0$.}
\label{fig:UconstNegRbranch3}
\end{figure}

\subsubsection*{Solutions with $\mathcal{R} <0$}
The space of solutions with negative UV curvature is even richer than in the positive curvature case. We plot $\f_{\star}$ vs.~$\mathcal{R}$ for $\mathcal{R} <0$ in fig.~\ref{phistarvsRAdS}. We see that for a given value of $\mathcal{R}$ up to six solutions exist, three of which exhibit $\f_- < 0$ (denoted in blue) and three $\f_- > 0$ (red). Note that branch denoted by the dashed red line only exists when the UV curvature is sufficiently negative, i.e.~$\mathcal{R} < \mathcal{R}_{c,2}$ with $\mathcal{R}_{c,2} \simeq -3.5$. The other branches exist for all $\mathcal{R} <0$.

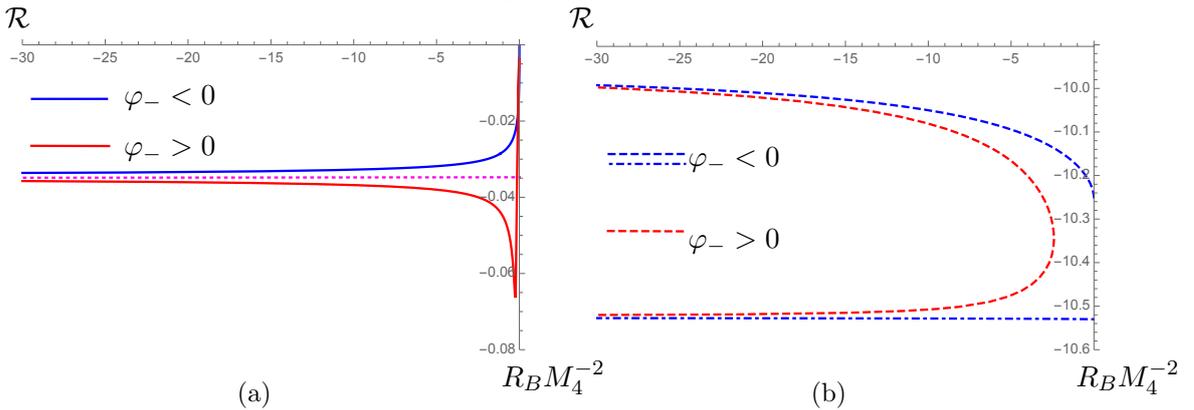
\begin{figure}[t]
\centering
\begin{subfigure}{.5\textwidth}
 \centering
\begin{overpic}
[width=0.9\textwidth]{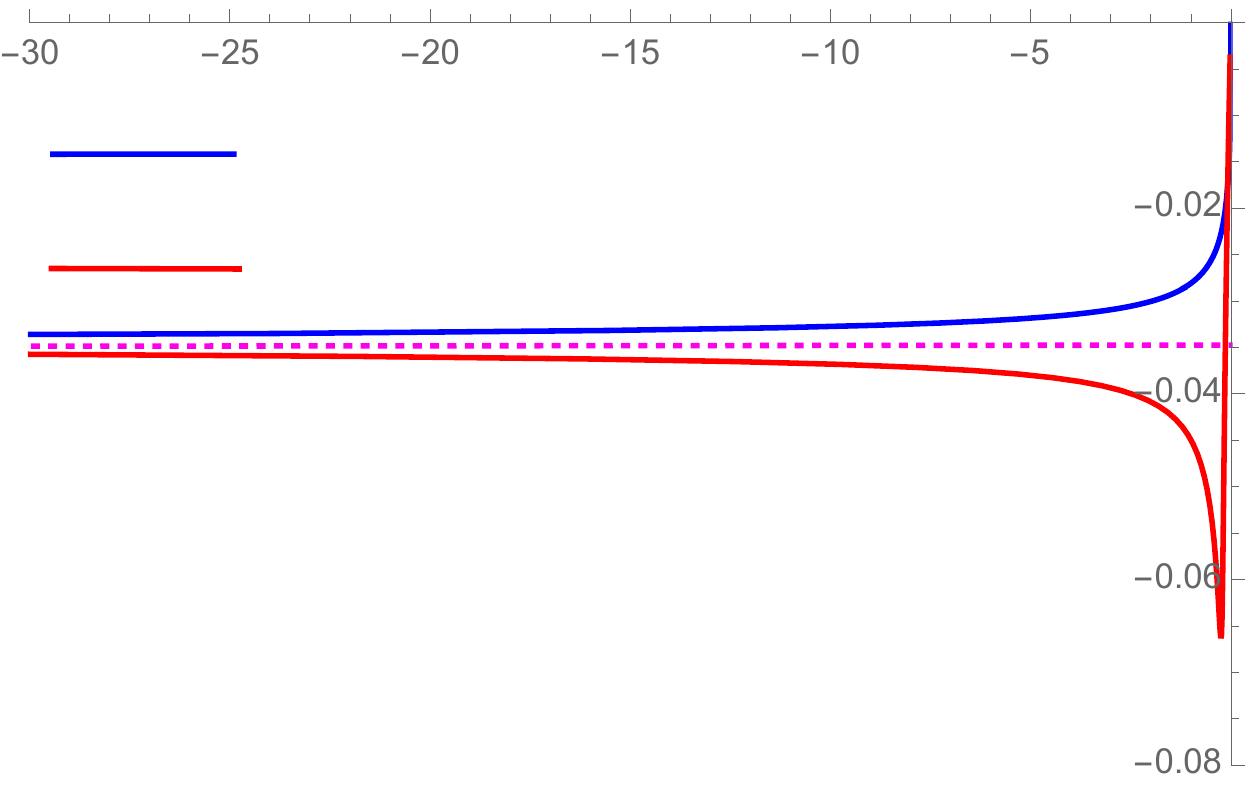}
\put(95,-5){$R_B M_4^{-2}$}
\put(-1,65){$ \mathcal{R} $}
\put(22,50){$\f_-<0$ }
\put(22,40){$\f_->0$ }
\end{overpic}
 \caption{\hphantom{A}}
  \label{RbvsCurlyRAdS1_new}
\end{subfigure}%
\begin{subfigure}{.5\textwidth}
  \centering
\begin{overpic}
[width=.9\textwidth]{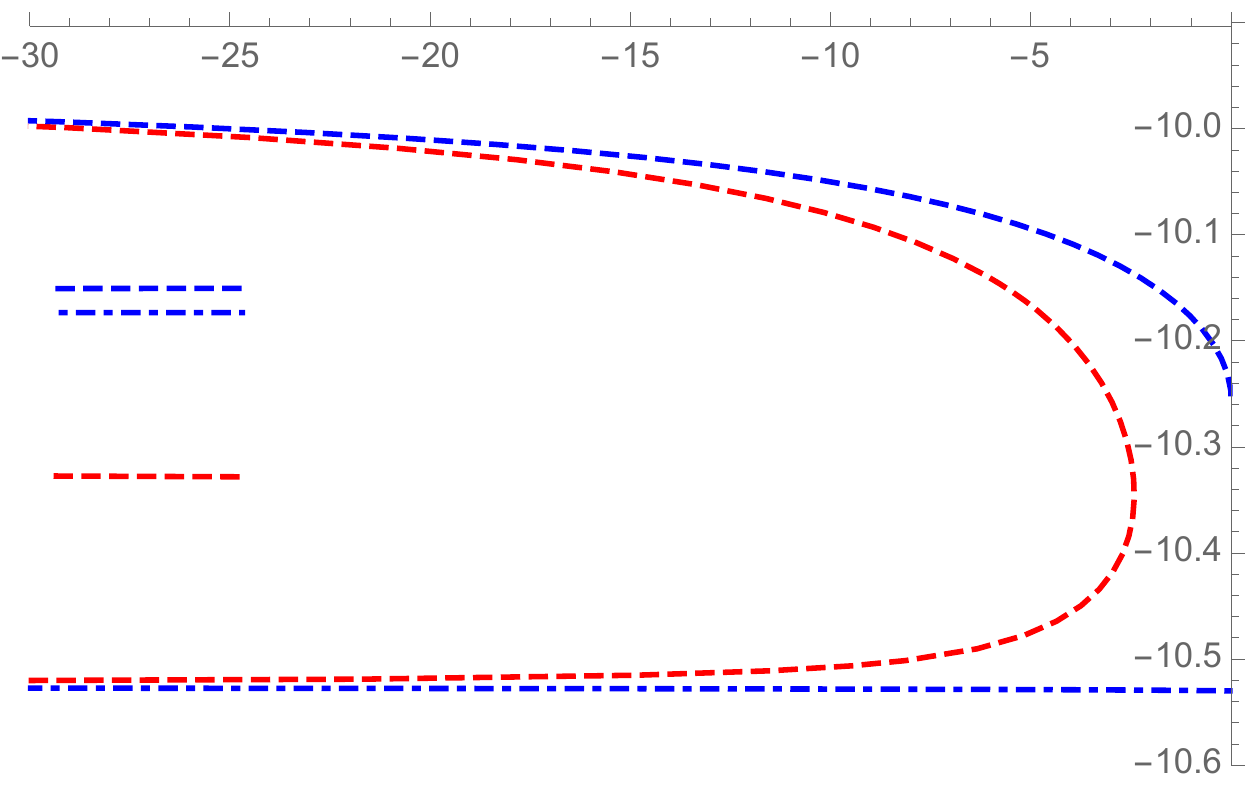}
\put(95,-5){$R_B M_4^{-2}$}
\put(-3,65){$ \mathcal{R}$}
\put(20,38){$\f_-<0$}
\put(20,22){$\f_->0$}
\end{overpic}
\caption{\hphantom{A}}
\label{RbvsCurlyRAdS2_new}
\end{subfigure}
\caption{Brane curvature in 4d Planck units $R_B M_4^{-2}$ vs.~$\mathcal{R}$ with $\mathcal{R}<0$ for bulk potential \protect\eqref{eq:bulk1}, brane quantities \protect\eqref{eq:brane1} and parameter values \protect\eqref{eq:para1}. In fig.~(a) we show results corresponding to the branches denoted by solid lines in fig.~\protect\ref{phistarvsRAdS}. In fig.~(b) the results for the dashed and dot-dashed branches in fig.~\protect\ref{phistarvsRAdS} are displayed.}
\label{RbvsRdS}
\end{figure}

We display example solutions for $W(\f)$ with $\mathcal{R}<0$ in figures \ref{fig:VpolUconstWAdS}. For one, we find solutions without a bounce, i.e.~ without a change of direction, as shown in fig.~\ref{fig:VpolUconstWAdSnb}. This is observed, for example, on the solid red branch in fig.~\ref{phistarvsRAdS}. Also note that for $\mathcal{R}<0$ we have $W=0$ in the IR. A representative solution for $W(\f)$ with a bounce is shown in fig.~\ref{fig:VpolUconstWAdSnb}. This occurs, for example, on the solid blue branch with $\f_{\star} >0$.

Before moving on, we will  have a closer look at the four branches in fig.~\ref{phistarvsRAdS} that admit a limit $\mathcal{R} \rightarrow 0$:
\begin{itemize}
\item We find that the solid red branch is continuously connected to a solution with $\mathcal{R}=0$ exactly. The brane position approaches and finally coincides with that of a flat brane when letting $\mathcal{R} \rightarrow 0$.
\item For the solid blue branch the brane is pushed towards the minimum of the potential at $\f_{\textrm{min}} =-1$ when letting  $\mathcal{R} \rightarrow 0$. While it can be arbitrarily close to $\f_{\textrm{min}}$, it can never coincide with it. The reason is that for $\f \rightarrow \f_{\textrm{min}}$ the bulk geometry shrinks to a point. Thus the strict limit $\f_\star \rightarrow \f_{\textrm{min}}$ is equivalent to the absence of the brane.
\item A more interesting phenomenon is observed for $\mathcal{R}$ for the solutions described by the dashed and dot-dashed blue lines in fig.~\ref{phistarvsRAdS}. This is best seen by plotting $W(\f)$ for the dashed blue branch for $\mathcal{R} \rightarrow 0$. This is shown in fig.~\ref{fig:UconstNegRbranch3}. Note that this solution exhibits a bounce, and for $\mathcal{R} \rightarrow 0$ the location of the bounce approaches the minimum of the potential at $\f_{\textrm{min}} =-1$. We can then understand why the solution does not exist in the strict limit $\mathcal{R} = 0$. In this case the flow leaving the UV fixed point at $\f_{\textrm{max}}=0$ passes through $\f_{\textrm{min}}$ exactly, which becomes the IR end point for the flow. The geometry shrinks to a point there and the flow ends before reaching the brane, thus excluding a solution with $\mathcal{R} = 0$.\footnote{The remaining part of the flow, i.e.~the part beyond  $\f_{\textrm{min}} =-1$ can then be understood as a solution where now the minimum  at $\f_{\textrm{min}} =-1$ plays the role of UV fixed point. Hence it is not part of the space of solutions associated UV fixed point $\f_{\textrm{max}}=0$.}
\end{itemize}
To summarise, while we have many solutions that permit arbitrarily small $\mathcal{R}$, only the branch denoted by the solid red line is continuously connected to a solution with $\mathcal{R} =0$.

We briefly return to fig.~\ref{fig:UconstNegRbranch3} and make another comment. Note that this solution exhibits a region with $W < 0$ and hence $W$ changes sign. As $W \sim \dot{A}$ this implies that $A$ is not monotonic in $u$. In the case of holographic RG flows such solutions would be excluded. The reason is that $A$ or rather $e^A$ is interpreted as the RG scale, which should be monotonic along the flow. Here we do not rely on the interpretation in terms of RG flows and hence we see no reason to exclude such solutions.

We now examine the brane curvature $R_B M_4^{-2}$ in 4d Planck units. In particular, in fig.~\ref{RbvsCurlyRAdS2_new} we plot $R_B M_4^{-2}$ vs.~$\mathcal{R}$ for solutions with $\f_{\star} >0$. For better visibility, we split the plot into two, covering different ranges in $R_B M_4^{-2}$. We make the following observations:
\begin{itemize}
\item Of all the branches of solutions, only the solid red and blue branches admit solutions with hierarchically small $R_B$, i.e.~they exhibit $R_B \rightarrow 0$ for $\mathcal{R} \rightarrow 0$. However, as remarked before, only the solid red branch connects continuously to a solution with $R_B=0$. On the solid blue branch the strict limit $R_B=0$ cannot be reached. All other solutions in this model exhibit $R_B M_4^{-2} \sim \mathcal{O}(1)$.
\item For all the branches of solutions we find that $R_B M_4^{-2}$ stays finite as $|\mathcal{R}| \rightarrow \infty$. This is the same qualitative behaviour as observed for $\mathcal{R} >0$ before.
\item We again find something new for the dashed and dot-dashed blue branches. Here we observe that $R_B M_4^{-2}$ stays finite for $\mathcal{R} \rightarrow 0$. This can be understood by looking again at fig.~\ref{fig:UconstNegRbranch3}. While the bounce locus is pushed towards a minimum for $\mathcal{R} \rightarrow 0$, the brane remains located at a generic point thus exhibiting finite curvature.
\item Finally note that $R_B$ is bounded, i.e.~we cannot obtain solutions with $R_B M_4^{-2} < -10.25$
\end{itemize}

\subsubsection*{Summary}
This concludes our description of the full space of solutions of our bulk-brane system for the model described by the bulk potential \protect\eqref{eq:bulk1}, brane quantities \protect\eqref{eq:brane1} and parameter values \protect\eqref{eq:para1}. We have made four main qualitative observations.
\begin{enumerate}
\item Our setup allows for solutions with a flat brane, i.e.~with $R_B=0$. However, we find that these can only be obtained if the UV curvature is also chosen to vanish, i.e.~$\mathcal{R}=0$. These are the flat self-tuning solutions of \cite{1704.05075}.
\item Solutions with a hierarchically small $R_B$ (that is $R_B M_4^{-2} \ll 1$) exist, but require $\mathcal{R} \ll 1$ to be tuned small.
\item At the same time, solutions exist with finite but arbitrarily small $\mathcal{R}$ that nevertheless exhibit a finite $R_B M_4^{-2}$.
\item While solutions with positive and negative $R_B$ exist, the brane curvature is bounded. In particular, solutions populate an interval $- |R_{B,1}| < R_B < |R_{B,2}|$.
\item Finally, we observe that the brane curvature $R_B$ remains finite even if $\mathcal{R} \rightarrow \infty$. Thus large hierarchies between $R_B M_4^{-2}$ and $\mathcal{R}$ can be achieved.
\end{enumerate}

There is a further caveat regarding the solutions in this section. In particular, we should also check explicitly, whether the solution portrayed here satisfy the stability criteria listed in \eqref{inst1}--\eqref{inst4}. One finds that our choice $W_B<0$ is in conflict with criterion \eqref{inst2}, which in practice often reduces to the requirement $W_B (\f_{\star}) >0$. However, note that \eqref{inst2} is only a sufficient condition and its violation thus does not automatically imply an instability. To check stability of our solutions would thus require a more detailed analysis, which goes beyond the scope of this paper. However, in section \ref{sec:numexpV} we will consider a different model based on different choices for $V(\f)$, $W_B(\f)$ and $U(\f)$, which will exhibit $W_B (\f_{\star}) >0$.


\subsection{IR-AdS with exponential $U(\f)$ }
\label{sec:expU}
So far we have considered a model with $U(\f)=1 =const$. In this section we will depart from this choice in a drastic way. In particular, we now choose $U(\f)$ to be exponentially sensitive to the brane position. Thus we will now consider a model with bulk potential \eqref{eq:bulk1} and brane quantities
given by
\begin{align}
\label{eq:brane2} W_B(\f)= \omega  \exp (\gamma  \f ) \, , \qquad U(\f)= |\omega|  \exp (\gamma  \f ) \, ,
\end{align}
with
\begin{align}
\label{eq:para2} \Delta_- = 1.2 \, , \qquad \omega=-0.015 \, , \qquad \gamma=5 \, .
\end{align}
To simplify the analysis we have chosen both $W_B(\f)$ and $U(\f)$ to be described by the same exponential function.\footnote{Furthermore, we expect  $\f$-dependence of the brane potential $W_B(\f)$ and of $U(\f)$ to have a common physical origin. Thus choosing related functions for $W_B$ and $U$ may not be unrealistic.}

\begin{figure}[t]
\centering
\begin{overpic}[width=.75\textwidth,tics=10]{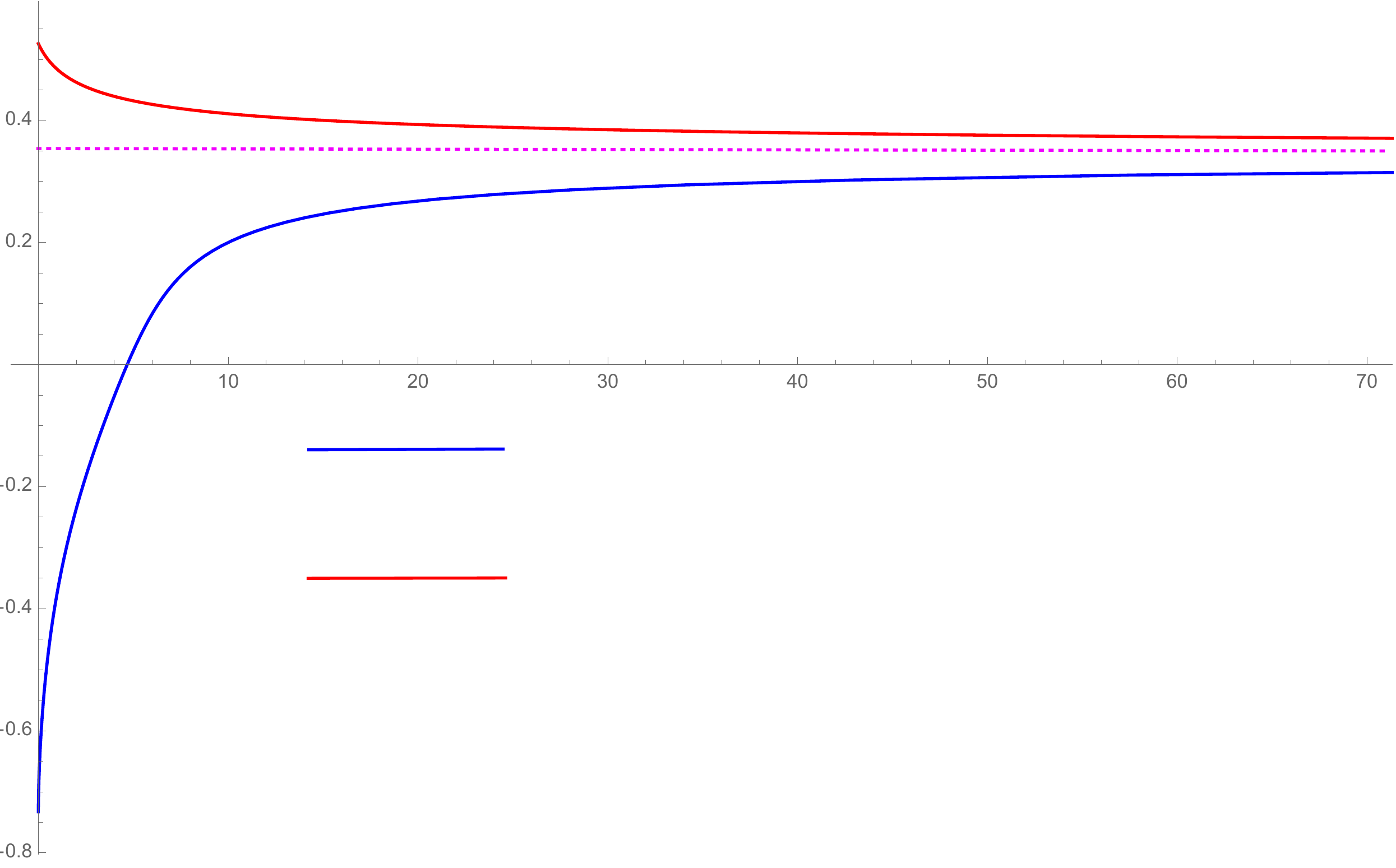}
\put (97,30) {$\mathcal{R}$}
\put (0,65) {$\f_\star$}
\put (38,29){$\f_-<0$}
\put (38,20){$\f_->0$}
\end{overpic}
\caption{Equilibrium brane position $\f_\star$ vs.~$\mathcal{R}$ for $\mathcal{R} > 0$. Results are obtained for bulk potential \protect\eqref{eq:bulk1}, brane quantities \protect\eqref{eq:brane2} and parameter values \protect\eqref{eq:para2}. For $\mathcal{R} \rightarrow \infty$ both the red and blue branch asymptote to the value for $\f_{\star}$ indicated by the dotted magenta line.}
\label{phistardSUphi}
\end{figure}

We will now explore the space of solutions of our bulk-brane system for this model. As before, we will distinguish between setups with positive and negative UV curvature.

\subsubsection*{Solutions with $\mathcal{R}>0$}
We display this part of the space of solutions by plotting $\f_{\star}$ vs.~$\mathcal{R}$ which is shown in fig.~\ref{phistardSUphi}. We find that for a given value of $\mathcal{R}$ there are up to two branches of solutions, one with $\f_->0$ (red) and one with $\f_- <0$ (blue). This is similar to the case with $U(\f)=1$ (see fig.~\ref{phistarvsRdS}) studied before. However, we find the following differences to the model  $U(\f)=1$. For one, here both branches of solutions exist for arbitrarily small $\mathcal{R}$. The most important difference is that there are now solutions with $\f_{\star} <0$ while such solutions were absent for $U(\f)=1$. The remaining differences are just quantitative, i.e.~how $\f_{\star}$ precisely behaves as a function of $\mathcal{R}$.

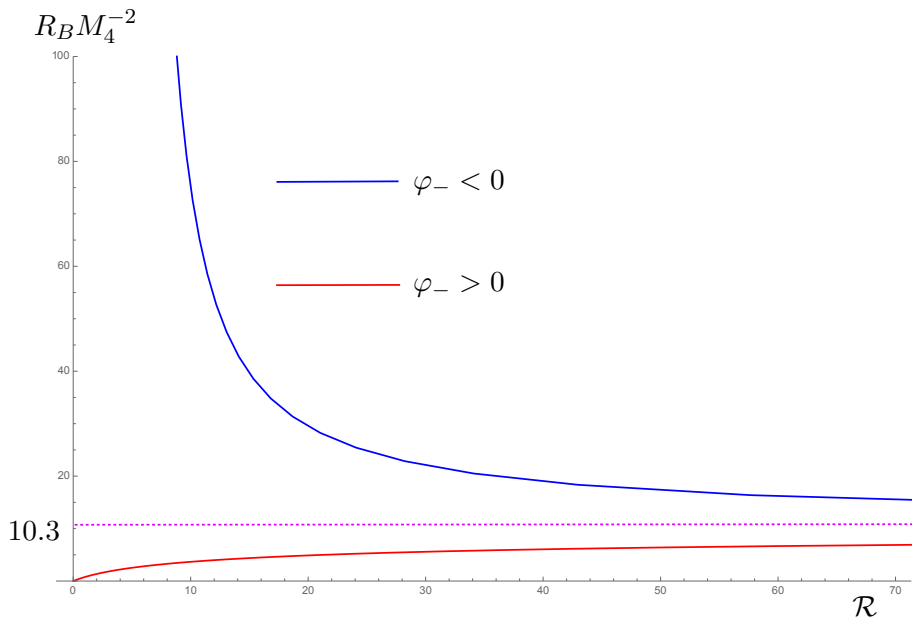
\begin{figure}[t]
\centering
\begin{overpic}[width=.75\textwidth,tics=10]{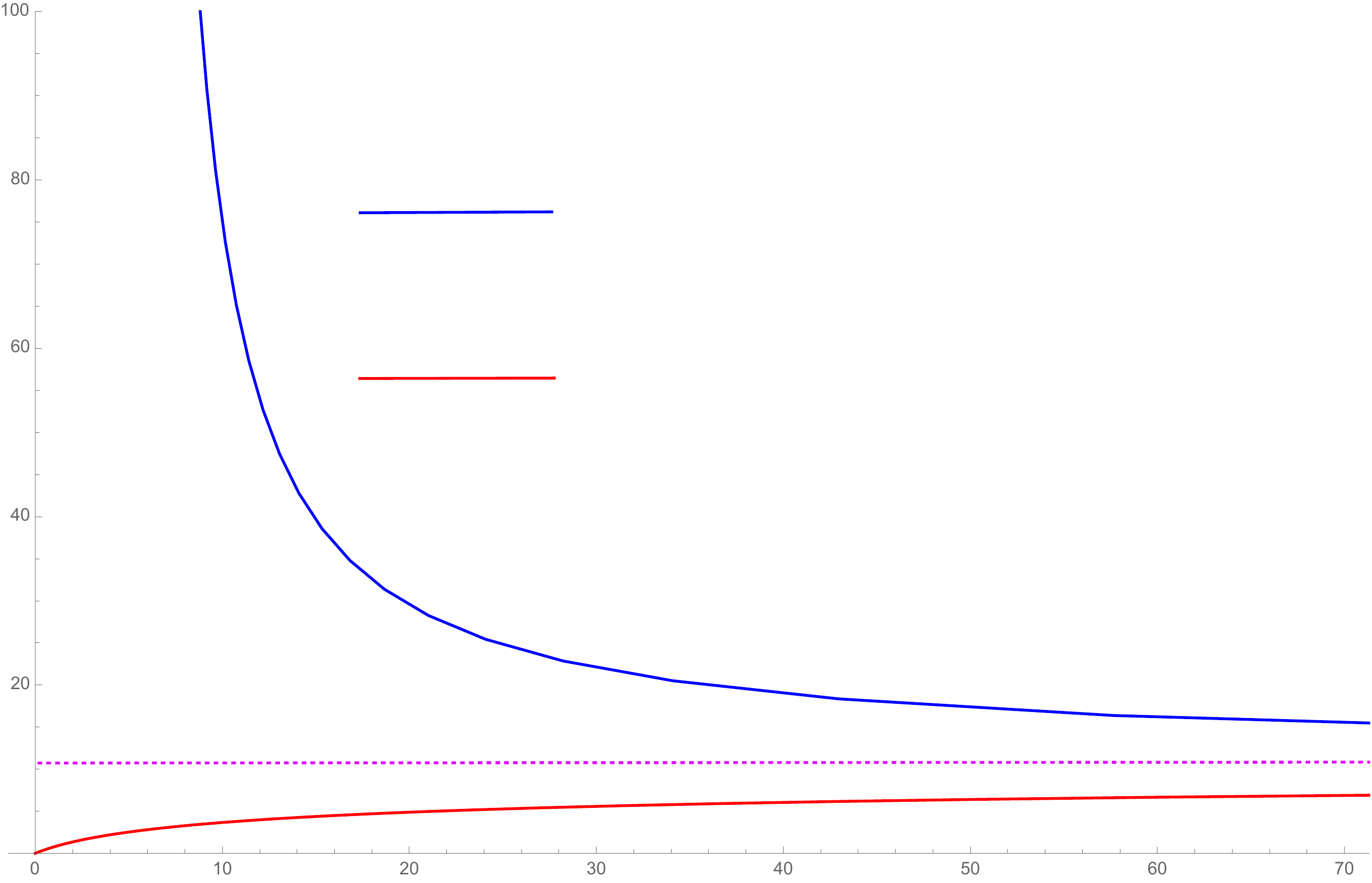}
\put (93,-2) {$\mathcal{R}$} \put (-2,66) {$R_B M_4^{-2}$}
\put (42,48){$\f_-<0$}
\put (42,36){$\f_->0$}
\put(-5,7){10.3}
\end{overpic}
\caption{Brane curvature $R_B M_4^{-2}$ vs.~$\mathcal{R}$ for $\mathcal{R} > 0$. Results are obtained for bulk potential \protect\eqref{eq:bulk1}, brane quantities \protect\eqref{eq:brane2} and parameter values \protect\eqref{eq:para2}. For $\mathcal{R} \rightarrow \infty$ both the red and blue branch asymptote to the value for $R_B M_4^{-2}$ indicated by the dotted magenta line.}
\label{RBvsphi0dSUphi}
\end{figure}

In fig.~\ref{RBvsphi0dSUphi} we plot $R_B M_4^{-2}$ as a function of $\mathcal{R}$ for the model with exponential $U(\f)$. This is to be compared with the corresponding findings for the model with $U(\f)=1$ which are plotted in fig.~\ref{RbvsRdS}. Here, while there are numerical differences between the two models, they exhibit similar qualitative results. Again, we find that $R_B \rightarrow 0$ can only occur when simultaneously $\mathcal{R} \rightarrow 0$. However, here the brane curvature $R_B$ is not necessarily bounded. The value of $R_B M_4^{-2}$ diverges on the blue branch for $\mathcal{R} \rightarrow 0$.

\subsubsection*{Solutions with $\mathcal{R}<0$}
We also find solutions with $\mathcal{R}<0$. In fact, here the space of solutions with negative UV curvature is even more involved than in the case with constant $U(\f)$. However, the picture that emerges is similar to what we have seen for $\mathcal{R}>0$: while the results change quantitatively, the main qualitative observations persist. For example, we again observe that solutions with arbitrarily small $|\mathcal{R}|$ but finite $R_B M_4^{-2}$. As we do not find any great qualitative differences we refrain from showing any explicit results.


\section{IR exponential potential}
\label{sec:numexpV}
In the previous section, we studied self-stabilisation in models with a finite range in $\f$ between the UV and IR. This constraint arose from choosing a bulk potential with one maximum surrounded by two minima. Solutions were confined to the region between the two minima. Here we will consider a model without this constraint and where a priori $\f$ has infinite range. This can be done by choosing a bulk potential with a maximum, but no minima. To be specific, we choose:
\begin{equation}
V(\f)=\frac{1}{\ell^2} \left[-12- \left(\frac{1}{2} \Delta  (4-\Delta )-\frac{b^2}{4} V_1 \right)\f ^2 -V_1 \sinh ^2\left(\frac{b \f  }{2}\right) \right] \, ,\label{bulkcaseIII}
\end{equation}
with $2 < \Delta < 4$ and $V_1 >0$ and $b$ another (dimensionless) parameter. Note that we still work in $d=4$. We will also again set $\ell=1$ in the following. The maximum of $V$ is at $\f_{\textrm{max}}=0$, which, in the language of holographic RG flows, is a UV fixed point with corresponding UV CFT. In this framework the parameter $\Delta$ is interpreted as the dimension of the operator perturbing this UV CFT. However, as there are no minima the IR is only reached for $|\f| \rightarrow \infty$. Solutions that flow all the way to $|\f| \rightarrow \infty$ are singular, but, as briefly reviewed in sec.~\ref{RG}, they can be acceptable if certain conditions are satisfied.

One reason for choosing a potential with unlimited range in $\f$ is as follows. In \cite{1704.05075}, for the case of a flat brane, it was observed that bulk potentials with a finite range for $\f$ do not easily exhibit self-tuning solutions satisfying the stability criterion \eqref{inst2}, at least not without some fine-tuning of parameters. However, for potentials with infinite range in $\f$ this difficulty can be overcome, as was shown for the case of a flat brane in \cite{1704.05075} using an example based on bulk potential \eqref{bulkcaseIII}.

To allow comparisons with the results in \cite{1704.05075} we choose a brane potential with the same mathematical form. Hence we will work with the following brane quantities:
\begin{align}
\label{branecaseIII} W_B (\f) = \Lambda^4 \left[ -1 - \frac{\f}{s} + {\left( \frac{\f}{s} \right)}^2 \right] \, , \qquad U(\f) = const \, ,
\end{align}
where $\Lambda$ and $s$ are numerical parameters (as we have set $\ell=1$). The brane potential is chosen such that it has at least one zero for $\f >0$. The position of the zero is controlled by $s$. Again, absent any prior knowledge regarding $U(\f)$ we take it to be constant for simplicity.

\subsection{Analytical results}
\label{sec:expanalytical}
Before moving on to numerical studies, we collect analytical results for the asymptotic region $\f \rightarrow \infty$. Note that apart from a region in the vicinity of $\f_{\textrm{max}}=0$ the potential is well-approximated by an exponential. As a result, an analytical understanding of solutions for an exponential potential will turn out to be very helpful for the interpretation of our numerical findings.

Therefore, we  collect analytical solutions for an exactly exponential bulk potential. To be specific, we will consider
\begin{align}
\label{eq:expV} V= - V_{\infty} \, \exp \left(b \f \right) \, ,
\end{align}
which is the asymptotic form of \eqref{bulkcaseIII} for $\f \rightarrow \infty$ if we identify $V_{\infty} = V_1 /4$. For $\f \rightarrow \infty$ the bulk solutions fall into three classes.
\begin{enumerate}
\item \textbf{Continuous branch}: \newline
For one there exists a family of solutions of the form
\begin{align}
\label{eq:expcontsol} W \simeq W_0 \, e^{Q \f  } \, , \quad S \simeq W' \, , \quad T \simeq T_0 \, e^{Q \f  } \, , \quad \textrm{where} \quad Q  = \sqrt{\frac{2}{3}} \, .
\end{align}
In this case $W_0$ and $T_0$ are free parameters. The existence of this branch of solutions requires
\begin{align}
b < 2Q  = 2\sqrt{\frac{2}{3}} \, .
\end{align}
This branch also exists for flat solutions (albeit with $T_0=0$) and was already discussed in
subsection \ref{RG}. For $\f \rightarrow \infty$ these solutions exhibit an unacceptable singularity according to Gubser's criterion \cite{0002160}. For more details on this class of solutions see e.g.~\cite{exotic}.
\item \textbf{A special solution with} $S=W'$:\newline
There exists an isolated solution of the form
\begin{align}
W= W_0 \, e^{b \f / 2} \, , \quad S = W' \, , \quad T=0 \, , \quad
\textrm{with} \quad W_0 = \sqrt{\frac{8 V_{\infty}}{4Q^2 - b^2}} \, ,
\end{align}
with $Q$ defined as in \eqref{eq:expcontsol}. Again, this solution only exists for
\begin{align}
b < 2Q = 2\sqrt{\frac{2}{3}} \, .
\end{align}
This is the special flat solution satisfying Gubser's criterion,
giving rise to an acceptable IR singularity, as discussed in sec.~\ref{RG}.
\item \textbf{A special solution with} $S = W/(3b)$:\newline
Finally, the equations of motion \eqref{eq:EOM4}--\eqref{eq:EOM6} also admit the solution
\begin{align}
&W= W_0 \, e^{b \f /2}  , \quad S= \frac{W}{3b} \, , \quad T = T_0 \, e^{b \f} \, , \\
\nonumber & \textrm{with} \quad W_0 = \sqrt{6 V_{\infty}} \, , \quad T_0 = \left[2 b^2 - \frac{4}{3} \right] V_\infty \, .
\end{align}
This solution exists for any value of $b$, but we observe that the sign of the function $T$ depends on $b$ as follows:
\begin{align}
\nonumber b > \sqrt{\frac{2}{3}} & \quad \Leftrightarrow \quad T >0 \, , \\
\nonumber b < \sqrt{\frac{2}{3}} & \quad \Leftrightarrow \quad T <0 \, , \\
\nonumber b = \sqrt{\frac{2}{3}} & \quad \Leftrightarrow \quad T =0 \, .
\end{align}
For $b^2=2/3$ the solution of type 3 discussed and the one of type 2
discussed above are identical.

Here we see that non-zero  curvature gives rise to a new solution
reaching the asymptotic IR region $\f\to + \infty$, for which $W(\f)$ has the same
exponential growth but different overall magnitude as the special,
flat solution, and for which $S\neq W'$. Depending on the value of $b$
these solution are found either for $R>0$ only, or for $R<0$
only.

 Intriguingly, the critical value separating these cases,
$b=\sqrt{2/3}$, is the same which separates confining from
non-confining theories. This may signal interesting consequences in
regards to confining holographic theories on curved manifolds, whose
analysis we leave for further investigation.
\end{enumerate}

\begin{figure}[t]
  \begin{subfigure}{.5\textwidth}
\centering
\begin{overpic}[width=1.0\textwidth,tics=10]{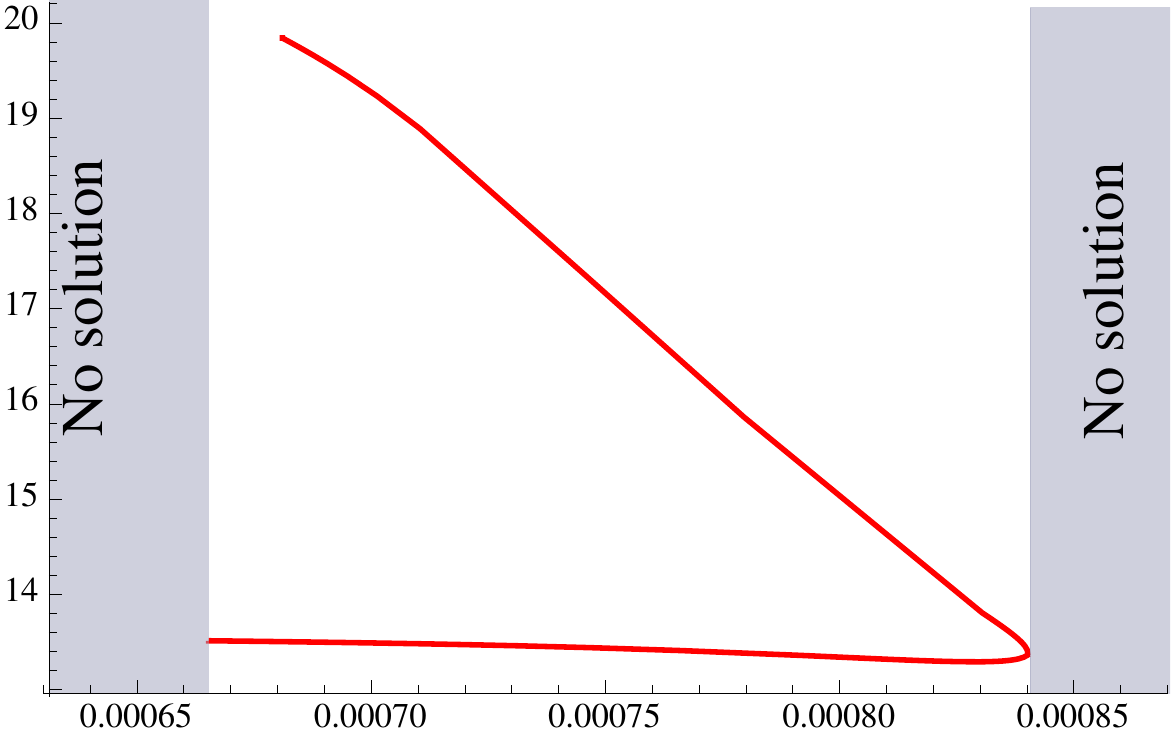}
\put(100,0){$\mathcal{R}$}
\put(0,67){$\f_\star$}
\end{overpic}
\caption{}
\label{phi0vRdSWbpos11}
\end{subfigure}
  \begin{subfigure}{.5\textwidth}
\centering
\begin{overpic}[width=1.0\textwidth,tics=10]{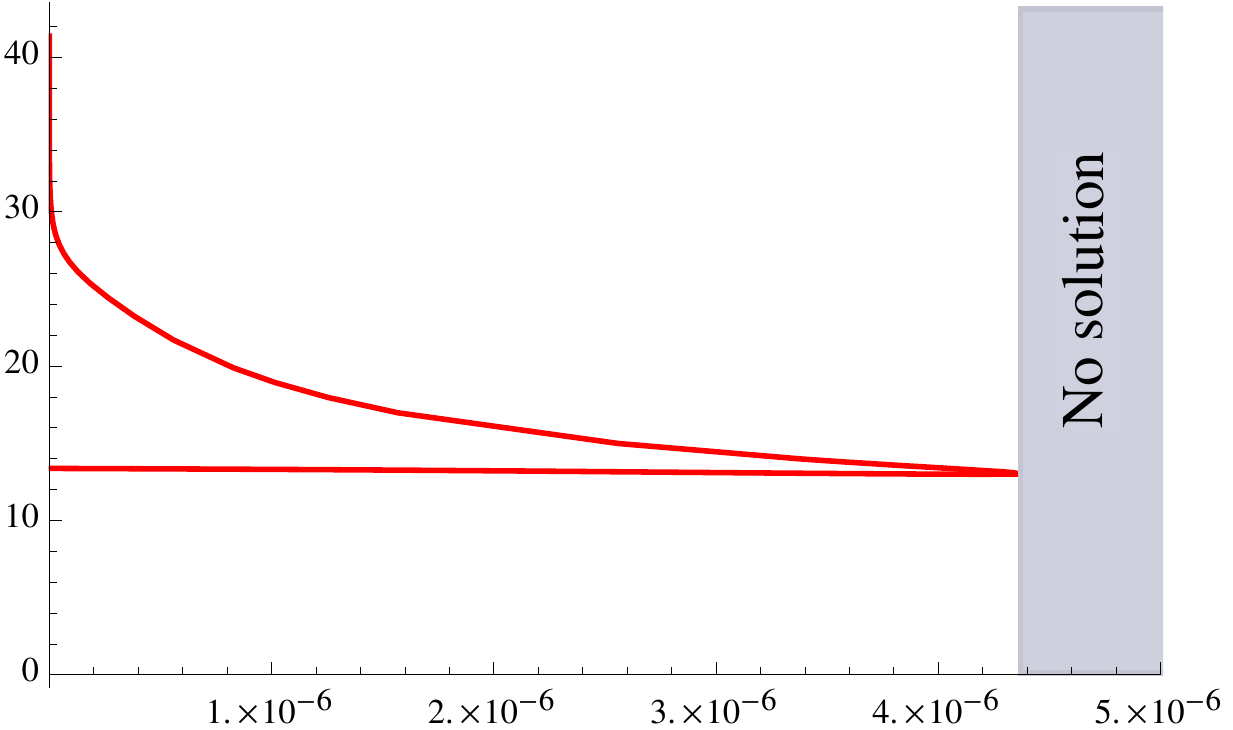}
\put(100,0){$\mathcal{R}$}
\put(0,64){$\f_\star$}
\end{overpic}
\caption{}
\label{phi0vRdSWbpos09}
\end{subfigure}
\caption{Equilibrium brane position $\f_\star$ vs.~$\mathcal{R}$ for $\mathcal{R} > 0$. Results are obtained for bulk potential \protect\eqref{bulkcaseIII}, brane quantities \protect\eqref{branecaseIII} and parameter values \protect\eqref{eq:para3}. The left figure (a) is for  $b=1.1\times \sqrt{2/3}$ and the right figure (b) is for $b=0.9\times \sqrt{2/3}$.} \label{phi0vRdSWbpos}
\end{figure}

\subsection{Numerical studies}
We now return to a study of the model with the full bulk potential \eqref{bulkcaseIII} and brane quantities \eqref{branecaseIII}. In particular, we now solve numerically for solutions of the bulk-brane system. For definiteness, we will choose the following values for the parameters:
\begin{align}
\label{eq:para3} \Delta &= 2.9 \, , \quad \Lambda= 3 \, , \quad s=8 \, , \quad V_1 = 1 \, , \quad U = 10^{-4} \, ,
\end{align}
but these values are in no way special. However, a small numerical value for $U$ will turn out to be favourable for satisfying the stability criteria \eqref{inst2} and \eqref{inst3}. We then perform the analysis for two different values of $b$. In particular, we will study the cases
\begin{align}
b &= 1.1 \, \sqrt{2/3} \, \ \textrm{ and } \ b= 0.9 \, \sqrt{2/3} \, .
\end{align}

We will restrict our analysis to solutions with positive UV curvature $\mathcal{R} >0$, as this will exhibit all the phenomena that we wish to illustrate with this example. In fig.~\ref{phi0vRdSWbpos} we show the space of solutions by plotting the equilibrium brane position $\f_{\star}$ vs.~$\mathcal{R}$. In fig.~\ref{phi0vRdSWbpos11} we show the results for $b= 1.1 \, \sqrt{2/3}$ while the results for $b= 0.9 \, \sqrt{2/3}$ are displayed in fig.~\ref{phi0vRdSWbpos09}.

One common observation for both values of $b$ is that solutions only exist for a very narrow range in $\mathcal{R}$. In particular, we find that solutions only exist for the following values of $\mathcal{R}$:
\begin{align}
\nonumber b&= 1.1 \, \sqrt{2/3} : \quad \textrm{solutions exist for} \quad 0.00066 \lesssim \mathcal{R} \lesssim 0.00084 \quad \textrm{and} \quad \mathcal{R}=0 \, , \\
\nonumber b&= 0.9 \, \sqrt{2/3} : \quad \textrm{solutions exist for} \quad 0 \leq \mathcal{R} \lesssim 4.5 \cdot 10^{-6} \, .
\end{align}
That is, for $b= 0.9 \, \sqrt{2/3}$ we only find solutions with very small absolute values of $\mathcal{R}$. For $b= 1.1 \, \sqrt{2/3}$ solutions with finite $\mathcal{R}$ can only exist in a very narrow band of width $\Delta \mathcal{R} \sim 2 \cdot 10^{-4}$ about the central value $\mathcal{R} \sim 7.5 \cdot 10^{-4}$.\footnote{Here we collect further, but less important observations. For one, we find that there are typically two solutions for the equilibrium position $\f_{\star}$ if $\mathcal{R}$ permits a solution. Also, for completeness, note that for $\mathcal{R} \rightarrow 0$ the lower branch in fig.~\ref{phi0vRdSWbpos09} is continuously connected to a solution with $\mathcal{R}=0$, whereas this is not the case for the upper branch.}

\begin{figure}[t]
\centering
\begin{overpic}[width=0.75\textwidth,tics=10]{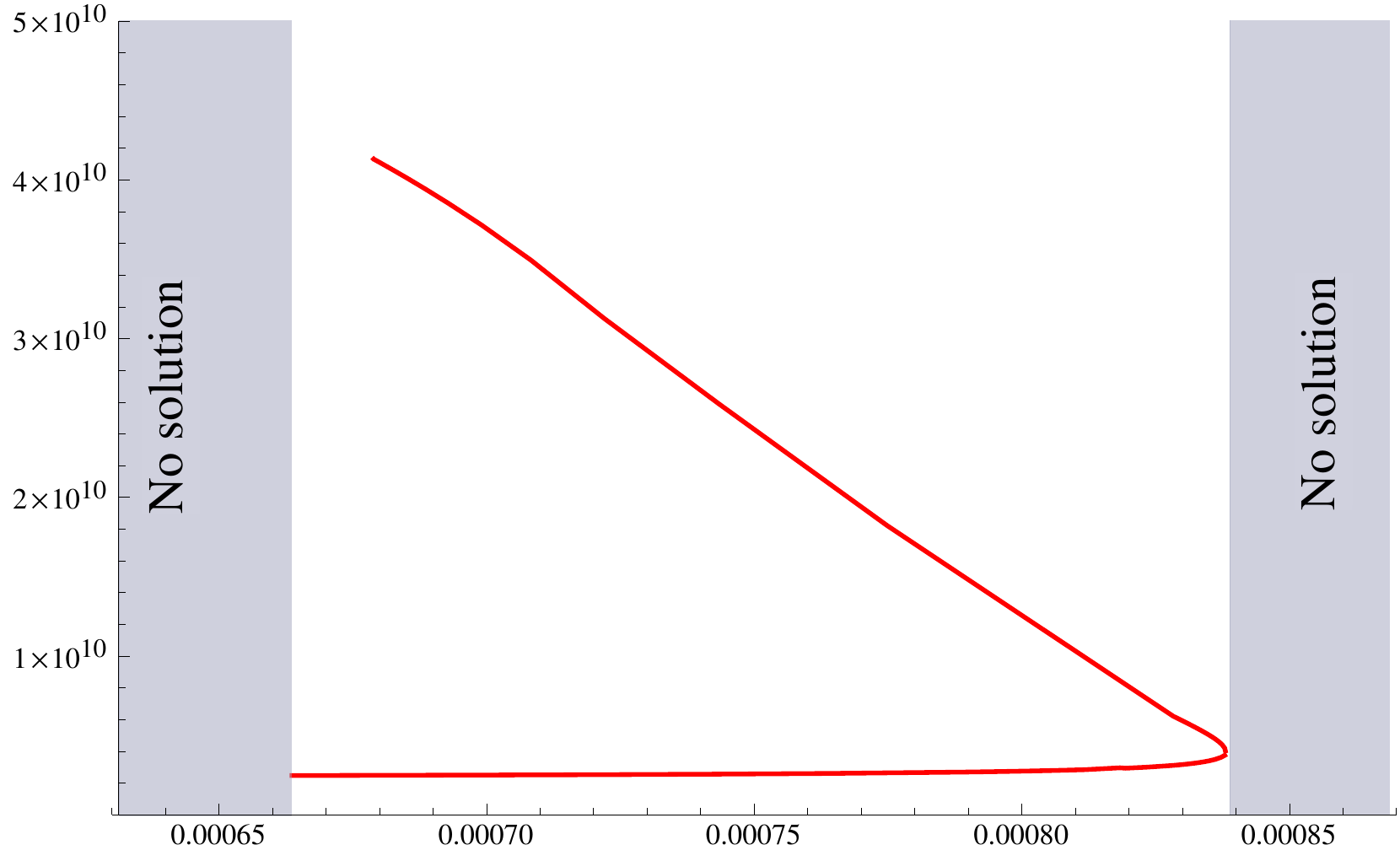}
\put(100,0){$\mathcal{R}$}
\put(0,62){$R_B M_4^{-2}$}
\end{overpic}
\caption{The brane curvature $R_B M_4^{-2}$ vs.~$\mathcal{R}$ for $b=1.1\times \sqrt{2/3}$. Results are obtained for bulk potential \protect\eqref{bulkcaseIII}, brane quantities \protect\eqref{branecaseIII} and parameter values \protect\eqref{eq:para3}.}
\label{fig:RBvsRexpb11}
\end{figure}

We can understand all these observations with the help of the analytical results collected in the previous section. To this end note that, for all the solutions, the brane finds it equilibrium position at a value of $\f$ where the bulk potential is well approximated by an exponential.\footnote{Note from fig.~\ref{phi0vRdSWbpos} that for all the solutions obtained the brane equilibrium position takes values $\f_{\star} \gtrsim 13$. There the bulk potential is well-approximated by $$ V = - \frac{V_1}{4} e^{b \f} \Big( 1+ \mathcal{O} \big(\f^2 e^{-b \f} \big) \Big) \, .$$}  As a result, immediately to the left and the right of the brane the bulk solutions will, at leading order, be given by the solutions collected in section \ref{sec:expanalytical}. More precisely, as the potential is not exactly exponential, the solutions in the full potential will be given by those in sec.~\ref{sec:expanalytical} up to some small corrections. With this we can explain the results in fig.~\ref{phi0vRdSWbpos} as follows.

\begin{itemize}
\item On the IR side of the brane all solutions for $b= 1.1 \sqrt{2/3}$ are small perturbations of the special solution of type $S=W/(3b)$ (case 3) in the classification of sec.~\ref{sec:expanalytical}. As this is a unique solution only a small subset of solution leaving the UV fixed point will asymptote to this solution. This explains the narrow range in $\mathcal{R}$ for which solutions exist. In addition, for $b= 1.1 \sqrt{2/3}$ this type of solution has $T \neq 0$, which implies that $\mathcal{R} \neq 0$. Hence we do not expect these solutions to exist for arbitrarily small values of $\mathcal{R}$, which is exactly what we observe. In other words, there is a gap in solutions for $\mathcal{R} >0$.
\item In contrast for $b= 0.9 \sqrt{2/3}$ the solutions on the IR side of the brane are small perturbations of the special solution of type $S =W'$ (case 2) in the classification of sec.~\ref{sec:expanalytical}.\footnote{For $b= 0.9 \sqrt{2/3}$ solutions of type 3 have $T<0$ and hence $\mathcal{R} <0$. As we restrict our attention to configurations with $\mathcal{R} >0$ we cannot find solutions of type 3 for $b= 0.9 \sqrt{2/3}$.} Note that this type of solution has $T=0$ exactly, which would imply $\mathcal{R}=0$. Since the potential is not exactly exponential, the solutions are only approximately of type 2 and finite but small values of $\mathcal{R}$ are allowed. This is exactly what we observe in fig.~\ref{phi0vRdSWbpos}.
\end{itemize}

\begin{figure}[t]
  \begin{subfigure}{.5\textwidth}
\centering
\begin{overpic}[width=1.0\textwidth,tics=10]{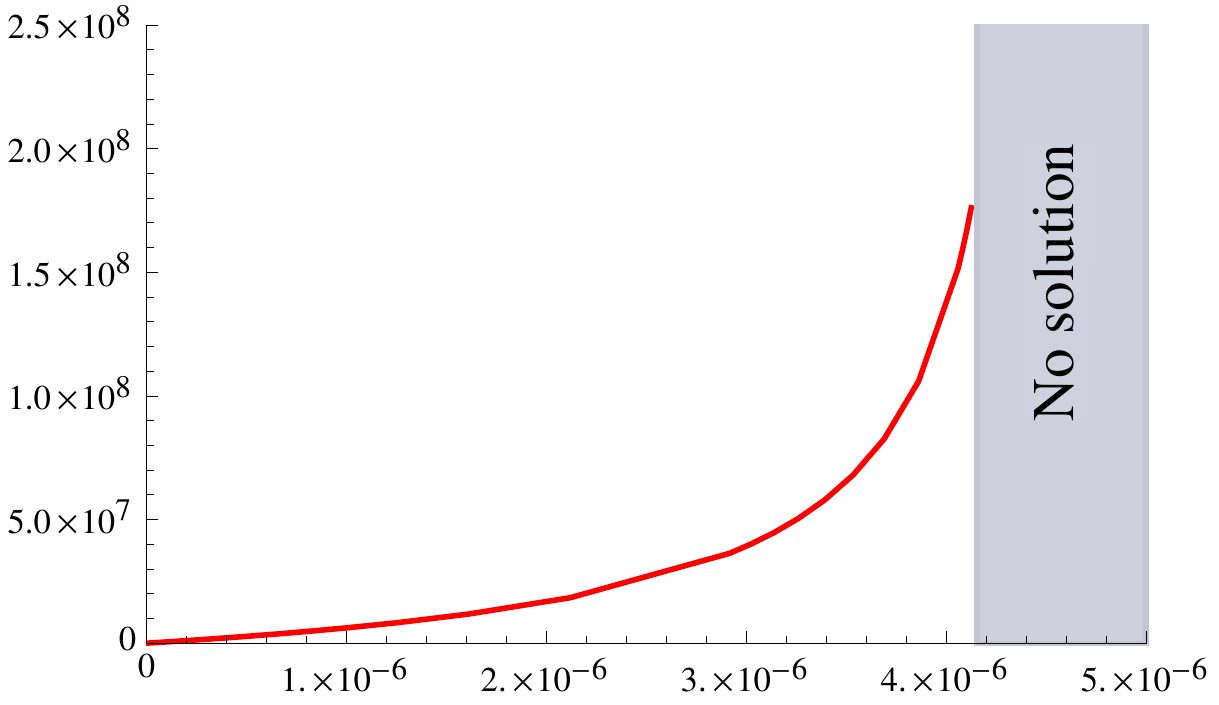}
\put(100,0){$\mathcal{R}$}
\put(0,62){$R_B M_4^{-2}$}
\end{overpic}
\caption{}
\label{RbvRdSWbpos099lower}
\end{subfigure}
  \begin{subfigure}{.5\textwidth}
\centering
\begin{overpic}[width=1.0\textwidth,tics=10]{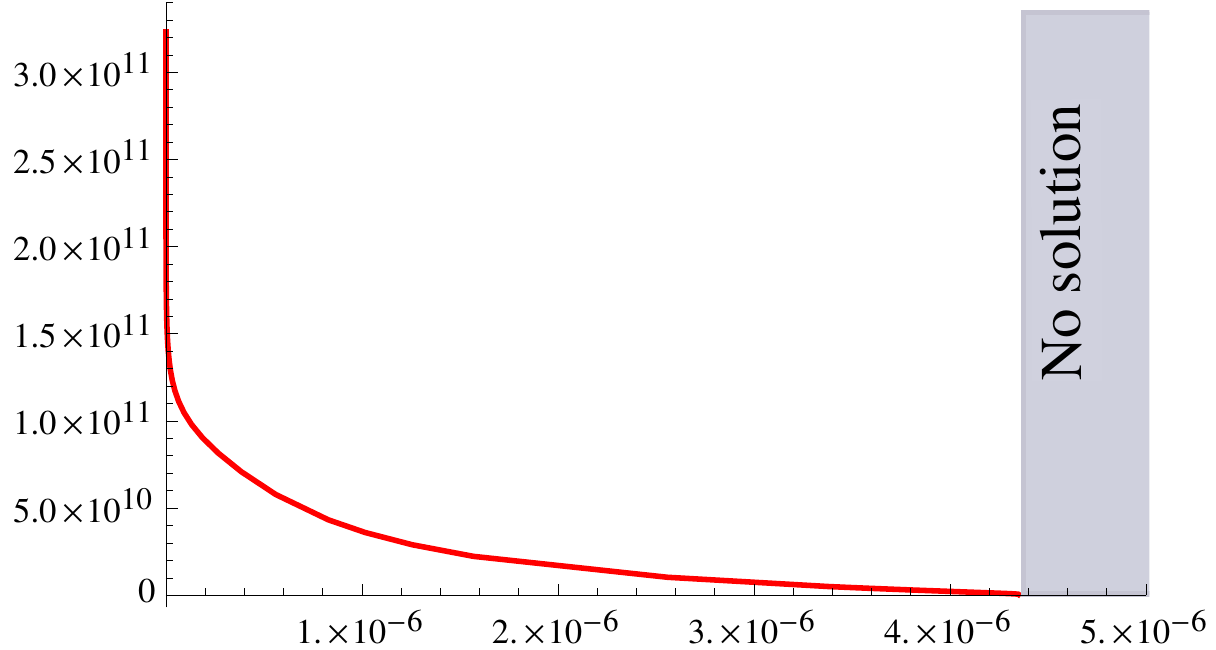}
\put(100,0){$\mathcal{R}$}
\put(0,62){$R_B M_4^{-2}$}
\end{overpic}
\caption{}
\label{RbvRdSWbpos099upper}
\end{subfigure}
\caption{The brane curvature $R_B M_4^{-2}$ vs.~$\mathcal{R}$ for $b=0.9\times \sqrt{2/3}$. Results are obtained for bulk potential \protect\eqref{bulkcaseIII}, brane quantities \protect\eqref{branecaseIII} and parameter values \protect\eqref{eq:para3}. The plots (a) and (b) exhibit results for the two branches of solutions separately.} \label{RbvRdSWbpos099}
\end{figure}

Next, we study the brane curvature across our space of solutions. For $b= 1.1 \sqrt{2/3}$ we plot $R_B M_{4}^{-2}$ vs.~$\mathcal{R}$ in fig.~\ref{fig:RBvsRexpb11}. The main observation is that, unless $\mathcal{R}=0$, the brane curvature $R_B$ is always finite and never small. (In fact, for our unrealistic choice of brane parameters it is also extremely super-Planckian.) There is no continuous limit where $R_B M_{4}^{-2} \rightarrow 0$.

For $b= 0.9 \sqrt{2/3}$ the findings are qualitatively different. We display the corresponding results for $R_B M_4^{-2}$ vs.~$\mathcal{R}$ in fig.~\ref{RbvRdSWbpos099}. In particular, in fig.~\ref{RbvRdSWbpos099lower} we show results for the lower branch in fig.~\ref{phi0vRdSWbpos11}, while fig.~\ref{RbvRdSWbpos099upper} contains the data for the upper branch of fig.~\ref{phi0vRdSWbpos11}. Most importantly, the lower branch exhibits a limit $R_B \rightarrow 0$ for $\mathcal{R} \rightarrow 0$ which is continuously connected to a solution with $R_B = 0$ and $\mathcal{R} = 0$. On the other branch of solutions (fig.~\ref{RbvRdSWbpos099upper}) $R_B$ is never zero and potentially diverges for $\mathcal{R} \rightarrow 0$.\footnote{We could not determine this decisively in our numerical analysis. While we observe that both $\f_\star$ and $R_B M_4^{-2}$ increase on this branch when $\mathcal{R}$ is decreased, we can neither exclude nor confirm whether this continues for arbitrarily small values of $\mathcal{R}$.}

Last, we comment on the stability of the solutions obtained here. In \cite{1704.05075} a set of sufficient criteria \eqref{inst1}--\eqref{inst4} was derived which guarantee the perturbative stability of a \emph{flat} brane solution. Here we do find a branch which in the limit $\mathcal{R} \rightarrow 0$ is connected continuously to a flat brane solution. By an explicit calculation we find that this solution satisfies the stability criteria \eqref{inst1}--\eqref{inst3}, but not \eqref{inst4}. This is not necessarily fatal, as \eqref{inst4} is only a sufficient condition for stability. However, a more detailed analysis is necessary to conclusively determine the stability of this solution, which is beyond the scope of this paper. We further expect the stability properties of the flat solution also to extend to the branch of curved brane solutions connected to the flat solution. The reason is that the solutions exhibit \emph{positive} boundary or brane curvature, which we do not expect to aversely affect stability.

We are now in a position to summarise our findings for the model studied in this section.
\begin{enumerate}
\item We find the space of solutions in $\mathcal{R}$ to be highly restricted. The reason is that for large $\f$ for all solutions have to asymptote to one of the two unique solutions. In our case these are the two special solutions for an exponential bulk potential described in sec.~\ref{sec:expanalytical}. Only a small subset of solutions departing from the UV fixed point will asymptote to such a solution and all lie within a narrow range in $\mathcal{R}$.
\item We again find a branch of solutions that in the limit $\mathcal{R} \rightarrow 0$ is connected continuously to a flat brane solution with $\mathcal{R}=0$ and $R_B=0$. Here we made sure that flat brane limit satisfies the criteria \eqref{inst1}--\eqref{inst3} for perturbative stability. It does not satisfy the sufficient condition \eqref{inst4}.
\item Note that to find solutions satisfying \eqref{inst1}--\eqref{inst3} in the flat limit we required a bulk potential with infinite range (no minima), a brane potential exhibiting a zero for some value $\f > 0$ and a sufficiently small value of $U(\f_{\star})$. No further tuning of parameters beyond this choice is required.
\end{enumerate}

\section{Conclusions and Outlook}
\label{sec:conclusion}
In this work we studied self-stabilising solutions of a 4-dimensional brane embedded into a 5-dimensional bulk, where the curvature of the brane is adjusted dynamically. This is in the spirit of self-tuning mechanisms of the cosmological constant in braneworld scenarios \cite{ArkaniHamed:2000eg,Kachru:2000hf,Csaki:2000wz,1704.05075}, with the difference that we are not exclusively interested in solutions where the brane is flat.

In particular, the (curved) brane is embedded in a bulk described by
Einstein-dilaton gravity with a potential for the dilaton. Following
the ideas laid out in \cite{1704.05075}, our braneworld scenario
offers a holographic interpretation. More precisely, the brane-bulk
system is dual to a weakly coupled sector (e.g.~the Standard Model)
interacting with a strongly coupled large $N$ CFT, with the CFT
residing on the boundary of the 5d bulk, \cite{smgrav}. This is
similar to \cite{1704.05075}, but a new aspect of this analysis is
that we also allow the boundary supporting the CFT to be curved. For
simplicity, we take the background of the CFT to be a  (locally)
maximally symmetric 4-dimensional space-time\footnote{ As mentioned in
  previous sections, when the curvature is negative the holographic
  dictionary is more subtle, and the boundary theory must also includes a
  defect corresponding to the boundary of the radial slices} (dS$_4$, $\mathcal{M}_4$ or AdS$_4$) which we characterise by its scalar curvature $R^{(\zeta)}$. The bulk metric can then be written in domain-wall form as
\begin{align}
\label{eq:bulkconclusion} ds^2 = du^2 + e^{2 A(u)} \zeta_{\mu \nu} d x^{\mu} d x^{\nu} \, ,
\end{align}
where $\zeta_{\mu \nu}$ is a metric describing the maximally symmetric 4d space-time with curvature $R^{(\zeta)}$. Note that $R^{(\zeta)}$ is not dynamical, but fixed as a boundary condition at the conformal boundary of the bulk space-time. The dilaton is also chosen to be constant on the boundary, which in holography corresponds to a constant scalar operator source.

The first observation is that generic self-stabilising solutions with $R^{(\zeta)}=0$ (i.e.~$\zeta_{\mu \nu} = \eta_{\mu \nu}$) only exist if the world-volume of the brane is also flat. That is, if the boundary CFT resides on Minkowski space, the world-volume of the brane is also given by Minkowski space, which is the scenario studied previously in \cite{1704.05075}. Exceptions exist, but are non-generic as they require a tuning of model-parameters (e.g.~a precise choice of the dilaton potential on the brane, see appendix \ref{app:evanescent} for details). This no-go result can also be overcome if the 5d bulk is not static as in \eqref{eq:bulkconclusion}, but time-dependent. However, in this case the dual interpretation in terms of RG flows does not apply any more. In fact,  in such a case it corresponds to time dependent dynamics of vevs, associated to non-trivial cosmological evolution on the brane, and we leave this possibility for future work.

To find self-stabilising solutions with a curved brane in a static bulk, one is hence forced to modify the UV boundary conditions of the bulk fields. In this work we mainly did so by choosing $R^{(\zeta)} \neq 0$. We then worked with a simple brane embedding in which the brane geometry is inherited from the boundary. For a bulk described by \eqref{eq:bulkconclusion} this amounts to locating the brane at some fixed $u = u_{\star}$. This choice is equivalent to restricting to branes with maximally symmetric world-volume with scalar curvature $R_B$. The brane curvature is then related to $R^{(\zeta)}$ as
\begin{align}
R_B = R^{(\zeta)} \, e^{-2 A(u_{\star})} \, .
\end{align}
To find solutions for a brane with world-volume given by (A)dS$_4$, the boundary CFT has to reside on (A)dS$_4$\footnote{As shown in \cite{Ghosh:2017big} such AdS-sliced bulk solutions have twin boundary singularities that also affect the brane. They could be resolved by lower codimension branes.}.

There exists an alternative realisation of the same solutions, which can be obtained via a bulk coordinate transformation. In this formulation the boundary metric is flat ($R^{(\zeta)}=0$), but the scalar sources are no longer constant on the boundary: they now vary in space or time. As a result, the holographic interpretation is also modified. Instead of the boundary QFT living on a curved space-time, we then have a flat-space QFT driven by a time-varying (in the dS case) or space-dependent (in the AdS case) source. In the dS case one requires the source $j$ to vary as $j \sim |t|^{\Delta-d}$, where $t$ is the de Sitter conformal time on the brane and $\Delta$ is the dimension of the relevant operator deforming the CFT.

We then studied how the brane curvature $R_B$ depends on $R^{(\zeta)}$ quantitatively. This was done mostly numerically by searching for self-stabilising solutions while scanning over all possible values of $R^{(\zeta)}$. To perform a numerical analysis, we have to specify a particular model by choosing a bulk dilaton potential $V(\f$), a brane dilaton potential $W_B(\f)$, and the (dilaton-dependent) Newton's `constant' term $U(\f)$ on the brane. Given a UV completion of our model, these functions can in principle be determined, but this goes beyond the scope of this work. Here we chose generic functions for $V(\f)$, $W_B(\f)$ and $U(\f)$ and studied the consequences for self-stabilisation. To assess how model-dependent our findings are, we performed our analysis for three different combinations of $V(\f)$, $W_B(\f)$ and $U(\f)$. To be specific, we compared setups with restricted vs.~unlimited dilaton range. We also contrasted the effect of choosing $U(\f)$ to be fast-varying vs.~constant in $\f$. Rather than trying to construct a phenomenologically viable model, we were more interested in exploring the scope of effects that can arise in this brane-bulk system.

While the precise numerical results are model-dependent, we observed similar qualitative features in all three examples studied:
\begin{itemize}
\item For any given value of $R^{(\zeta)}$ there typically exists at least one self-stabilising solution. Frequently, there exist several branches of solutions as a function of $R^{(\zeta)}$, which differ in the values for the brane position $\f_{\star}= \f(u_\star)$ and the brane curvature $R_B$.

\item For $R^{(\zeta)} \rightarrow 0$ one of the branches connects continuously to a solution with $R^{(\zeta)}=0$ and $R_B=0$, i.e.~a flat brane as studied in \cite{1704.05075}. Hence there exists a limit where $R_B$ can be made arbitrarily small by  letting $R^{(\zeta)} \rightarrow 0$ in a controlled way. Thus we can obtain solutions with parametrically small $R_B$ in this setup.

\item Interestingly, there also exist solutions where $R_B$ stays finite for $R^{(\zeta)} \rightarrow 0$. In this case there is no solution with $R^{(\zeta)} = 0$ exactly. The reason is that the `would-be solution' with $R^{(\zeta)}= 0$ can be shown to be a solution with finite $R^{(\zeta)}$ associated with a different UV fixed point. This effect is also observed for holographic RG flows in absence of a brane \cite{Ghosh:2017big,Gursoy:2018umf}.

\item Last, for solutions with $R^{(\zeta)} \rightarrow \infty$,  the brane curvature does not diverge, but rather asymptotes to a finite value.

\item However, for the models studied here, the typical scale of the brane curvature is the 4d Planck scale $M_4$ on the brane, i.e.~$R_B \sim M_4^2$. Only on the branch connected to the flat brane solution we can achieve $R_B \ll M_4^2$ by tuning $R^{(\zeta)}$ to a sufficiently small value.
\end{itemize}

These findings make the bulk-brane setup studied here a promising framework for further phenomenological investigation. However, before a realistic model can be constructed,  there are several open questions that should be addressed. For one, it is important to determine to what extent our results above are general, or model-dependent artifacts. In particular, is it a generic feature of this construction that $R_B \ll M_4^2$ \emph{only} occurs when perturbing a solution for a flat brane?

In addition, more work is also needed regarding the theoretical foundations of the model. It would be desirable if the quantities $V(\f)$, $W_B(\f)$ and $U(\f)$ could be constrained, either by direct calculation or from physical principles. For example, it is expected that consistency with quantum gravity gives stringent constraints on the physics of scalar fields \cite{Vafa:2005ui, Ooguri:2006in, 1610.00010, 1705.04328, 1708.06761, 1802.08264, Obied:2018sgi}, restricting their field range and even constraining the shape of the potential. It would be interesting to study to what extent these conditions, also known as `swampland conjectures', can be used to constrain this model.

Another important question regards the stability of the solutions obtained here. The perturbative stability of self-stabilising solutions for a \emph{flat} brane was analysed in \cite{1704.05075}. The result of this analysis is a set of five sufficient conditions for stability involving the bulk solutions and the brane quantities evaluated at the position of the brane. Here, we expect that the presence of \emph{positive} brane curvature will not introduce any additional instabilities (see \cite{Ghosh:2017big} for more details). Hence we expect any solutions, which can be obtained from a stable flat brane solution by turning on positive (boundary and brane) curvature to be stable. On the other hand, the presence of \emph{negative} (boundary and brane) curvature may introduce new instabilities.  In this case, perturbative stability has to be checked explicitly case by case, which goes beyond the scope of this work. As our priority in this work was to explore the space of self-stabilising solutions rather than to perform realistic model-building, the solutions explored in this work do not always satisfy all of the sufficient conditions for stability. For example, the solutions in sec.~\ref{sec:numexpV} satisfy all but one of the sufficient stability conditions. This does not imply that these solutions are necessarily unstable, but a more detailed analysis is required.

Last, some of our findings may have interesting applications in the study of holographic RG flows. In particular, certain bulk solutions studied in the theory with $V(\f) \sim - \exp(\f)$ should be relevant for the study of the RG behaviour of confining theories \cite{0707.1349} on curved backgrounds in holography. We plan to report on this in a separate work in the future.

\section*{Acknowledgements}\label{ACKNOWL}
\addcontentsline{toc}{section}{Acknowledgements}

We thank Christos Charmousis, Jerome Gauntlett, Dieter L\"ust, Hiroshi
Ooguri, Ioannis Papadimitriou, Cumrun Vafa for discussions and comments.

\noindent This work was supported in part  by the Advanced ERC grant SM-grav, No 669288.

\newpage
\appendix
\renewcommand{\theequation}{\thesection.\arabic{equation}}
\addcontentsline{toc}{section}{Appendix\label{app}}
\section*{Appendix}

\section{Junction conditions for curved brane embeddings in a
  flat-sliced bulk} \label{Israel2}

Here we  consider a   brane
describing a curve  $u=u_\star(\tau)$, which constitutes the interface
between  solutions of the form
\be \label{flatmapp}
ds^2 =  du^2 + e^{2A(u)}\eta_{\mu\nu}dx^\mu dx^\nu, \quad \f= \f(u)
\ee
in which the
scale factor and scalar field profile are, a priori,  different on each side of
the interface,
\be\label{flat1}
(A, \f) = \left\{\begin{array}{ll} \left(A_-(u), \f_-(u)\right) &
    \quad u <
      u_\star(\tau) \\  & \\ \left( A_+(u), \f_+(u)\right) & \quad u > u_\star(\tau) \end{array}\right.
\ee
The connection across the brane is specified by Israel's junction
conditions:
\begin{enumerate}
\item The metric and scalar field are continuous:
\be\label{FE3app}
\Big[g_{ab}\Big]^{UV}_{IR} = 0,   \qquad \Big[\f\Big]^{IR}_{UV} =0
\ee
\item The extrinsic curvature and normal derivative of $\f$ are discontinuous:
\be\label{FE4app}
\Big[K_{\mu\nu} - \gamma_{\mu\nu} K \Big]^{IR}_{UV} =   {1\over \sqrt{-\gamma}}{\delta S_{brane} \over \delta \gamma^{\mu\nu}}  ,  \qquad \Big[n^a\de_a \f\Big]^{IR}_{UV} =- {1\over \sqrt{-\gamma}}{\delta S_{brane} \over \delta \f} ,
\ee
where $\gamma_{\mu\nu}=e^{2 A(u)}\zeta_{\mu\nu}$ is the induced metric, $K_{\mu\nu}$ is the extrinsic curvature of the brane with trace $K = \gamma^{\mu\nu}K_{\mu\nu}$, and $n^a$ a unit vector normal to the brane with orientation towards the IR.
\end{enumerate}

 The first of these  conditions, the continuity of the
metric and scalar field across the interface, implies
\be\label{flat2app}
A_-(u_\star(\tau)) = A_+(u_\star(\tau)), \qquad \f_-(u_\star(\tau)) = \f_+(u_\star(\tau)).
\ee

If $u_\star(\tau)$ is a non-trivial function, equation
(\ref{flat2app}) implies the identity of the functions $A_-(u)$ and
$A_+(u)$,  and of  $\f_-(u)$ and $\f_+(u)$, over a continuous set of
values. Since in the bulk these functions satisfy a system of
 {\em ordinary} differential equations, this implies that the
 solutions on each side must coincide,
\be
A_-(u) = A_+(u), \quad \f_-(u) = \f_+(u), \qquad \forall \, u.
\ee
Therefore,  not only $A$ and $\f$ but
 also their derivatives must be continuous. Then, the second junction
 conditions require
\be \label{flat3app}
{\delta S_{brane} \over \delta \gamma^{\mu\nu}} = 0, \qquad {\delta
  S_{brane} \over \delta \f} =0.
\ee
In other words, the induced metric and the scalar on the brane must
satisfy their lower-dimensional field equations, as dictated by the
brane action alone. Recall however that the induced metric
$\gamma_{\mu\nu}$ and the brane scalar field $\f$ are not independent
quantities, but they are determined by the bulk metric and scalar
field, via the embedding function $u_\star(\tau)$: therefore, generically the
solution of equations (\ref{flat3app}) will be incompatible with the
bulk solution.

To illustrate this more explicitly, we write the induced metric
and scalar field  for a general embedding $u_*(t)$:
\be\label{ind1app}
ds^2_{ind} = \left[\left({d u_\star \over d\tau}\right)^2 - e^{2A(u_\star(\tau))}\right] d\tau^2 +
e^{2A(u_\star(\tau))} dx_i dx^i, \qquad \phi(\tau)= \f(u_\star(\tau))
\ee
where we have used a different notation $\phi(\tau)$ to denote the
induced scalar field. We can change coordinates on the brane to proper
time $\eta$, where the induced metric takes the canonical FRW form
\be\label{ind2app}
ds^2_{ind} = -d\eta^2 +a^2(\eta)  dx_i dx^i,  \qquad a(\eta) \equiv
e^{A(u_\star(\eta))}.
\ee

Because of (\ref{flat3app}), the induced scale factor $a(\eta)$  and  scalar
field $\phi(\eta)$ must satisfy the brane  Einstein-scalar equations,
whose solution is determined purely by the brane potentials without
reference to the bulk.

Given a  solution $(a(\eta), \phi(\eta))$  of the brane Einstein's equations and knowing the bulk geometry $A(u)$ we can
determine the embedding $u_\star(\eta)$  by inverting the implicit relation
\be\label{avsa}
A(u_\star(\eta)) = \log a(\eta)
\ee
Having found $u_\star(\eta)$ we can go back to the bulk time coordinate
$\tau$ by solving the differential equation
\be \label{tvstau}
{d \eta \over d\tau }  = {a(\eta) \over \left[1+ \left({du_\star \over
        d\eta}\right)^2\right]^{1/2}},
\ee
which follows from the change of coordinates between (\ref{ind1app}) and
(\ref{ind2app}).

The embedding $u_\star(\eta)$ must be such that, at the same time as
(\ref{avsa}), one  must also
satisfy the relation
\be \label{phivsphi}
\f(u_\star(\eta)) = \phi(\eta)
\ee
On the other hand, the functions $A(u)$ and $\f(u)$ are
determined by the {\em bulk} Einstein equations, which generically
know nothing about the brane potentials.  Therefore, if we determine
$u_\star(\tau)$ from knowledge of $a(\eta)$ and $A(u)$ as explained above, generically the relation
(\ref{phivsphi}) will {\em not} hold, and we are forced to conclude
that the ansatz we started from does not lead to a solution of the
full system.

The argument above assumes generic (and unrelated) bulk and brane
potentials. However, if we abandon genericity, it may be possible to
tune the model such that equations \eqref{avsa}--\eqref{phivsphi} are
indeed compatible, and a solution exists. This leads to the curious
case which we call an {\em evanescent brane}, i.e.~an exact solution of the
bulk-brane system in which
the brane has no backreaction on the bulk.

\subsection{Evanescent branes}
\label{app:evanescent}

As we have seen in the previous discussion, embedding  a non-trivial
brane  trajectory in a flat-slicing is possible if the induced
quantities on the brane satisfy their lower-dimensional field
equations governed by the brane potentials. If that is the case, the
bulk is smooth across the brane, and the interface is transparent (or
invisible), although all bulk equations and junction conditions are
exactly satisfied: curiously, we have  a fully backreacted system where the
backreaction is exactly vanishing.

A simple example of such a situation is given by a bulk solution which is
Poincar\'e-AdS with constant scalar field (realised e.g.~at an extremum of
$V(\f)$, say at  $\f=0$),
\be \label{ghost1}
A(u) = -{u \over \ell}, \quad \f(u) = 0,
\ee
and a brane action of the form (\ref{sbrane}) with constant $U$ and
$Z$ and  a potential
$W_B(\f)$  such that it {\em also} has an extremum at $\f=0$, with
$W_B(0) >0$.  In this case, the brane field equations (\ref{flat3app})
admit a de Sitter solution with constant scalar $\phi=0$ and
Hubble constant $H = \sqrt{W_B(0)}/M_p^2$, where $M_p^2 = M^3 U$,
\be\label{ghost2}
a(\eta) =  e^{H\eta}, \quad \phi(\eta)=0.
\ee
Comparing equations (\ref{ghost1}) and (\ref{ghost2}) we can read-off
the trajectory using
equation (\ref{avsa}),
\be
u_\star(\eta) = -\ell H \eta.
\ee
Equation   (\ref{tvstau}) becomes
\be
{d \eta \over d\tau } = {e^{H\eta} \over \left[1+ H^2 \ell^2
  \right]^{1/2}},
\ee
 and by  integrating it we can find the trajectory in the
 original bulk coordinates,
\be
u_\star(\tau) = \ell \log\left[- {H \over \left(1+ H^2 \ell^2\right)^{1/2}}
  \, \tau\right], \qquad -\infty< \tau< 0.
\ee
From the brane point of view, $\tau$ is the de Sitter conformal time.  Finally, and crucially, $\f(u(\tau)) = \phi(\tau)$ since both sides
vanish identically, by equations (\ref{ghost1}-\ref{ghost2}). Therefore, we have an exact solution  of the full
system, including the junction conditions. This was possible because
we have tuned the brane theory such that an  extremum   of the brane
potential coincides with an extremum of the bulk potential. It is
likely that similar examples can be constructed with a non-trivial
bulk scalar field profile, e.g. by appropriate combinations of bulk
and brane exponential potentials.

We stress that in these solutions the bulk does not detect at all the
presence of the brane: the bulk AdS solution would be the same were
the brane absent. What we have here is a non-trivial generalization of
the fact that, if the world-volume action has only a potential term,
then a tensionless brane produces no backreaction. In our case
instead, we have a non-vanishing tension, but induced kinetic terms
for gravity and the scalar. The corresponding statement is that a
brane satisfying its own world-volume Einstein equation behaves (from
the point of view of the bulk) as if it were tensionless.

\section{Perturbative analysis near the maximum of the potential}  \label{app:UV}
Here we record expressions for the functions $W$, $S$ and $T$ defined in \eqref{defW}--\eqref{defT} in the vicinity of a maximum of the potential. Without a loss of generality, we can consider the maximum to be located at $\f=0$ and near this maximum the potential can be written as
\begin{equation}
 V(\f)= -\frac{d(d-1)}{\ell^2} +\frac{m^2}{2}\f^2+\mathcal{O}(\f^3)
 \end{equation}
where $m^2<0$. We can now solve eqs.~\eqref{eq:EOM7}--\eqref{eq:EOM8} in a series expansion in $\f$. There exist two types of solutions which are distinguished by the subscripts $(+)$ and $(-)$, respectively.  The $(-)$ solutions are:
 \begin{align}
\nonumber W_{-}(\f) & = \frac{1}{\ell} \left[2(d-1) + \frac{\Delta_-}{2} \f^2 + \mathcal{O}(\f^3) \right] + \frac{\mathcal{R}}{d \ell} \, |\f|^{\frac{2}{\Delta_-}} \ [1+ \mathcal{O}(\f) + \mathcal{O}(\mathcal{R})] \\
\label{eq:Wmsol} & \hphantom{=} \,  + \frac{C}{\ell} \, |\f|^{\frac{d}{\Delta_-}} \ [1+ \mathcal{O}(\f)+ \mathcal{O}(C) + \mathcal{O}(\mathcal{R})] \, , \\
 \nonumber \\
\label{eq:Smsol} S_{-}(\f) & = \frac{\Delta_-}{\ell} \f \ [1+ \mathcal{O}(\f)] + \frac{Cd}{\Delta_- \ell} \, |\f|^{\frac{d}{\Delta_-}-1} \ [1+ \mathcal{O}(\f) + \mathcal{O}(C)] \, , \\
\nonumber & \hphantom{=} \,  + \frac{1}{\ell} \mathcal{O}\left( \mathcal{R} |\f|^{\frac{2}{\Delta_-}+1} \right) + \frac{1}{\ell} \mathcal{O}\left(\mathcal{R} C |\f|^{\frac{2+d}{\Delta_-}-1} \right) \\
\nonumber \\
\label{eq:Tmsol} T_{-}(\f) &= \ell^{-2} \, \mathcal{R} \, |\f|^{\frac{2}{\Delta_-}} [1+ \mathcal{O}(\f) + \mathcal{O}(C) + \mathcal{O}(\mathcal{R})] \, ,
\end{align}
where $C$ and ${\mathcal R}$ are two integration constants and
\be
\label{eq:Deltadef} \Delta_{\pm}  = \frac{1}{2}\left( d \pm \sqrt{d^2+  4 m^2 \ell^2} \right) \, \qquad \textrm{with} \quad  -\frac{d^2}{4 \ell^2} <m^2 < 0\, .
\ee
The mass bound is precisely the BF bound for a scalar in $d$ dimensions. On the other hand, the $(+)$ solutions are:
\begin{align}
\label{eq:Wpsol} W_{+}(\f) & = \frac{1}{\ell} \left[2(d-1) + \frac{\Delta_+}{2} \f^2 + \mathcal{O}(\f^3) \right] + \frac{\mathcal{R}}{d \ell} \,
|\f|^{\frac{2}{\Delta_+}} \ [1+ \mathcal{O}(\f) +
\mathcal{O}(\mathcal{R})] \, , \\
\nonumber \\
\label{eq:Spsol} S_{+}(\f) & = \frac{\Delta_+}{\ell} \f \ [1+
\mathcal{O}(\f)] + \mathcal{O}\left( \mathcal{R}
  |\f_-|^{\frac{2}{\Delta_+}+1} \right) \, ,\\
\nonumber \\
\label{eq:Tpsol} T_{+}(\f) &=\ell^{-2} \, \mathcal{R} \, |\f|^{\frac{2}{\Delta_+}} [1+ \mathcal{O}(\f) + \mathcal{O}(\mathcal{R})] \, .
\end{align}
For the $(-)$ solutions both integration constants $\mathcal{R}$ and $C$ appear, but for the $(+)$ solutions $\mathcal{R}$ is the only integration constant.

The solutions $(W, S, T)$ consist of two parts. One is an analytic series in $\f$ and the other consists of non-analytic expansions in $\f$ also containing the integration constants. In fact we can write the $W_-$ solution as a triple expansion as
 \begin{align}
W_{-}(\f) = \frac{1}{\ell} \sum_{l=0}^{\infty} \sum_{m=0}^{\infty} \sum_{n=0}^{\infty} A_{l,m,n} \left(C \, |\f|^{d/\Delta_-} \right)^l \, \left(\mathcal{R} \, |\f|^{2/\Delta_-} \right)^m \, \f^n .
\end{align}
On the other hand, the $W_+$ solution can be written as a double expansion, which is schematically given by
 \begin{align}
W_{+}(\f) = \frac{1}{\ell} \sum_{m=0}^{\infty} \sum_{n=0}^{\infty} A_{m,n}  \, \left(\mathcal{R} \, |\f|^{2/\Delta_-} \right)^m \, \f^n .
\end{align}
Given the expressions of the functions $W(\f)$ and $S(\f)$, we can use the definitions \eqref{defW} and \eqref{defS} to find the scale factor $A(u)$ and the scalar field profile $\f(u)$. These are recorded below. For the $(-)$ solutions they are:
\begin{align}
\label{eq:phimsol} \f(u) &= \f_- \ell^{\Delta_-}e^{\Delta_-u / \ell} \left[ 1+ \mathcal{O} \left(\mathcal{R} |\f_-|^{2/\Delta_-} e^{2u/\ell} \right) + \ldots \right] \\
\nonumber & \hphantom{=} \ + \frac{C d \, |\f_-|^{\Delta_+ / \Delta_-}}{\Delta_-(d-2 \Delta_-)} \, \ell^{\Delta_+} e^{\Delta_+ u \ell} \left[ 1+ \mathcal{O} \left(\mathcal{R}|\f_-|^{2/\Delta_-} e^{2u/\ell} \right) + \ldots \right] + \ldots \, , \\
\label{eq:Amsol} A(u) &= \bar{A}_- -\frac{u}{\ell} - \frac{\f_-^2 \, \ell^{2 \Delta_-}}{8(d-1)} e^{2\Delta_- u / \ell}  -\frac{\mathcal{R}|\f_-|^{2/\Delta_-} \, \ell^2}{4d(d-1)} e^{2u/\ell} \\
\nonumber & \hphantom{=} \ - \frac{\Delta_+ C |\f_-|^{d/\Delta_-} \, \ell^d}{d(d-1)(d-2 \Delta_-)}e^{du/\ell} +\ldots \, .
\end{align}
where $\f_-$ and $\bar{A}_-$ are integration constants.  For the $(+)$ solutions one obtains:
\begin{align}
\label{eq:phipsol} \f(u) &= \f_+ \ell^{\Delta_+}e^{\Delta_+ u / \ell} \left[ 1+ \mathcal{O} \left(\mathcal{R} |\f_+|^{2/\Delta_+} e^{2u/\ell} \right) + \ldots \right] + \ldots \, , \\
\label{eq:Apsol} A(u) &= \bar{A}_+ -\frac{u}{\ell} - \frac{\f_+^2 \, \ell^{2 \Delta_+}}{8(d-1)} e^{2\Delta_+ u / \ell}  -\frac{\mathcal{R}|\f_+|^{2/\Delta_+} \, \ell^2}{4d(d-1)} e^{2u/\ell} +\ldots \, ,
\end{align}
with integration constants $\f_+$ and $\bar{A}_+$.

As we are solving the equations of motion close to $\f=0$, from eqs.~\eqref{eq:phimsol} and \eqref{eq:phipsol} we can see that these solutions are valid near $u\to-\infty$. Choosing $A_{\pm}=0$, which can be done by a redefinition of the boundary coordinates, is equivalent to the choice $\zeta_{\mu \nu} = \zeta_{\mu \nu}^{UV}$ made in \eqref{fid}.

The integration constant $\f_-$ is interpreted as the source for the operator $\mathcal{O}$ of the boundary QFT. The dimension of this operator is $\Delta_+$ and it is related to the mass parameter $m$ by Eq. \eqref{eq:Deltadef}. The integration constant $C$ is related to the vacuum expectation value of the scalar operator and it is given by
\begin{align}
\langle \mathcal{O} \rangle = \frac{Cd}{\Delta_-} \, |\f_-|^{\Delta_+ / \Delta_-} \, .
\end{align}
For the $(+)$ solution, the source vanishes. The flow corresponding to the $(+)$ solution is thus driven purely by the vev. In this case the vev of the operator $\mathcal{O}$ is related to the integration constant $\f_+$ by
\begin{align}
\langle \mathcal{O} \rangle_+ = (2\Delta_+-d) \, \f_+ \, .
\end{align}
The integration constant $\mathcal{R}$ is dimensionless. This is the dimensionless combination of the parameters of the theory namely source $\f_-$ and the boundary curvature $R^{UV}$. More precisely the relation is
\begin{equation}
\mathcal{R}=R^{(\zeta)} |\f_{\mp}|^{-2/\Delta_{\mp}}.
\end{equation}
To conclude this section, we see that the integration constants are
related to the different parameters of the boundary field theory. We
also see that the bulk geometry asymptotes to   AdS$_{d+1}$ near the
maximum of the potential. From the boundary QFT point of view, this
corresponds to the  UV fixed point of the RG flows.

\section{Regular IR geometries} \label{app:IR}
After analyzing the solution near the maximum of the potential which corresponds to the UV fixed point of the RG flows of the boundary field theory, we now analyze the regular solutions in the interior. We are interested in how the scale factor $A(u)$ can obtain its minimum value. For the positively curved case, this will correspond to the regular IR end points. On the other hand, for the negatively curved case, this will correspond to a turning point. In the course of obtaining the solution, a third possibility also arises which we call a bounce and where the flow reverses direction in $\f$.

We want to analyze the solution near $\f=\f_0$ where $S(\f_0)=0$. As explained in \cite{Ghosh:2017big}, in the vicinity of $\f_0$ we can expand the functions $W$, $S$ and $T$ in powers of the square root of $x=\f_0-\f$:
\begin{align}
S(x)&=\sqrt{x}\left(S_0+S_1\sqrt{x}+\cdots \right), \label{expansionS}\\
W(x)&=\frac{1}{\sqrt{x}}\left(W_0+W_1\sqrt{x}+\cdots \right), \label{expansionW}\\
T(x)&=\frac{1}{x}\left(T_0+T_1\sqrt{x}+\cdots \right). \label{expansionT}
\end{align}
As $\f_0$ is a generic point the potential $V(\f)$ has a regular series expansion:
\begin{equation}
V(\f)= V_0+V_1 x+ V_2 x^2 +\cdots
\end{equation}
The unknown coefficients can be determined by plugging expansions \eqref{expansionS} and \eqref{expansionW} into eqs.~\eqref{eq:EOM5}, \eqref{eq:EOM7} and \eqref{eq:EOM8}. We find three classes of solutions.

\subsubsection*{Case (a): IR endpoints}
In this case the leading order terms in the functions $S(\f),W(\f)$ and $T(\f)$ are:
\begin{align}
S(\f)&\approx S_0 \sqrt{\f_0-\f} , \label{expansionSa}\\
S(\f)&\approx \frac{W_0}{ \sqrt{\f_0-\f}}, \label{expansionWa}\\
T(\f)&\approx \frac{W_0}{ \f_0-\f}. \label{expansionTa}
\end{align}
where the coefficients are given by:
\be\label{coeff}
S_0^2=\frac{2 V_1}{d+1}, \qquad W_0= (d-1) S_0, \qquad T_0=\frac{d(d-1)
}{4(d+1)}\, S_0^2 \ .
\ee
Note that the function $T(\f)$ diverges when $\f\to\f_0$. From the definition of $T=R^{(\zeta)} e^{-2 A(u)}$, it means that the scale factor, $A(u)$, is shrinking to zero. As the value of $T_0$ is positive for $d>1$,  this case corresponds to the IR end point of the corresponding RG flow in positively curved space.

\subsubsection*{Case (b): AdS throat}
In this case, the leading order terms of the functions $S,W$ and $T$ are:
\begin{align}
S(\f)&\simeq S_0 \sqrt{\f_0-\f}+\cdots , \\
W(\f)&\simeq W_0 \sqrt{\f_0-\f}+\cdots , \\
T(\f)&\simeq T_2+T_3 \sqrt{\f_0-\f}+\cdots,
\end{align}
where \bea
&& S_0^2 = 2V_1, \quad  W_2 = {4V_0\over d S_0}, \quad  T_2 = V_0.
\eea
In this case, both the functions $S(\f)$ and $W(\f)$ are going to zero as $\f\to\f_0$. This means that  both the scalar field and the scale factor reach an extremum. As $\ddot{A}=W' S>0$ when $\f\to\f_0$, the scale  factor is attaining its minimum value at this point.  On the other hand, the function $T(\f)$ obtains a constant negative value. Hence, this case corresponds to the turning point of the scale factor for the negatively curved space.
\subsubsection*{Case (c): bouncing points}
In the third case the leading order terms of the functions $S,W$ and $T$ are:
\begin{align}
S(\f)&\simeq S_0 \sqrt{\f_0-\f}+\cdots , \\
W(\f)&\simeq W_1+W_2 \sqrt{\f_0-\f}+\cdots , \\
T(\f)&\simeq T_2+T_3 \sqrt{\f_0-\f}+\cdots,
\end{align}
where
\be\label{bounce2}
S_0^2 = 2V_1, \quad  T_2 = V_0 + {d W_1^2  \over 4(d-1)},
\ee
and in this case $W_1$ is arbitrary. The function $S(\f)$ goes to zero when $\f\to\f_0$. This means that the scalar field is obtaining its extremum value. As $\ddot{\f}=SS' \sim V_1$, the scalar field can obtain its maximum or minimum value depending on the sign of $V_1$.  On the other hand the function $W(\f)$ obtains a finite value $W_1$ when $\f\to\f_0$ which indicates that the scale factor is not attaining its minimum value at this point. This means that the scalar field changes its direction and continues to flow. This refers to the bouncing point and the bouncing flows are expected for both the positively and the negatively curved space.

\section{Perturbative solution of the junction conditions near a
  flat equilibrium point}

 We  assume that the equilibrium brane position $\f_\star$ has the following curvature expansion:
 \begin{equation}
 \f_\star=\f_{\star,0}+R^{(\zeta)} \f_{\star,1}+\cdots
 \end{equation}
where $\f_{\star,0}$ is the flat brane position and $R^{(\zeta)} \f_{\star,1}$ is the leading-order curvature correction. To remove clutter, we will drop the superscript on $R^{(\zeta)}$ in the following and simply write $R$. Then we can solve eq.~\eqref{JC} order by order in $R$. The functions $W,S$ and $T$ have the following curvature expansion:
 \begin{align}
 W(\f)&=W_0(\f)+R W_1(\f)+\cdots \\
 S(\f)&=S_0(\f)+R S_1(\f)+\cdots \\
 T(\f)&=R T_0(\f)+\cdots
 \end{align}
\textbf{Zeroth order}\\
 The junction condition at zeroth order in $R$ is given by
 \begin{equation}
\left[  -2 Q^2 W_0 W_B +Q^2 W_B^2+2 W_0' W_B'-(W_B')^2\right]_{\f_{\star,0}}=0 \, , \label{JCflat}
  \end{equation}
 where we have used $S_0=W_0'$. Note that the function $U(\f)$ does not appear in the zeroth order equation. This is because in eq.~\eqref{JC} the term with the function $U(\f)$ is first order in $R$. The solution to eq.~\eqref{JCflat} will determine the brane position in the case of flat slicing. \\ \ \\
  \textbf{Linear order}\\
Equating the coefficients of $R$ in eq.~\eqref{JC} we find
  \begin{align}
& \left[ Q^2 W_1 W_B+Q^2 \f_{\star,1} W_B W_0'+T_0\left\{\frac{2-d}{d}Q^2 U(W_0-W_B)+U'(S_0-W_B') \right\} \right. \nonumber \\
& -S_1 W_B' +Q^2 \f_{\star,1} W_0W_B' -Q^2 \f_{\star,1} W_B W_B' -\f_{\star,1} S_0' W_B' -\f_{\star,1} S_0 W_B'' \nonumber \\
& \left. +\f_{\star,1} W_B' W_B''\right]_{\f_{\star,0}}=0 \ .
 \end{align}
 This equation is linear in $\f_{\star,1}$ and can be easily inverted to get an expression for $\f_{\star,1}$. This is recorded below
 \begin{equation}
\f_{\star,1}= \frac{\frac{(2-d)Q^2}{d}\left(T_0 U W_0 -T_0 U W_B \right)+Q^2 W_1 W_B+S_0 T_0 U'-S_1W_B'-T_0 U' W_B'  }{-Q^2\left(W_B W_0' +W_0 W_B' -W_B W_B'  \right) +S_0' W_B' +S_0 W_B''-W_B' W_B'' }\  \vline_{\f_{\star,0}} \ .
 \end{equation}
 This is the first leading curvature correction to the equilibrium brane position $\f_\star$.
\addcontentsline{toc}{section}{References}
 \bibliography{references.bib}

 \bibliographystyle{JHEP}

\end{document}